\documentclass[fleqn,usenatbib]{mnras}

\usepackage{newtxtext,newtxmath}
\usepackage{pdflscape}

\usepackage[T1]{fontenc}

\DeclareRobustCommand{\VAN}[3]{#2}
\let\VANthebibliography\thebibliography
\def\thebibliography{\DeclareRobustCommand{\VAN}[3]{##3}\VANthebibliography}
\usepackage[english]{babel}
\usepackage{hyperref}
\addto\extrasenglish{
    
}
\addto\extrasenglish{
    
}

\usepackage{graphicx}	
\usepackage{amsmath}	

\title[The multi-planet system TOI-4311]{An Ultra-Short Period Super-Earth and a Sub-Neptune Orbiting the K dwarf TOI-4311\thanks{This article uses data from the CHEOPS Guaranteed Time Observation programme CH\_PR120054 and CH\_PR140065. }}

\author[Y. N. E. Eschen et al.]{
Yoshi~Nike~Emilia~Eschen$^{1}$\,$^{\href{https://orcid.org/0009-0006-6397-2503}{\protect\includegraphics[height=0.19cm]{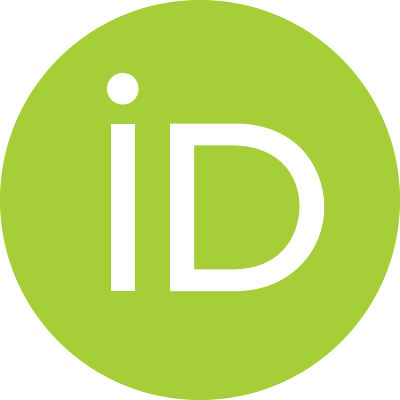}}}$, 
Thomas~G.~Wilson$^{1}$\,$^{\href{https://orcid.org/0000-0001-8749-1962}{\protect\includegraphics[height=0.19cm]{figures/orcid.pdf}}}$, 
Andrew~Collier~Cameron$^{2}$\,$^{\href{https://orcid.org/0000-0002-8863-7828}{\protect\includegraphics[height=0.19cm]{figures/orcid.pdf}}}$, \newauthor
Alexander~James~Mustill$^{3}$\,$^{\href{https://orcid.org/0000-0002-2086-3642}{\protect\includegraphics[height=0.19cm]{figures/orcid.pdf}}}$, 
Jo~Ann~Egger$^{4}$\,$^{\href{https://orcid.org/0000-0003-1628-4231}{\protect\includegraphics[height=0.19cm]{figures/orcid.pdf}}}$, 
Solène~Ulmer-Moll$^{5,6}$\,$^{\href{https://orcid.org/0000-0003-2417-7006}{\protect\includegraphics[height=0.19cm]{figures/orcid.pdf}}}$, 
Davide~Gandolfi$^{7}$\,$^{\href{https://orcid.org/0000-0001-8627-9628}{\protect\includegraphics[height=0.19cm]{figures/orcid.pdf}}}$, \newauthor
Alexis~M.~S.~Smith$^{8}$\,$^{\href{https://orcid.org/0000-0002-2386-4341}{\protect\includegraphics[height=0.19cm]{figures/orcid.pdf}}}$, 
Olivier~D.~S.~Demangeon$^{9,10}$\,$^{\href{https://orcid.org/0000-0001-7918-0355}{\protect\includegraphics[height=0.19cm]{figures/orcid.pdf}}}$, 
Sérgio~G.~Sousa$^{9,10}$\,$^{\href{https://orcid.org/0000-0001-9047-2965}{\protect\includegraphics[height=0.19cm]{figures/orcid.pdf}}}$, 
Andrea~Bonfanti$^{11}$\,$^{\href{https://orcid.org/0000-0002-1916-5935}{\protect\includegraphics[height=0.19cm]{figures/orcid.pdf}}}$, \newauthor
Alexandre~C.~M.~Correia$^{12}$\,$^{\href{https://orcid.org/0000-0002-8946-8579}{\protect\includegraphics[height=0.19cm]{figures/orcid.pdf}}}$, 
Vardan~Adibekyan$^{9}$\,$^{\href{https://orcid.org/0000-0002-0601-6199}{\protect\includegraphics[height=0.19cm]{figures/orcid.pdf}}}$, 
Gaia~Lacedelli$^{13,14}$\,$^{\href{https://orcid.org/0000-0002-4197-7374}{\protect\includegraphics[height=0.19cm]{figures/orcid.pdf}}}$, 
Alexis~Brandeker$^{15}$\,$^{\href{https://orcid.org/0000-0002-7201-7536}{\protect\includegraphics[height=0.19cm]{figures/orcid.pdf}}}$, \newauthor
Camilla~Pezzotti$^{6}$, 
Babatunde~Akinsanmi$^{16}$\,$^{\href{https://orcid.org/0000-0001-6519-1598}{\protect\includegraphics[height=0.19cm]{figures/orcid.pdf}}}$, 
Yann~Alibert$^{17,18}$\,$^{\href{https://orcid.org/0000-0002-4644-8818}{\protect\includegraphics[height=0.19cm]{figures/orcid.pdf}}}$, 
Roi~Alonso$^{13,14}$\,$^{\href{https://orcid.org/0000-0001-8462-8126}{\protect\includegraphics[height=0.19cm]{figures/orcid.pdf}}}$, 
Tamas~Bárczy$^{19}$\,$^{\href{https://orcid.org/0000-0002-7822-4413}{\protect\includegraphics[height=0.19cm]{figures/orcid.pdf}}}$, \newauthor
David~Barrado$^{20}$\,$^{\href{https://orcid.org/0000-0002-5971-9242}{\protect\includegraphics[height=0.19cm]{figures/orcid.pdf}}}$, 
Susana~C.~C.~Barros$^{9,10}$\,$^{\href{https://orcid.org/0000-0003-2434-3625}{\protect\includegraphics[height=0.19cm]{figures/orcid.pdf}}}$, 
Wolfgang~Baumjohann$^{11}$\,$^{\href{https://orcid.org/0000-0001-6271-0110}{\protect\includegraphics[height=0.19cm]{figures/orcid.pdf}}}$, 
Willy~Benz$^{18,17}$\,$^{\href{https://orcid.org/0000-0001-7896-6479}{\protect\includegraphics[height=0.19cm]{figures/orcid.pdf}}}$, \newauthor
Nicolas~Billot$^{16}$\,$^{\href{https://orcid.org/0000-0003-3429-3836}{\protect\includegraphics[height=0.19cm]{figures/orcid.pdf}}}$, 
Luca~Borsato$^{21}$\,$^{\href{https://orcid.org/0000-0003-0066-9268}{\protect\includegraphics[height=0.19cm]{figures/orcid.pdf}}}$, 
Christopher~Broeg$^{18,17}$\,$^{\href{https://orcid.org/0000-0001-5132-2614}{\protect\includegraphics[height=0.19cm]{figures/orcid.pdf}}}$, 
Amadeo~Castro-González$^{16}$, \newauthor
Szilard~Csizmadia$^{8}$\,$^{\href{https://orcid.org/0000-0001-6803-9698}{\protect\includegraphics[height=0.19cm]{figures/orcid.pdf}}}$, 
Patricio~E.~Cubillos$^{11,22}$, 
Melvyn~B.~Davies$^{23}$\,$^{\href{https://orcid.org/0000-0001-6080-1190}{\protect\includegraphics[height=0.19cm]{figures/orcid.pdf}}}$, 
Magali~Deleuil$^{24}$\,$^{\href{https://orcid.org/0000-0001-6036-0225}{\protect\includegraphics[height=0.19cm]{figures/orcid.pdf}}}$, \newauthor
Adrien~Deline$^{16}$, 
Brice-Olivier~Demory$^{17,25,18}$\,$^{\href{https://orcid.org/0000-0002-9355-5165}{\protect\includegraphics[height=0.19cm]{figures/orcid.pdf}}}$, 
Aliz~Derekas$^{26}$, 
Giuseppe~Di~Persio$^{27}$, 
Billy~Edwards$^{28}$, \newauthor
David~Ehrenreich$^{16,29}$\,$^{\href{https://orcid.org/0000-0001-9704-5405}{\protect\includegraphics[height=0.19cm]{figures/orcid.pdf}}}$, 
Anders~Erikson$^{8}$, 
Andrea~Fortier$^{18,17}$\,$^{\href{https://orcid.org/0000-0001-8450-3374}{\protect\includegraphics[height=0.19cm]{figures/orcid.pdf}}}$, 
Luca~Fossati$^{11}$\,$^{\href{https://orcid.org/0000-0003-4426-9530}{\protect\includegraphics[height=0.19cm]{figures/orcid.pdf}}}$, \newauthor
Malcolm~Fridlund$^{30,31}$\,$^{\href{https://orcid.org/0000-0002-0855-8426}{\protect\includegraphics[height=0.19cm]{figures/orcid.pdf}}}$, 
Kosmas~Gazeas$^{32}$\,$^{\href{https://orcid.org/0000-0002-8855-3923}{\protect\includegraphics[height=0.19cm]{figures/orcid.pdf}}}$, 
Michaël~Gillon$^{33}$\,$^{\href{https://orcid.org/0000-0003-1462-7739}{\protect\includegraphics[height=0.19cm]{figures/orcid.pdf}}}$, 
Manuel~Güdel$^{34}$, \newauthor
Maximilian~N.~Günther$^{4}$\,$^{\href{https://orcid.org/0000-0002-3164-9086}{\protect\includegraphics[height=0.19cm]{figures/orcid.pdf}}}$, 
Alexis~Heitzmann$^{16}$\,$^{\href{https://orcid.org/0000-0002-8091-7526}{\protect\includegraphics[height=0.19cm]{figures/orcid.pdf}}}$, 
Ch.~Helling$^{11,35}$, 
Kate~G.~Isaak$^{4}$\,$^{\href{https://orcid.org/0000-0001-8585-1717}{\protect\includegraphics[height=0.19cm]{figures/orcid.pdf}}}$, 
Jon~M.~~Jenkins$^{36}$, \newauthor
Tatiana~Keller$^{18,17}$, 
Laszlo~L.~Kiss$^{37,38}$, 
Daniel~Kitzmann$^{18,17}$, 
Judith~Korth$^{16}$\,$^{\href{https://orcid.org/0000-0002-0076-6239}{\protect\includegraphics[height=0.19cm]{figures/orcid.pdf}}}$, 
Kristine~W.~F.~Lam$^{8}$\,$^{\href{https://orcid.org/0000-0002-9910-6088}{\protect\includegraphics[height=0.19cm]{figures/orcid.pdf}}}$, \newauthor
Jacques~Laskar$^{39}$\,$^{\href{https://orcid.org/0000-0003-2634-789X}{\protect\includegraphics[height=0.19cm]{figures/orcid.pdf}}}$, 
Alain~Lecavelier~des~Etangs$^{40}$\,$^{\href{https://orcid.org/0000-0002-5637-5253}{\protect\includegraphics[height=0.19cm]{figures/orcid.pdf}}}$, 
Adrien~Leleu$^{16,18}$\,$^{\href{https://orcid.org/0000-0003-2051-7974}{\protect\includegraphics[height=0.19cm]{figures/orcid.pdf}}}$, 
Monika~Lendl$^{16}$\,$^{\href{https://orcid.org/0000-0001-9699-1459}{\protect\includegraphics[height=0.19cm]{figures/orcid.pdf}}}$, \newauthor
Demetrio~Magrin$^{21}$\,$^{\href{https://orcid.org/0000-0003-0312-313X}{\protect\includegraphics[height=0.19cm]{figures/orcid.pdf}}}$, 
Pierre~F.~L.~Maxted$^{41}$\,$^{\href{https://orcid.org/0000-0003-3794-1317}{\protect\includegraphics[height=0.19cm]{figures/orcid.pdf}}}$, 
Bruno~Merín$^{42}$\,$^{\href{https://orcid.org/0000-0002-8555-3012}{\protect\includegraphics[height=0.19cm]{figures/orcid.pdf}}}$, 
Christoph~Mordasini$^{18,17}$, \newauthor
Valerio~Nascimbeni$^{21}$\,$^{\href{https://orcid.org/0000-0001-9770-1214}{\protect\includegraphics[height=0.19cm]{figures/orcid.pdf}}}$, 
Göran~Olofsson$^{15}$\,$^{\href{https://orcid.org/0000-0003-3747-7120}{\protect\includegraphics[height=0.19cm]{figures/orcid.pdf}}}$, 
Hugh~P.~Osborn$^{17,43}$\,$^{\href{https://orcid.org/0000-0002-4047-4724}{\protect\includegraphics[height=0.19cm]{figures/orcid.pdf}}}$, 
Roland~Ottensamer$^{34}$, \newauthor
Isabella~Pagano$^{44}$\,$^{\href{https://orcid.org/0000-0001-9573-4928}{\protect\includegraphics[height=0.19cm]{figures/orcid.pdf}}}$, 
Enric~Pallé$^{13,14}$\,$^{\href{https://orcid.org/0000-0003-0987-1593}{\protect\includegraphics[height=0.19cm]{figures/orcid.pdf}}}$, 
Gisbert~Peter$^{8}$\,$^{\href{https://orcid.org/0000-0001-6101-2513}{\protect\includegraphics[height=0.19cm]{figures/orcid.pdf}}}$, 
Glen~Petitpas$^{45}$, 
Daniele~Piazza$^{46}$, \newauthor
Giampaolo~Piotto$^{21,47}$\,$^{\href{https://orcid.org/0000-0002-9937-6387}{\protect\includegraphics[height=0.19cm]{figures/orcid.pdf}}}$, 
Don~Pollacco$^{1}$, 
Didier~Queloz$^{43,48}$\,$^{\href{https://orcid.org/0000-0002-3012-0316}{\protect\includegraphics[height=0.19cm]{figures/orcid.pdf}}}$, 
Roberto~Ragazzoni$^{21,47}$\,$^{\href{https://orcid.org/0000-0002-7697-5555}{\protect\includegraphics[height=0.19cm]{figures/orcid.pdf}}}$, 
Nicola~Rando$^{4}$, \newauthor
Heike~Rauer$^{49,50}$\,$^{\href{https://orcid.org/0000-0002-6510-1828}{\protect\includegraphics[height=0.19cm]{figures/orcid.pdf}}}$, 
Ignasi~Ribas$^{51,52}$\,$^{\href{https://orcid.org/0000-0002-6689-0312}{\protect\includegraphics[height=0.19cm]{figures/orcid.pdf}}}$, 
Nuno~C.~Santos$^{9,10}$\,$^{\href{https://orcid.org/0000-0003-4422-2919}{\protect\includegraphics[height=0.19cm]{figures/orcid.pdf}}}$, 
Gaetano~Scandariato$^{44}$\,$^{\href{https://orcid.org/0000-0003-2029-0626}{\protect\includegraphics[height=0.19cm]{figures/orcid.pdf}}}$, \newauthor
Damien~Ségransan$^{16}$\,$^{\href{https://orcid.org/0000-0003-2355-8034}{\protect\includegraphics[height=0.19cm]{figures/orcid.pdf}}}$, 
Avi~Shporer$^{45}$, 
André~M.~Silva$^{9,10}$\,$^{\href{https://orcid.org/0000-0003-4920-738X}{\protect\includegraphics[height=0.19cm]{figures/orcid.pdf}}}$, 
Attila~E.~Simon$^{18,17}$\,$^{\href{https://orcid.org/0000-0001-9773-2600}{\protect\includegraphics[height=0.19cm]{figures/orcid.pdf}}}$, 
Richard~Southworth$^{53}$, \newauthor
Manu~Stalport$^{6,33}$, 
Sophia~Sulis$^{24}$\,$^{\href{https://orcid.org/0000-0001-8783-526X}{\protect\includegraphics[height=0.19cm]{figures/orcid.pdf}}}$, 
Gyula~M.~Szabó$^{26}$\,$^{\href{https://orcid.org/0000-0002-0606-7930}{\protect\includegraphics[height=0.19cm]{figures/orcid.pdf}}}$, 
Stéphane~Udry$^{16}$\,$^{\href{https://orcid.org/0000-0001-7576-6236}{\protect\includegraphics[height=0.19cm]{figures/orcid.pdf}}}$, 
Bernd~Ulmer$^{8}$, \newauthor
Valérie~Van~Grootel$^{6}$\,$^{\href{https://orcid.org/0000-0003-2144-4316}{\protect\includegraphics[height=0.19cm]{figures/orcid.pdf}}}$, 
Julia~Venturini$^{16}$\,$^{\href{https://orcid.org/0000-0001-9527-2903}{\protect\includegraphics[height=0.19cm]{figures/orcid.pdf}}}$, 
Eva~Villaver$^{13,14}$, 
Nicholas~A.~Walton$^{54}$\,$^{\href{https://orcid.org/0000-0003-3983-8778}{\protect\includegraphics[height=0.19cm]{figures/orcid.pdf}}}$, 
Sebastian~Wolf$^{18}$, \newauthor
Carl~Ziegler$^{55}$, and
Tiziano~Zingales$^{47,21}$\,$^{\href{https://orcid.org/0000-0001-6880-5356}{\protect\includegraphics[height=0.19cm]{figures/orcid.pdf}}}$
\\
Affiliations are listed at the end of the paper
}

\date{Accepted XXX. Received YYY; in original form ZZZ}

\pubyear{\the\year{}}

\begin{document}
\label{firstpage}
\pagerange{\pageref{firstpage}--\pageref{lastpage}}
\maketitle

\begin{abstract}
We report the discovery and characterisation of the multi-planetary system around TOI-4311, a K dwarf kinematically between the Galactic thick disk and Hercules stream. TOI-4311 hosts an ultra-short-period super-Earth (P$\sim$0.99\,d, $1.376\substack{+0.077\\-0.080}$\,R$_\oplus$) and a longer period sub-Neptune (P$\sim$15\,d, $2.47\substack{+0.12\\-0.11}$\,R$_\oplus$) that was first detected in the \textit{TESS} photometry. Using follow-up observations with \textit{CHEOPS} and HARPS, we refine the planetary radius of both planets, derive the mass of planet b ($4.5\substack{+1.5\\-1.4}$\,M$_\oplus$) and confirm the planetary nature of planet c. Intriguingly, a third periodic signal is clearly detected in our HARPS RVs that we cannot link to stellar activity. This signal could be attributed to a third planet (P$\sim$38\,d, Msin(i)=$26.4\substack{+6.3\\-6.8}$\,M$_\oplus$) in the system, however with the current photometric dataset we do not find a transit. Our dynamical analysis highlights that this potential outer planet would remain stable.
Using the precise radius and mass for TOI-4311\,b we model its interior structure and find that it is very dense given the host star's galactic kinematics and chemistry. Hence this system could challenge current formation theories and provide insights into planet formation across the galaxy.

\end{abstract}




\section{Introduction}
With the field of exoplanets rapidly growing since the detection of the first exoplanet orbiting a main sequence star, 51 Peg\,b \citep{Mayor_Queloz_51Peg_1995}, more than 6000 exoplanets have been detected to date \citep[][accessed on 27 December 2025]{Exoplanet_Archive}. Photometric space-based surveys have significantly added to the sample of known exoplanets to date. Especially the \textit{Kepler} mission \citep{Borucki_Kepler_2010} discovered many new exoplanets and provided insights into the diversity of these planets. One of the main findings of this mission was the detection of Super-Earths and Sub-Neptunes \citep{Batalha_Kepler_Candidates_2013}, planets that are not found in our own Solar System. However, occurrence studies have shown that they are very common \citep{Howard_Kepler_Planet_Occurrence_2012,Petigura_super_earths_plateau_2013}.  These two planet types are separated by the radius valley, which is a lack of planets between 1.5-2\,R$_\oplus$ \citep{Fulton_Radius_Valley_2017,Remo_Radius_Valley_2024}. This gap in the population can be explained by atmospheric mass loss \citep{VanEylen_Radius_Valley_2018}, photo-evaporation or core-powered mass loss \citep{Owen_Schlichting_Photoevaporation_2024} or by formation and evolution models \citep{Venturini_Radius_Valley_2020}. 

As small planets have been precisely characterised, their interior structures and compositions can be modelled, e.g. \citep{Zeng_Growth_Model_2019,Baumeister_ExoMDN_2023,Egger_TOI469_plaNETic_2024}. Since stars and planets are formed from the same material, their composition should be directly linked, which is also seen for the refractory elements of the proto-Sun and Earth \citep{Wang_Protosun_Abundances_2019}. Compositional links for exoplanets were explored and proposed by \citet{Plotnykov_Stellar_Link_SuperEarths_2020} and \citet{Adibekyan_compositional_link_2021, Adibekyan_compositional_link_2024} who identified a trend between stellar abundances and the planetary density showing that the planet density decreases with decreasing metallicity.  However, their sample of planets orbiting metal-poor stars was sparse. Hence this trend is still being investigated and small planets orbiting metal-poor stars are of great interest. In addition to this, two clusters of small planets were identified in this previous sample; the Super-Mercuries at high densities and Super-Earths having a lower density. This could give insights into different formation processes of the systems. If the planets had undergone collisional stripping the population density distribution is expected to be more continuous compared to formation-driven mechanisms that might result in a density gap  \citep{Adibekyan_compositional_link_2021}.

Metal-poor stars are mainly found in the thick disk of the Milky Way and hence old. This is due to massive stars (M$>$8M$_\odot$) exploding in type II supernova and enriching the metal-poor ISM in $\alpha$-elements \citep{Gondoin_chemical_abundances_history_milkyway_2024}. However, as low-mass stars later evolved into white dwarfs, type Ia supernovae enriched the ISM with iron-peak elements and hence more recently formed stars, i.e. stars of the thin disk, are found to be more metal-rich and with lower $\alpha$/Fe abundance ratios. Thus, to explore compositional links for planets orbiting metal-poor stars, planet-hosting stars of the thick disk provide great targets. 

In addition to stars of the thin and the thick disk of the Milky Way, stars in galactic streams can show different kinematics and chemistry. One of the most studied streams in the Milky Way is the Hercules stream. While stars within this stream are chemically not clearly distinguishable from stars of the thin disk, they show different kinematic behaviour. The origin and properties of this stream are still investigated \citep{Quillen_GALAH_Hercules_2018}. However, hunting for planets orbiting stars of this stream can provide interesting insights due to the lack of known exoplanets in stellar streams and the effect of stellar chemistry and kinematics on their formation and evolution.

We present the discovery of two transiting planets as well as a radial-velocity planet candidate orbiting TOI-4311, a star kinematically between the thick disk and Hercules stream. We present the data taken from \textit{TESS} \citep{Ricker_TESS_2015}, \textit{CHEOPS} \citep{Benz_CHEOPS_2021} and HARPS \citep{Pepe_HARPS_2000} to characterise these planets as well as additional ground-based observations in \autoref{sec:observations}, characterise the host star in \autoref{sec:stellar_charactersation} and model the planetary parameters in \autoref{sec:planet_fitting}. Finally, we discuss the signal of the third planet candidate, model the interior structure of the transiting inner planet and discuss the compositional links and galactic membership of the host star in \autoref{sec:discussion}.

\section{Observations}
\label{sec:observations}
TOI-4311 has been observed by space-based photometric instruments, \textit{TESS} and \textit{CHEOPS} and the ground-based radial velocity spectrograph HARPS which allow us to characterise the planets orbiting the star. Additionally, the target has been monitored by the ground-based photometric instruments, WASP \citep{Pollaco_WASP_2006} and ASAS-SN \citep{Shappee_ASAS_SN_2014,Kochanek_ASAS_SN_2017} and imaged by SOAR \citep{Tokovinin_SOAR_2018} to characterise the star.

\subsection{TESS}
The Transiting Exoplanet Survey Satellite \citep[\textit{TESS};][]{Ricker_TESS_2015} has been performing an all-sky survey searching for transiting exoplanets since 2018. In order to monitor stars all over the sky, \textit{TESS} initially split the Southern and Northern hemisphere into 13 sectors each. \textit{TESS}' observing strategy started with observing a sector for 27.4 days while data are downlinked every 13.7 days. This results in observational gaps in the data. Due to the shape of the sectors, they overlap at the Southern and Northern pole. This results in targets at the poles being continuously monitored for up to one year before \textit{TESS} moves to the opposite hemisphere. 

\textit{TESS} consists of four cameras, each containing 4 CCDs. From these, full frame images are obtained. In \textit{TESS}' Primary Mission (PM), the  cadence was at 30\,min, followed by 10\,min in the First Extended Mission (EM1) and 200\,s in the Second Extended Mission (EM2) and current Third Mission Extension (EM3). However, selected targets \citep{Stassun_TIC_2018, Stassun_TIC_2019} are additionally processed into shorter cadence data of 120\,s and later on also 20\,s. The downlinked data is prepared by the Science Processing Operations Centre \citep[SPOC;][]{Jenkins_TESS_SPOC_2016,Caldwell_TESS_SPOC_2020} into lightcurves. This follows the procedures previously adapted by the \textit{Kepler} mission \citep[][]{Jenkins_Kepler_Pipeline_2010}, resulting in lightcurves containing flux values from Simple Aperture Photometry \citep[SAP;][]{Twicken_Kepler_Pipeline_2010, Morris_Kepler_Data_Processing_Handbook_2020} as well as a detrended SAP flux using Co-trending Basis Vectors, the Pre-search Data Conditioning SAP \citep[PDCSAP;][]{Twicken_PDCSAP_2010,Smith_PDCSAP_Kepler_2012, Stumpe_PDCSAP_2012}. The latter contain less systematic trends due to the detrending instrumental artefacts as well as correcting for contamination from other sources and is hence used for further analysis. Additionally quality flags for all timestamps, the respective flux values and a noise metric, the combined differential photometric precision (CDPP) computed over 2\,hours \citep{Christiansen_CDPP_2012}, are reported.

\begin{figure*}
    \centering
    \includegraphics[width=\linewidth]{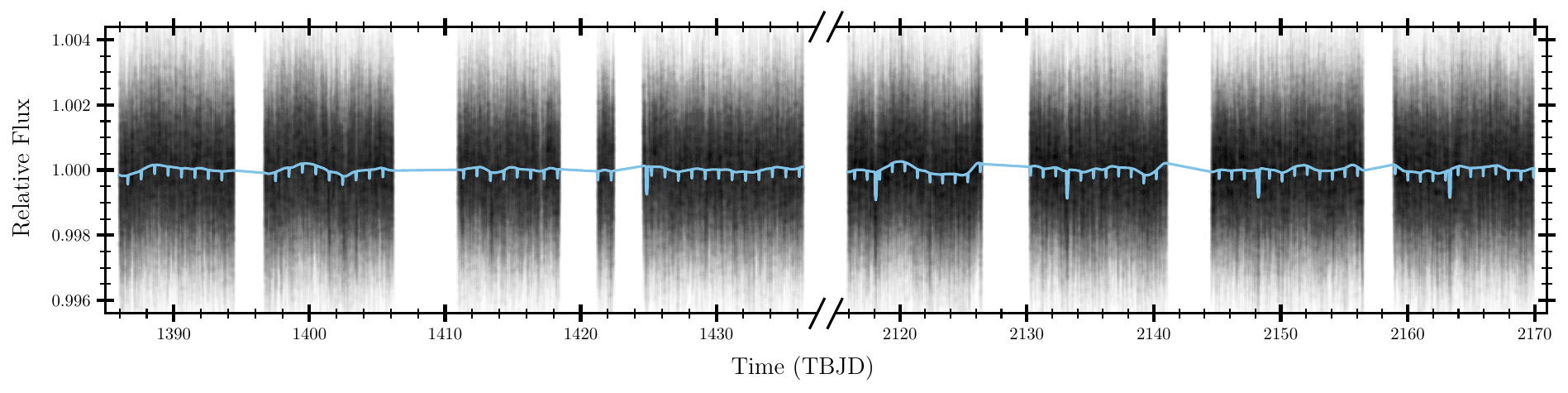}
    \caption{\textit{TESS} data of TOI-4311 observed in sectors 3, 4, 30 and 31 with the transit and GP model from \textsc{juliet} plotted in blue on top of the \textit{TESS} data in black.}
    \label{fig:tess_data_gp}
\end{figure*}

TOI-4311 was observed by \textit{TESS} in sectors 3, 4, 30 and 31. Sectors 3 and 4 were observed from 20 September 2018 to 14 November 2018 at a cadence of 1800\,s, while sectors 30 and 31 were observed as part of the First Extended Mission at a cadence of 600\,s from 23 September 2020 to 16 November 2020. Additionally, all sectors were processed at the shorter cadence of 120\,s. Using \textsc{tpfplotter} \citep{Aller_tpfplotter_2020} and \textsc{tess-cont} \citep{Castro_Gonzales_TOI5005_2024}, we check for contaminating sources in the aperture, but do not find any significant contamination by nearby stars. We summarise the available \textit{TESS} data in \autoref{tab:tess_obs} and show it in \autoref{fig:tess_data_gp}. 


The signatures of two planet candidates at orbital periods of 0.99 days and 15.1 days were detected by the SPOC in a search over these data \citep{Jenkins_Detectability_Transiting_Planets_2002, Jenkins_Kepler_Transit_Search_2010, Jenkins_Kepler_Transit_Search_2020} and the signatures passed all the data validation tests \citep{Twicken_Kepler_Validation_Vetting_2018, Li_Kepler_Validation_2019} including the difference image centroid analysis which constrained the location of the host star to within 7$\pm$5\," and 9$\pm$17\," of the transit source, respectively. The results were reviewed by the TESS Science Office and released on 28 and 19 July 2021 as TOI 4311.02 and TOI 4311.01, respectively \citep{Guerrero_TOI_Catalogue_2021}. From here onwards we will refer to them as TOI-4311\,b and TOI-4311\,c, respectively.

For our analysis we use the PDCSAP-flux from the 4 sectors that were observed with a cadence of 120\,s, we remove bad quality points (\texttt{QUALITY>0}) and those with Not-a-Number fluxes.

\begin{table*}
    \centering
    \caption{\textit{TESS} Observations of TOI-4311.}
    \begin{tabular}{ccccccccc}
        \hline
        \hline
        Sector & Camera & CCD & Start Date & End Date & Cadence & 2-hour CDPP & \#Transits of  & \#Transits \\
         &  &  & (UTC) & (UTC) & (s) & (ppm) & Planet b &  Planet c \\
        \hline
        3 & 2 & 2 & 2018-09-20T12:52:53.590 & 2018-10-17T21:17:34.435 & 120 & 225.2 & 17 & 0\\
        4 & 2 & 1 & 2018-10-19T09:41:34.796 & 2018-11-14T08:06:57.360 & 120 & 240.5 & 22 & 1\\
        30 & 2 & 2 & 2020-09-23T09:18:33.756 & 2020-10-20T12:29:04.797 & 120 & 255.8 & 22 & 2\\
        31 & 2 & 1 & 2020-10-22T00:25:04.978 & 2020-11-16T10:44:20.642 & 120 & 228.5 & 24 & 2\\
        \hline
        \hline
    \end{tabular}
    \label{tab:tess_obs}
\end{table*}

\subsection{CHEOPS}
Following \textit{TESS}, the CHaracterising ExOPlanet Satellite \citep[\textit{CHEOPS};][]{Benz_CHEOPS_2021, Fortier_CHEOPS_2024} was launched as an ESA S-class mission in 2019. \textit{CHEOPS}' observing strategy and instrumental setup enables it to determine planetary radii very precisely. Additionally it has shown to be able to nail down orbital periods of planets where only one or two transits are caught in \textit{TESS} data, e.g. HIP 9618\,c \citep{Osborn_HIP9618_2023} or in detecting further and smaller planets in known systems \citep[e.g.][]{Bonfanti_TOI1233_2021}. \textit{CHEOPS} is in a nadir-locked, Sun-synchronous, low-Earth orbit, causing gaps in the data which arise due to events such as the Earth's occultation or South Atlantic Anomaly. The fraction of data outside the gaps per visit is represented by the observational efficiency that we report in \autoref{tab:toi4311_cheops_obs}. 

We obtained 20 \textit{CHEOPS} visits of TOI-4311 over 2022, 2023, 2024 and 2025 as part of the GTO's XGAL and YGAL programs (PI: Wilson). We summarise these visits in \autoref{tab:toi4311_cheops_obs}. Within these visits we caught 17 transits of planet b and 7 transits of planet c. These data were processed with the \textit{CHEOPS} Data Reduction Pipeline \citep[DRP;][]{Hoyer_CHEOPS_DRP_2020} which uses aperture photometry. Additionally, \textit{CHEOPS} data has been shown to be more precise for faint stars (G>11\,mag) when extracting the data with Point Spread Function (PSF) photometry using \textsc{pipe} \citep[][]{Brandeker_PIPE_2024}. Since TOI-4311 is a rather faint star to observe with \textit{CHEOPS} (G=11.27\,mag), we re-extract the data using \textsc{pipe}. Comparing the RMS of each visit (see \autoref{tab:toi4311_cheops_obs}), we find that \textsc{pipe} produces lightcurves of lower RMS and hence we select \textsc{pipe} for our further analysis.

\begin{table*}
    \centering
    \caption{CHEOPS Observations of TOI-4311. The file keys of visits 1-9 begin with CH\_PR120054\_TG00; the file keys of visits 10-20 begin with CH\_PR140065\_TG00. }
    \begin{tabular}{ccccccccccc}
    \hline
    \hline
        Visit & Planets & Start Date & Duration & Data Points & File Key & Efficiency & Exp Time & Aperture & DRP RMS & PIPE RMS \\
        &  & (UTC) & (h) & (\#) &  & (\%) & (s) &  & (ppm) & (ppm)  \\
        \hline
        1 & c & 2022-10-18T17:50:42 & 14.87 & 694 & 2301 & 77.7 & 60 & RINF & 1332 & 919\\
        2 & b & 2022-10-25T02:34:42 & 13.59 & 757 & 1201 & 92.7 & 60 & R21 & 1626 &  862 \\
        3 & b & 2022-11-02T02:01:42 & 11.57 & 638 & 1203 & 91.8 & 60 & R21 & 1569 & 1041 \\
        4 & b,c & 2022-11-02T18:59:42 & 13.62 & 750 & 2302 & 91.6 & 60 & R21 & 1541 & 763\\
        5 & b & 2022-11-09T22:50:42 & 11.57 & 641 & 1204 & 92.2 & 60 & R21 & 2131 & 953 \\
        6 & b & 2022-11-13T23:18:42 & 12.54 & 592 & 3001 & 78.6 & 60 & R21 & 1369 & 827 \\
        7 & b & 2022-11-14T21:50:41 & 11.81 & 553 & 3002 & 78.0 & 60 & R21 & 1327 & 1073 \\
        8 & b & 2022-11-20T21:28:42 & 11.52 & 450 & 3003 & 65.0 & 60 & R21 & 1282 & 761 \\
        9 & b,c & 2022-12-02T22:51:42 & 13.44 & 451 & 2901 & 55.9 & 60 & R21 & 962 & 772 \\
        10 & c & 2023-09-15T09:11:43 & 11.62 & 421 & 0601 & 60.3 & 60 & R21 & 1420 & 811 \\
        11 & c & 2023-09-30T08:44:42 & 13.51 & 538 & 1201 & 66.3 & 60 & R21 & 1636 & 798 \\
        12 & b & 2023-10-12T15:12:43 & 12.14 & 544 & 0401 & 74.6 & 60 & R23 & 1449 & 936 \\
        13 & b & 2023-10-27T12:47:43 & 11.61 & 647 & 0402 & 92.8 & 60 & R21 & 1384 & 794 \\
        14 & b,c & 2023-10-30T13:14:43 & 13.51 & 763 & 1202 & 94.0 & 60 & R21 & 1455 & 1084 \\
        15 & b & 2023-11-03T10:54:43 & 11.61 & 640 & 0403 & 91.8 & 60 & R21 & 1663 & 963 \\
        16 & b & 2023-11-07T09:07:43 & 11.61 & 643 & 0404 & 92.2 & 60 & R22 & 1629 & 798 \\
        17 & b & 2023-11-11T08:12:43 & 11.84 & 569 & 0405 & 80.0 & 60 & R21 & 1904 & 734 \\
        18 & b,c & 2023-11-14T13:38:43 & 13.51 & 574 & 1203 & 70.7  & 60 & R21 & 1477 & 960 \\
        19 & b & 2024-10-19T01:37:43 & 11.61 & 569 & 1301 & 81.6 & 60 & R21 & 1655 & 596 \\
        20 & b & 2025-11-06T06:45:43 & 11.61 & 578 & 1302 & 100.0 & 60 & R21 & 7918 & 1023 \\
    \hline
    \hline
    \end{tabular}
    \label{tab:toi4311_cheops_obs}
\end{table*}

\subsection{HARPS}
The High Accuracy Radial Velocity Planet Searcher \citep[HARPS;][]{Pepe_HARPS_2000} is a high-resolution Echelle spectrograph on ESO's 3.6\,m telescope in La Silla. Its wavelength coverage ranges from 380\,nm to 680\,nm and it has a resolving power of 115,000. Spectra obtained by HARPS are processed by its Data Reduction Software \citep[DRS 3.2.5;][]{Lovis_Pepe_HARPS_DRS_2007} which obtains the RVs by computing the cross-correlation function (CCF) of the spectra using a mask dependent on the spectral type of the observed star. From the CCFs activity indicators such as the FWHM, BIS and contrast are obtained. Further activity indicators such as the S, H$\alpha$, Na and Ca\,II indices are computed directly from the spectra.

\begin{figure*}
    \centering
    \includegraphics[width=\linewidth]{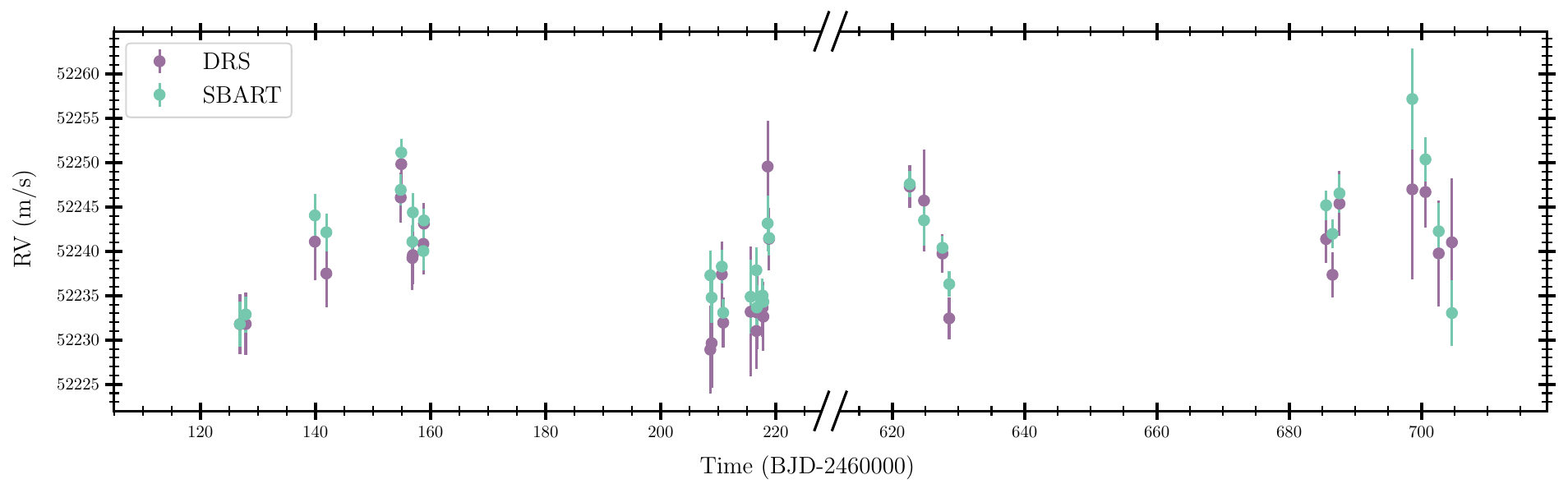}
    \caption{RV observations of TOI-4311 as processed by the HARPS DRS in purple and our \textsc{s-bart} extraction in green. }
    \label{fig:rv_timeseries}
\end{figure*}

TOI-4311 was monitored by HARPS from 1 July 2023 to 29 January 2025 as part of the 111.254R program (PI: Wilson). This resulted in 32 data points with an average uncertainty of 3.65\,m/s and RMS of 6.10\,m/s. While the DRS computes the RVs using a cross-correlation function with a K2 mask, we additionally reprocess the spectra using Semi-Bayesian Approach for RVs with Template-matching \citep[\textsc{s-bart};][]{Silva_sbart_2022}. \textsc{s-bart} uses template matching to extract the RVs which has been found to improve the uncertainties \citep[][]{Silva_sbart_2022}. Following \citet{Eschen_TOI2345_2025}, we reprocess the HARPS 2D spectra for different combinations of template matching parameters and spectra quality checks. We vary the template matching parameters in RV steps of 0.1, 0.5, 1.0\,m/s and RV limits of 200, 500, 1000\,m/s. For all of these we use the classical and Laplacian method from \textsc{s-bart}. Additionally, we vary the quality checks of the spectra that are included in the template creation. These are the minimum order SNRs, airmasses and RV errors. For these we apply cuts at 1.5, 5 and 10; 1.5, 2.0, 2.2 and 2.6; 5, 6, 7 and 10\,m/s respectively. This does remove spectra which do not fulfil the quality checks from the template creation, however does reprocess the RVs arising from these spectra using the created template. From all of these different checks, we compare the median errors of the resulting RVs. We get the lowest median error for a \textsc{s-bart} setup of a RV step of 0.1\,m/s, RV limit of 500\,m/s, minimum order SNR of 1.5, airmass of 1.5 and RV error cut of 5\,m/s using \textsc{s-bart}'s Laplacian method, which resulted in a median error of 2.15\,m/s and RMS of 6.02\,m/s. We used these RVs in the following analysis. We show the RVs obtained from the DRS and \textsc{s-bart} in \autoref{fig:rv_timeseries} for comparison. 

\subsection{ASAS-SN}
The All-Sky Automated Survey for Supernovae \citep[ASAS-SN][]{Shappee_ASAS_SN_2014, Kochanek_ASAS_SN_2017} is performing an all-sky survey monitoring targets down to a magnitude of V=17\,mag. For this, ASAS-SN consists of five stations spread over the world, each consisting of four 14\,cm telescopes and CCDs. The stations use either the V- or g-band and observe three dithered 90\,s exposures. The data is made publicly available and lightcurves can be computed using the ASAS-SN Sky Patrol (\url{https://asas-sn.osu.edu}). In addition to lightcurves for targets being available on this platforms, they can also be recomputed using the coordinates and proper motions of the respective target.
Due to its long-term photometric monitoring, ASAS-SN can be used to identify stellar rotations periods \citep{Turner_Gliese12_2025}. 

TOI-4311 has been monitored by ASAS-SN from 24 October 2013 to 27 August 2025. During this time TOI-4311 was observed from 24 October 2013 to 24 September 2018 in the V-band, resulting in 1133 data points and from 17 September 2017 to 27 August 2025 in the g-band for which 5231 data points were collected. For each of the two datasets we separately remove data points fainter than V=15\,mag and g=15\,mag and perform 5-sigma clipping which results in the removal of 43 and 254 outliers, leaving 1090 and 4977 data points in the V- and g-band respectively. We show the resulting timeseries in \autoref{fig:asas_sn_wasp_lightcurves}.

\subsection{WASP}
The Wide Angle Search for Planets \citep[WASP;][]{Pollaco_WASP_2006} has been photometrically monitoring bright stars from the Southern and Northern hemisphere. For this it has eight telescopes each at both of its sites, located at the Roque de los Muchachos Observatory on La Palma, Spain and the Sutherland Station located at the South African Astronomical Observatory. During its operational time, this survey detected nearly 200 planets \citep[][accessed on 27 December 2025]{Exoplanet_Archive}, the majority of which are Hot Jupiters. While the WASP survey does not reach the precision to detect Super-Earths or Sub-Neptunes, it provides long-term photometry which can be used to determine stellar rotation periods \citep[e.g.][]{Wilson_TOI1064_2022}.

TOI-4311 has been monitored by WASP from 10 August 2006 to 19 December 2014. During this observing time the survey collected 107399 photometric measurements of the target. We normalise the flux and remove points that have an uncertainty higher than 5\%. We then perform a 5-sigma clipping leaving us with 96694 data points. We show the resulting timeseries in \autoref{fig:asas_sn_wasp_lightcurves}.

\subsection{SOAR-Imaging}
\begin{figure}
    \centering
    \includegraphics[width=\linewidth]{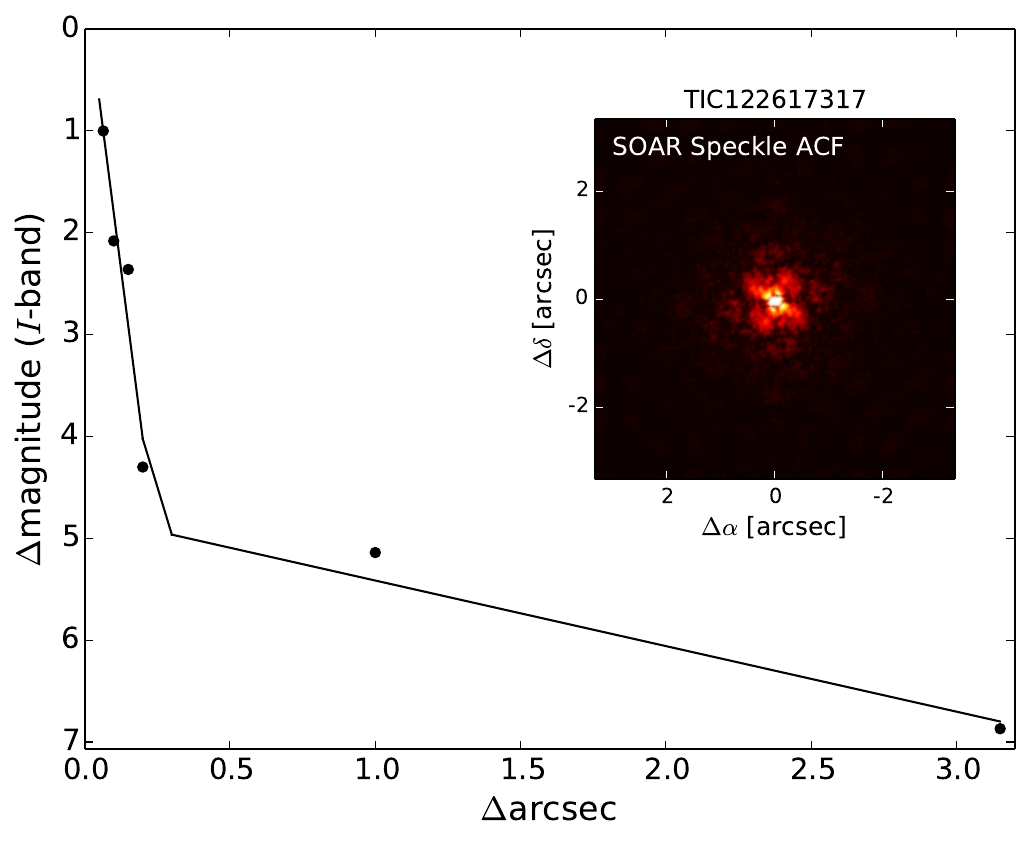}
    \caption{SOAR I-band observation of TOI-4311 with a 5-$\sigma$ detection sensitivity.}
    \label{fig:soar_imaging}
\end{figure}

High-angular resolution imaging is needed to search for nearby sources that can contaminate the TESS photometry, resulting in an underestimated planetary radius, or be the source of astrophysical false positives, such as background eclipsing binaries \citep[e.g.][]{Buchave_Kepler14_2011,Lillo_Box_High_Res_Imaging_2014, Lillo_Box_High_Res_Imaging_TESS_2024}. We searched for stellar companions to TOI-4311 with speckle imaging on the 4.1-m Southern Astrophysical Research (SOAR) telescope \citep{Tokovinin_SOAR_2018} on 1 October 2021 UT, observing in Cousins I-band, a similar visible bandpass as TESS. This observation was sensitive with 5-sigma detection to a 5.2-magnitude fainter star at an angular distance of 1\arcsec from the target. More details of the observations within the SOAR TESS survey are available in \citet{Ziegler_SOAR_TESS_2020}. The 5$\sigma$ detection sensitivity and speckle auto-correlation functions from the observations are shown in \autoref{fig:soar_imaging}. No nearby stars were detected closer than 3\arcsec within the sensitivity limits of TOI-4311 in the SOAR observations.

\section{Stellar Characterisation}
\label{sec:stellar_charactersation}

\begin{table}
\caption{Stellar properties of TOI-4311.}
\begin{center}
\begin{tabular}{lll} 
\hline\hline                 
\multicolumn{3}{c}{TOI-4311} \\    
\hline                        
2MASS & \multicolumn{2}{l}{J02464430-3110468} \\
\textit{Gaia} DR3 & \multicolumn{2}{l}{5064693746900436736} \\
TIC & \multicolumn{2}{l}{122617317} \\
LP & \multicolumn{2}{l}{886-41} \\ 
\hline
Parameter & Value & Note \\ 
\hline
   $\alpha$ [J2000] & 02$^{\rm h}$46$^{\rm m}$44.32$^{\rm s}$ & 1 \\
   $\delta$ [J2000] & -31$^{\circ}$10$^{'}$46.69$^{\arcsec}$ & 1 \\
   $\mu_{\alpha}$ [mas/yr] & 171.231$\pm$0.026 & 1\\
   $\mu_{\delta}$ [mas/yr] & 73.586$\pm$0.028 & 1 \\
   $\varpi$ [mas] & 7.413$\pm$0.028 & 1 \\
   $d$ [pc] & 134.90$\pm$0.51 & 1 \\
   RV [km~s$^{-1}$] & 51.84$\pm$0.22 & 1 \\
   U [km~s$^{-1}$] & $-$116.99$\pm$0.40 & 5$^{\rm a}$ \\
   V [km~s$^{-1}$] & $-$57.44$\pm$0.17 & 5$^{\rm a}$ \\
   W [km~s$^{-1}$] & $-$0.32$\pm$0.27 & 5$^{\rm a}$ \\
\hline
   $V$ [mag] & 11.39$\pm$0.10 & 2 \\
   $G_{\rm BP}$ [mag] & 11.7047$\pm$0.0028 & 1 \\
   $G$ [mag] & 11.2714$\pm$0.0028& 1 \\
   $G_{\rm RP}$ [mag] & 10.6714$\pm$0.0038 & 1 \\
   $J$ [mag] & 9.977$\pm$0.023 & 3 \\
   $H$ [mag] & 9.557$\pm$0.025 & 3 \\
   $K$ [mag] & 9.488$\pm$0.021 & 3 \\
   $W1$ [mag] & 9.430$\pm$0.023 & 4 \\
   $W2$ [mag] & 9.487$\pm$0.020 & 4 \\
\hline
   $T_{\mathrm{eff}}$ [K] & 5063$\pm$66 &  5; spectroscopy\\
   $\log{g}$ [cm~s$^{-2}$] & 4.442$\pm$0.034 & 5; spectroscopy \\\relax
   [Fe/H] [dex] & $-$0.065$\pm$0.043 & 5; spectroscopy \\\relax
   [Mg/H] [dex] & $-$0.015$\pm$0.053 & 5; spectroscopy \\\relax
   [Si/H] [dex] & $-$0.011$\pm$0.036 & 5; spectroscopy \\
   $v\sin{i}$ [km\,s$^{-1}$] & $1.10\pm0.60$ & 5; spectroscopy \\
   $\log{R'_{\mathrm{HK}}}$ & $-5.00\pm0.11$ & 5; spectroscopy \\
   $R_{\star}$ [$R_{\odot}$] & 0.8441$\pm$0.0062 & 5; IRFM  \\
   $M_{\star}$ [$M_{\odot}$] & $0.816_{-0.049}^{+0.047}$ & 5; isochrones \\  
   $t_{\star}$ [Gyr] & $6.6\pm2.3$ & 5; $\log{R'_{\mathrm{HK}}}$ \\
   $L_{\star}$ [$L_{\odot}$] & 0.422$\pm$0.023 & 5; from $R_{\star}$ and $T_{\mathrm{eff}}$\\
   $\rho_{\star}$ [$\rho_\odot$] & 1.357$\pm$0.086 & 5; from $R_{\star}$ and $M_{\star}$ \\
   $\rho_{\star}$ [$\mathrm{kg\,m^{-3}}$] & 1910$\pm$120 & 5; from $R_{\star}$ and $M_{\star}$ \\
\hline\hline                               
\end{tabular}
\label{tab:stellarParam}
\end{center}
[1] \cite{Gaia_DR3_Release_Summary_2023}, [2] \cite{Hog2000}, [3] \cite{Skrutskie2006}, [4] \cite{Wright2010}, [5] This work \\
$^{\rm a}$ Calculated via the right-handed, heliocentric Galactic spatial velocity formulation of \cite{Johnson_Soderblom_galactic_space_velocity_1987} using the proper motions, parallax, and radial velocity from [1].
\end{table}

\subsection{Atmospheric Properties and Abundances}

The stellar spectroscopic parameters ($T_{\mathrm{eff}}$, $\log g$, [Fe/H]) were estimated using the \textsc{ARES+MOOG} methodology, as described in more detail in \citet[][]{Sousa-21, Sousa-14} and \citet{Santos-13}. For this we used the latest version of \textsc{ARES}\footnote{The latest version, \textsc{ARES v2}, can be downloaded at \url{https://github.com/sousasag/ARES}.} \citep{Sousa-07, Sousa-15} to measure the equivalent widths (EW) of iron lines on a combined HARPS spectrum of TOI-4311. We used the list of FeI and FeII lines presented in \citet[][]{Tsantaki-2013} which is more appropriate for this target with temperature below 5200 K. A minimization process was used to find ionization and excitation equilibrium and converge to the set of best spectroscopic parameters. For this we made use of a grid of Kurucz model atmospheres \citep{Kurucz-93} and the radiative transfer code \textsc{MOOG} \citep{Sneden-73}. In this analysis we also derived a more accurate trigonometric surface gravity using GAIA DR3 data following the same procedure as described in \citet[][]{Sousa-21}. Stellar abundances of the elements were derived using the classical curve-of-growth analysis method assuming local thermodynamic equilibrium and with the same codes and models that were used for the stellar parameters determinations. For the derivation of chemical abundances of refractory elements we closely followed the methods described in e.g. \citet{Adibekyan-12, Adibekyan-15} and \citet{Delgado-17}.

\subsection{Radius, Mass, and Age}

We computed the stellar radius of TOI-4311 by utilising a MCMC modified infrared flux method \citep[IRFM --][]{Blackwell1977,Schanche2020} in combination with stellar atmospheric models \citep{Kurucz1993,Castelli2003} and our spectroscopically derived stellar parameters as priors. Broadband synthetic photometry were produced from constructed spectral energy distributions (SED) and compared to observed fluxes in the following bandpasses: 2MASS $J$, $H$, and $K$, WISE $W1$ and $W2$, and Gaia $G$, $G_\mathrm{BP}$, and $G_\mathrm{RP}$ \citep{Skrutskie2006,Wright2010,Gaia_DR3_Release_Summary_2023}. This resulted in the stellar bolometric flux from which we obtained the effective temperature and angular diameter of TOI-4311. We converted the angular diameter into the stellar radius using the offset-corrected Gaia parallax \citep{lindegren2021} and accounted for model uncertainties via a Bayesian Model Averaging of the stellar radius posteriors produced from both stellar atmospheric model catalogues.

Using $T_{\mathrm{eff}}$, [Fe/H], and $R_{\star}$ along with their uncertainties as input parameters, we determined the isochronal mass and age via two stellar evolutionary models. In detail, we employed the isochrone placement routine \citep{bonfanti2015,bonfanti2016} to interpolate the input set within pre-computed grids of PARSEC\footnote{\textsl{PA}dova and T\textsl{R}ieste \textsl{S}tellar \textsl{E}volutionary \textsl{C}ode: \url{https://stev.oapd.inaf.it/cgi-bin/cmd}} v1.2S \citep{marigo2017} isochrones and tracks and get a first pair of estimates for the mass and age of the star. A second pair of mass and age values, instead, was computed with the CLES \citep[Code Liègeois d'Évolution Stellaire;][]{scuflaire2008} code that builds the best-fit stellar evolutionary track based on the set of input parameters by following a Levenberg-Marquadt minimisation scheme \citep{salmon2021}.
As detailed in \citet{bonfanti2021}, we then checked the mutual consistency of the two pairs of mass and age distributions via a $\chi^2$-based criterion and finally merged them obtaining $M_{\star}=0.816_{-0.049}^{+0.047}\,M_{\odot}$ and $t_{\star}=7.1\pm6.7$\,Gyr. Since the stellar age is not well constrained by evolutionary models, we applied the $\log{R'_{\mathrm{HK}}}$-$t_{\star}$ relation from \citet{mamajek2008} and obtained $t_{\star}=6.6\pm2.3$\,Gyr as reported in \autoref{tab:stellarParam}.

\subsection{Rotation Period}
While photometric surveys such as WASP and ASAS-SN do not reach the precision to detect small planets, their long-term photometric monitoring is valuable to identify rotation periods of stars \citep[e.g.][]{Turner_Gliese12_2025}. Since TOI-4311 has been monitored by these ground-based surveys for more than 10 years, we run a Lomb-Scargle periodogram on each of the available datasets, differentiating between V- and g-band in ASAS-SN. Since we are using ground-based data, we are removing peaks due to cadence, Moon and season gaps. We do not identify any clear peaks indicating a rotation period, we acknowledge a peak in the ASAS-SN V-band at $\sim$160\,d which however is not found in the ASAS-SN g-band or by WASP. 

\begin{figure*}
    \centering
    \includegraphics[width=\linewidth]{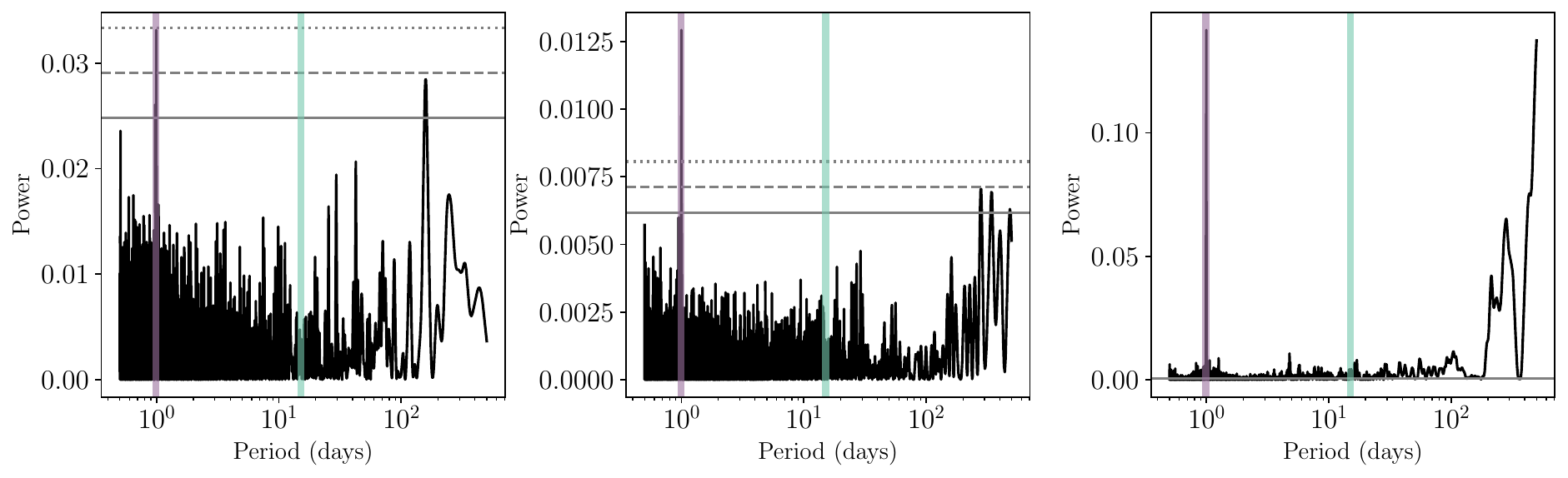}
    \caption{Lomb-Scargle periodograms of the long-term photometric surveys. The two planetary periods are highlighted in green and purple. False Alarm Probabilities of 1\%, 0.1\% and 0.01\% are shown by the gray continuous, dashed and dotted line respectively. Left: ASAS-SN V-band. Middle: ASAS-SN g-band. Right: WASP.}
    \label{fig:rotation_period}
\end{figure*}

In addition to the long-term ground-based observations, there are also four sectors of \textit{TESS} data available. Since we have two continuous sectors for each of the two \textit{TESS} missions (PM and EM1), we use these to check for a rotation period in \textit{TESS} as well. Using the non-detrended SAP flux (as shown in \autoref{fig:tess_sap_lightcurve}) after masking out the transits of planet b and planet c using \textsc{wotan} \citep{Hippke_wotan_2019}, we run two Lomb-Scargle periodograms, the first on \textit{TESS} sectors 3 and 4; the second on \textit{TESS} sectors 30 and 31. As shown in \autoref{fig:rotation_period} and \autoref{fig:tess_periodogram}, the ASAS-SN g-band, WASP and \textit{TESS} sectors do not show any significant peaks. 

However, there is a peak in the ASAS-SN V-band at $\sim$160\,d. Since this peak is not too well constrained we do not consider it a rotation period for the further analysis but acknowledge its presence here.
Since the analyses of long-term photometric monitoring from ASAS-SN and WASP as well as the more recent high-precision photometric monitoring from \textit{TESS} resulted in no clear detections of a rotation period, we conclude that TOI-4311 is not an active star.

\subsection{Kinematic Analysis}
\label{sec:kinematic_analysis}
We compute the kinematic properties of TOI-4311 following \citet{Johnson_Soderblom_galactic_space_velocity_1987} and \citet{Bensby_lsr_2003}. We report the galactic velocities U, V and W in \autoref{tab:stellarParam}.  We correct these for the Local Standards of Rest (LSRs) as reported in \citet{Koval_lsr_2009,Schoenrich_lsr_2010,Coskunoglu_RAVE_LSR_2011,Bobylev_LSR_2014,Francis_LSR_2014,Tian_Stellar_Kinematic_LAMOST_LSR_2015} and \citet{Almeida_Fernandes_Stellar_Ages_LSR_2018} and compute the weighted means for U, V and W.  Following \citet{Bensby_lsr_2003}, we compute the weighted kinematic probabilities of TOI-4311 belonging to each population. This results in 52.7$\pm$5.3\% and 47.3$\pm$5.3\% of the target belonging kinematically to the thin and thick disk respectively. Hence, TOI-4311 lies between the thin and thick disk of the Milky Way. 

However, \citet{Bensby_lsr_2003} does not include the normalisation fractions and velocity dispersions for the Hercules Stream. Recent publications \citep[][]{Bensby_lsr_2014, Chen_PAST_I_2021} include these and hence we perform an additional kinematic analysis in which we also weigh for the different LSRs. This results in a weighted kinematic membership of 21.7$\pm$3.3\% for the thin disk, 30.9$\pm$3.2\% for the thick disk and 47.4$\pm$3.1\% for the Hercules stream when using \citet{Bensby_lsr_2014}, as well as weighted kinematic probabilities of 17.3$\pm$3.0\%, 33.3$\pm$3.0\% and 49.4$\pm$2.9\% for the thin disk, thick disk and Hercules stream respectively when using the values from \citet{Chen_PAST_I_2021}.

\begin{figure*}
    \centering
    \includegraphics[width=\linewidth]{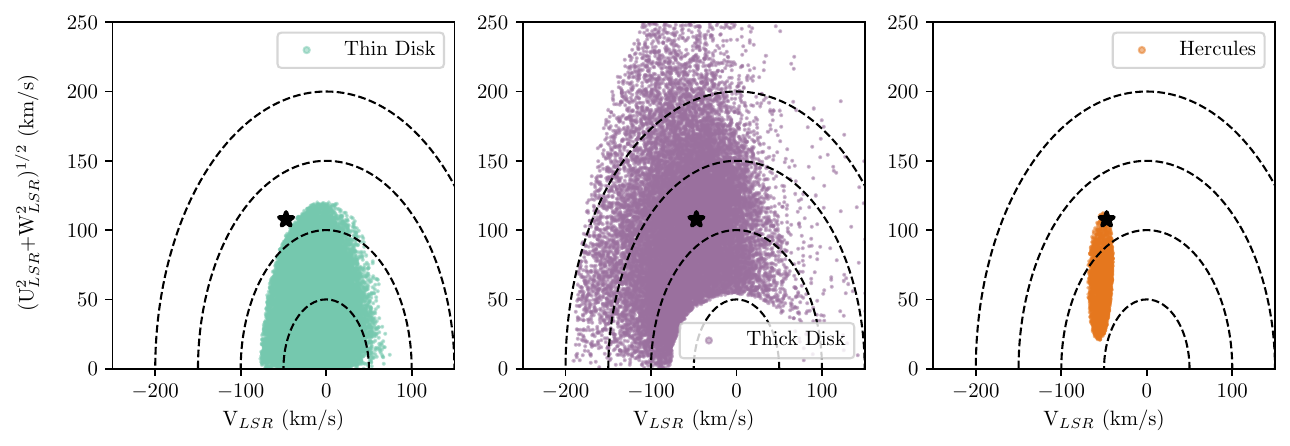}
    \caption{Location of TOI-4311 on the Toomre diagram compared to stars of different galactic memberships, thin disk (left), thick disk (middle) and Hercules stream (right).}
    \label{fig:toomre_divided}
\end{figure*}

Using the stars in the publicly available APOGEE DR17 catalogue \citep{Niveder_APOGEE_Data_Reduction_2015,Majewski_APOGEE_Overview_2017,Abdurrouf_APOGEE_DR17_Release_2022}, we compute the kinematic membership probabilities for every star in the catalogue that is passing the quality cuts based on the ASCAPFLAG and STARFLAG in APOGEE and has reported proper motions, parallax and radial velocity values in Gaia DR3 \citep[][]{Gaia_DR3_Release_Summary_2023}. Using the above methods, we are able to determine the galactic membership of each star using the normalisation fractions and velocity dispersions from \citet{Chen_PAST_I_2021} and LSRs from \citet{Almeida_Fernandes_Stellar_Ages_LSR_2018}. Selecting stars that have a thin disk, thick disk or Hercules stream probability above 50\% respectively, we plot them on a Toomre diagram. The kinematic properties and according galactic memberships allow us to compare the kinematics of TOI-4311 which is represented by the star on the Toomre diagram in \autoref{fig:toomre_divided} to a wider sample of stars. Since the kinematics of TOI-4311 point to an edge case between the galactic populations of thin disk, thick disk and Hercules stream and they overlap on a Toomre diagram, we plot the thin disk, thick disk and Hercules stream stars from APOGEE DR17 individually in \autoref{fig:toomre_divided}. This shows TOI-4311 lies outside of the thin disk as it is hinted by its probabilities, within thick disk stars but also overlaps with the edge of the Hercules stream.

\begin{figure*}
    \centering
    \includegraphics[width=\linewidth]{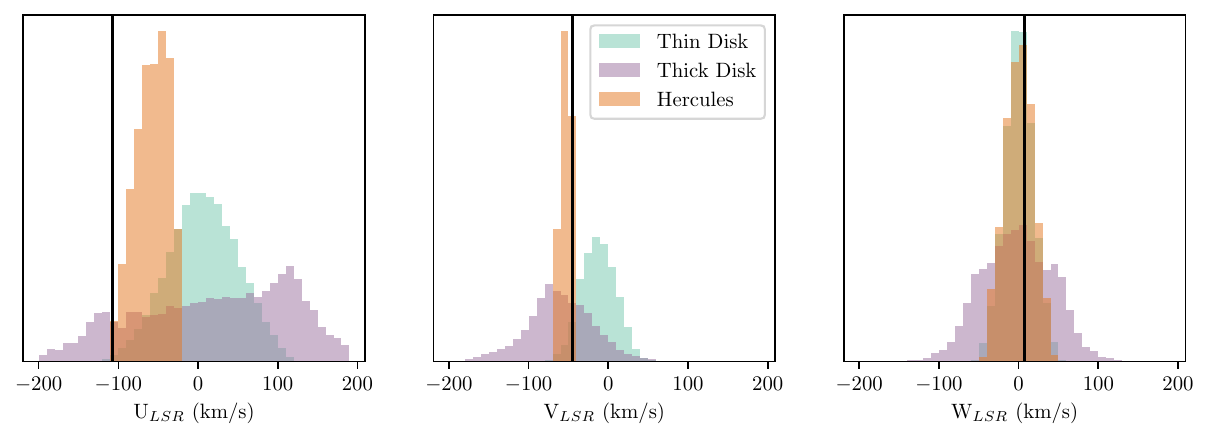}
    \caption{Histograms of the U, V and W velocities of the APOGEE sample of stars with a kinematic probability of above 50\% that belong to the thin disk, thick disk or Hercules stream. The membership is indicated by the colour. TOI-4311's velocities are marked by the black line. }
    \label{fig:apogee_velocities}
\end{figure*}

We show this further by plotting the normalised distributions of U, V and W of the stars in the APOGEE sample that belong to either population with a probability of 50\%. We show the velocities of TOI-4311 in comparison to these distributions in \autoref{fig:apogee_velocities} and find that TOI-4311 has a very low value of U for which it lies  below the narrow Hercules population. At this low value of U nearly no kinematic thin disk stars are found, however the kinematic thick disk population still covers this regime albeit in low numbers. This agrees with our kinematic probabilities which also find a low thin disk but reasonable thick disk probability. Within the V velocity, TOI-4311 overlaps mainly with the very narrow distribution of Hercules stream velocities but is also in the regime of thin and thick disk, while in W it overlaps well with all three populations. 

We also investigated the local Galactic phase space density of TOI-4311, following the approach of \cite{Winter_2020}, using the Mahalanobis metric \citep{Mahalanobis_1936} to scale its phase space density relative to neighbouring stars, and a Gaussian Mixture Model to decompose the distribution of phase space densities into high- and low-density components. We find that TOI-4311 has a low phase space density compared to most of its neighbours, placing it in the low-density population. \cite{Mustill_Phase_Space_2022} showed that this population is a mixture of halo stars, thick disc stars, and the dynamically hot end of the thin disc, and that the phase space density primarily depends on the magnitude of the peculiar velocity $|\mathbf{v}|$. We find that TOI-4311 does have a high phase space density when compared to stars of similar $|\mathbf{v}|$. We attribute this to the fact that the star's large peculiar velocity is excited almost entirely in the plane of the Galaxy, with very little vertical motion.


\section{Modelling and Determining Planet Properties}
\label{sec:planet_fitting}
Using the available photometry from \textit{TESS} and \textit{CHEOPS} and the radial velocity observations from HARPS, we fit the data to obtain the planetary parameters precisely in \textsc{juliet} \citep{Espinoza_juliet_2019}. 

\subsection{Photometry}
\label{sec:photometry}
\begin{figure}
    \centering
    \includegraphics[width=\linewidth]{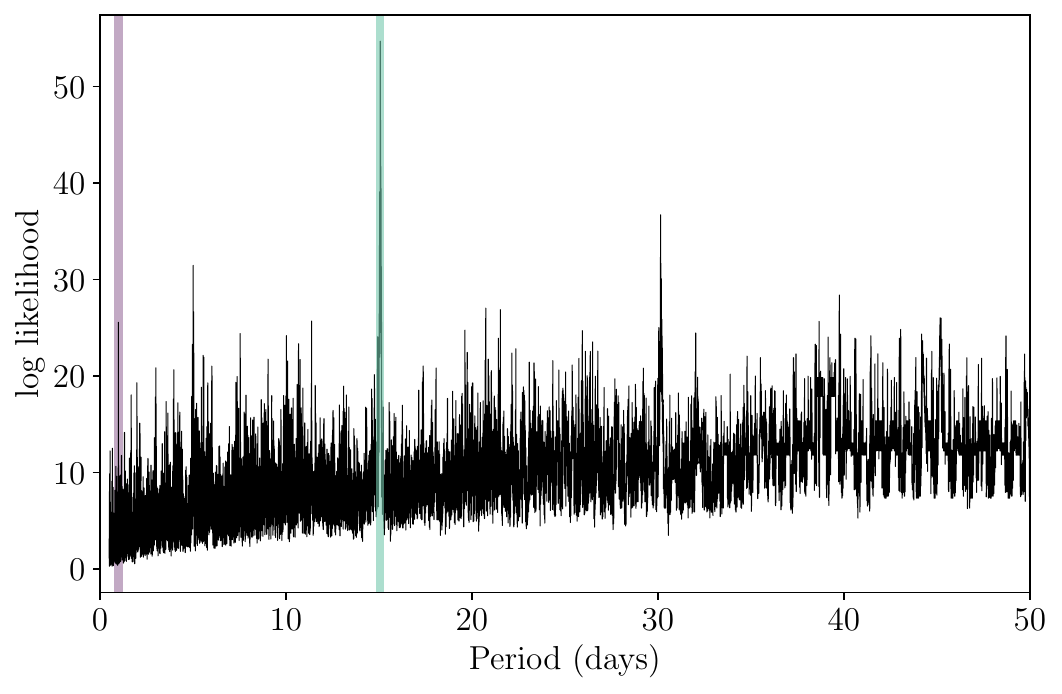}
    \caption{Box-Least-Squares periodogram of the 4 available \textit{TESS} sectors of TOI-4311. The period of the inner planet is highlighted in purple while the period of the outer planet is highlighted in green. }
    \label{fig:BLS}
\end{figure}

We run a Box Least Squares periodogram \citep[BLS;][]{Kovacs_BLS_2002} to identify the signals alerted by \textit{TESS} in our own analysis. The resulting periodogram in \autoref{fig:BLS} shows two significant peaks at 0.99\,d and 15\,d with log-likelihoods of 50 and 71 respectively. This is in agreement with the orbital periods found in the \textit{TESS} vetting team. We also find a peak at 30\,d which is a multiple of the peak at 15\,d and arises as an alias due to the limited number of transits for this planet in {\it TESS}. Additionally to the vetting performed by the \textit{TESS} team, we use the LEO-Vetter \citep{Kunimoto_LEOVetter_2025}, which is performing several tests based on the \textit{Kepler}-Robovetter \citep[][]{Coughlin_Robovetter_2016} and implemented further tests to optimise the vetting transits in \textit{TESS} by \citet{Eschen_Vetting_PC_MDwarfs_2024}. We run this vetting tool for TOI-4311\,b and TOI-4311\,c and both planets pass all the tests and get identified as planet candidates using the standard metric in Flux and Pixel Vetting. 

Using \textsc{juliet} we fit the \textit{TESS} and \textit{CHEOPS} data jointly for the planetary parameters including the period, P, the mid-transit time, t$_0$, the planet to star radius ratio, R$_\text{P}$/R$_*$, the impact parameter, b as well as the instrumental and stellar parameters including the limb darkening coefficients, $q1$ and $q2$ parametrised following \citet{Kipping_Limb_Darkening_2013}, the offset flux and jitter for each instrument and the stellar density. Additionally, we include a Gaussian Process (GP) with a Matern-3/2 kernel against time \citep[][]{Ambikasaran_george_2015, Foreman_Mackey_celerite_2017} for each instrument fitting for the GP amplitude and time-scale to account for any remaining noise in the \textit{TESS} data and instrumental trends such as variations in the roll angle in the \textit{CHEOPS} data. The distributions of these priors are summarised in \autoref{tab:fit_results_planet} and \autoref{tab:instr_fit}.  This fit results in precise planet radii of $1.372\substack{+0.086\\-0.087}$\,R$_\oplus$ and $2.46\substack{+0.11\\-0.12}$\,R$_\oplus$ for TOI-4311\,b and TOI-4311\,c respectively.

\subsection{Radial Velocities}
\label{sec:RV}
\begin{figure}
    \centering
    \includegraphics[width=\linewidth]{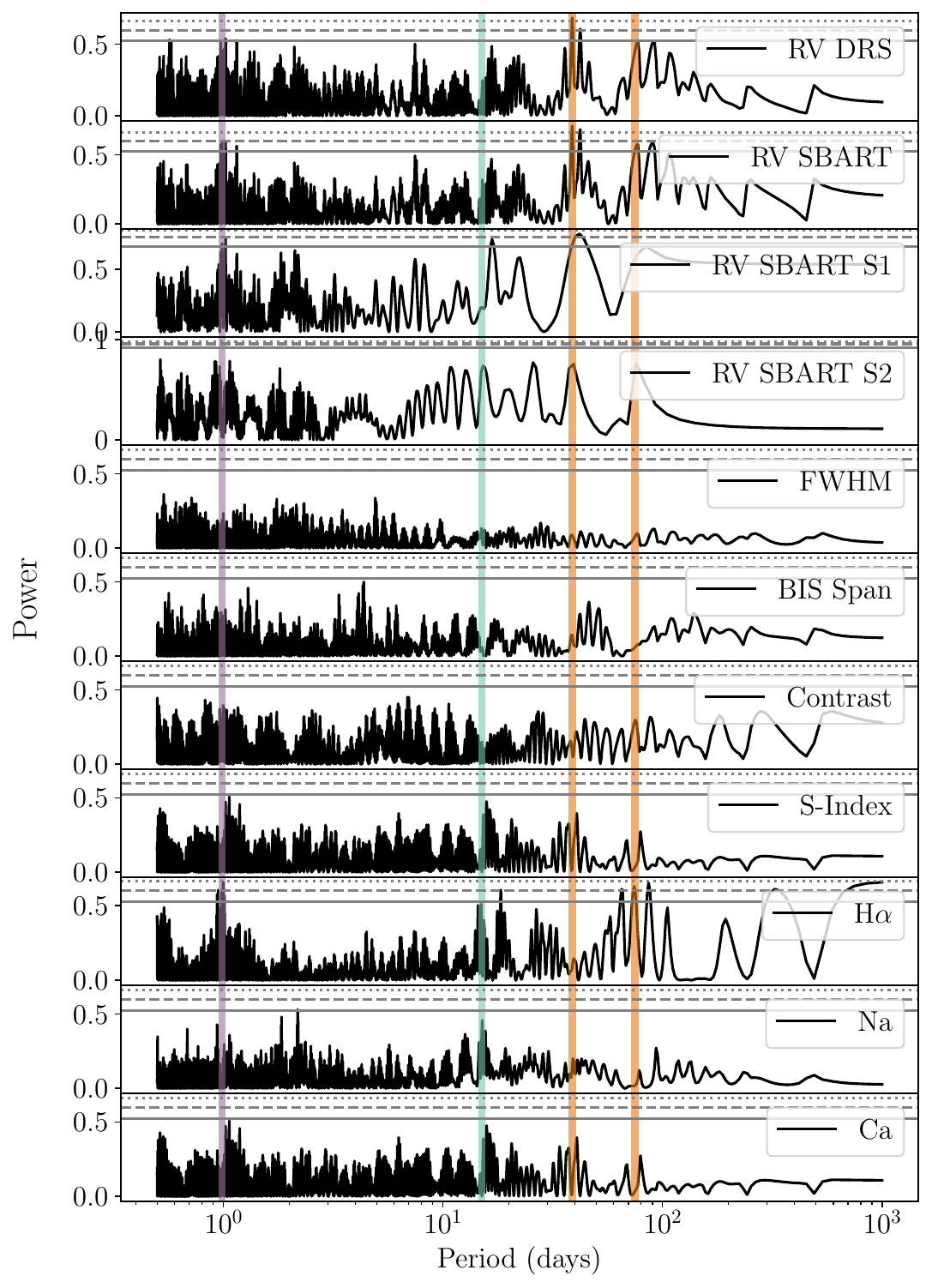}
    \caption{Lomb-Scargle periodograms of the RVs and activity indicators of TOI-4311 obtained by HARPS. From top to bottom: RVs from the HARPS DRS, RVs extracted from \textsc{s-bart}, RVs from 1 July 2023 to 1 October 2023 (S1) extracted from \textsc{s-bart} , RVs from 8 November 2024 to 29 January 2025 (S2) extracted from \textsc{s-bart}, activity indicators reported by the HARPS DRS. We highlight the periods of the two transiting planets, TOI-4311\,b and TOI-4311\,c, in purple and green respectively. We highlight the peaks at $\sim$38\,d and $\sim$75\,d in orange. False Alarm Probabilities of 1\%, 0.1\% and 0.01\% are shown by the gray continuous, dashed and dotted line respectively.}
    \label{fig:activity_indicators}
\end{figure}

Using the re-extracted RVs produced by \textsc{s-bart} and the activity indicators reported by the HARPS DRS, we use Lomb-Scargle periodograms to identify any significant signals in the RVs and activity indicators. These are shown in \autoref{fig:activity_indicators} and we find additional peaks at $\sim$38\,d and $\sim$75\,d in the RVs which we highlight in orange. These peaks are found in the RVs produced by the HARPS DRS as well as the RVs produced by \textsc{s-bart} but not in the activity indicators. To check whether these peaks can be recovered as coherent signals independently in different seasons confirming they are not due to observing biases or stellar activity, we separate the data into season 1 spanning 1 July 2023 to 1 October 2023, season 2 from 8 November 2024 to 29 January 2025. Season 1 and season 2 contain 21 and 11 observations respectively. We show the resulting Lomb-Scargle periodogram for season 1 and 2 on the \textsc{s-bart} RVs in \autoref{fig:activity_indicators}. Both seasons find a peak at $\sim$38\,d and $\sim$75\,d. The peak by season 1 is more significant due to more data being available within this timespan. The activity indicators do not show any significant peaks at this period. We hence perform further tests to determine the origin of this signal. 

Firstly, we use \textsc{scalpels} \citep{Cameron_scalpels_2021}, which computes the autocorrelation function from the CCFs to isolate shape-driven variations which arise from stellar activity. We include the two transiting planets at the periods and mid-transit times found by our photometric analysis of the \textit{TESS} and \textit{CHEOPS} data in our analysis. \textsc{scalpels} does not identify any principal components related to activity. To further assess our RV data, we apply \textsc{kima} \citep{Faria_Kima_2018,John_HD166620_HD144579_2023} and compare our results. In brief, \textsc{kima} allows us to model RV data of exoplanetary systems with the inclusion of decorrelation vectors in a linear model to account for  stellar activity. Importantly, \textsc{kima} takes the number of exoplanet Keplerian models as a free parameter meaning that we can statistically ascertain the optimal number of exoplanets represented in our RV data. In our analysis, we set the maximum number of additional planets to 2. First we decorrelate the activity signal with a maximum number of decorrelation vectors set to 3 as suggested by \textsc{scalpels}. The posterior distribution of the orbital periods gives us two significant peaks at $\sim$38\, and $\sim$75\,d, with the latter being more significant. We perform another analysis in \textsc{kima} without the decorrelation vectors. This also results in two significant peaks at the same periods, however this time both are similarly strong. Although these peaks are strong, \textsc{kima} statistically prefers a model with no additional planets in both cases. Hence we cannot confirm that this signal is due to stellar activity or rule out its planetary origin.

However, since \textsc{scalpels} identifies two peaks that are multiples of each other we fit a Keplerian in \textsc{juliet} for either a 38\,d (using $\mathcal{U}(37,40)$ as prior) or 75\,d (with $\mathcal{U}(70,80)$ as prior) signal. In both cases \textsc{juliet} clearly identifies a signal supporting our \textsc{kima} analysis.
As shown in the periodogram in \autoref{fig:activity_indicators}, the signal at 38\,d has a lower False Alarm Probability. This signal also has a better constrained period, transit time and semi-amplitude in the \textsc{juliet} analysis and we hence follow with this for our further analysis.

As a next step, we try to pick up this signal using (multi-dimensional) Gaussian Processes. For this, we use the \textsc{pyaneti} \citep{Barragan_pyaneti_2019, Barragan_pyaneti_2022} package and perform several RV analyses. For all, we keep the period and mid-transit time fixed at the values obtained in our photometric analysis as well as the eccentricity and argument of periastron at 0 and 90° respectively and fit for the semi-amplitude for these two planets. We compare these results to including a GP using a quasi-periodic kernel as well as multi-dimensional GPs in which we fit for H$\alpha$, the S-index or the full-width at half maximum (FWHM) in different runs. For all (multi-)GPs we compare priors for the stellar rotation period that are either uniformly distributed between (30,50) or (30,200).  We vary whether we include a third Keplerian for the $\sim$38\,d signal (with a prior uniformly distributed between 30 to 100 days) or not. We find that all our results agree within 1$\sigma$ for the inner planet's mass, except for the one run where we do not account for the signal at $\sim$38\,d through a GP or Keplerian. We find that the $\sim$38\,d signal can be modelled either with a third Keplerian or a GP and has to be accounted for in the analyses to improve the mass precision on the inner planet. Comparing the BIC values, we find that the 3 planet model is preferred over the 2 planet model with $\Delta$BIC$\sim$7. When using wide priors on the stellar rotation period in the GP, we find a rotation period with high uncertainties that overlap with the peak found in the ASAS-SN V-band data at around $\sim$160\,d. 

For our further analysis, we decided to treat this signal at 38\,d as a planet candidate and fit it with a third Keplerian. As this does not require multi-dimensional GPs, we perform the further analysis in \textsc{juliet} since this also allows us to model noise in the photometry with a GP as shown in \autoref{sec:photometry}, a feature that is not included in \textsc{pyaneti}. In \textsc{juliet}, we first perform further RV-only analysis also fitting for the eccentricity. Since the inner planet is on an Ultra-Short Period orbit, it is unlikely to be eccentric \citep{vanEylen_eccentricity_small_planets_2018, Pu_Lai_USP_low_eccentricity_2019} and hence we only fit the eccentricity for the outer two signals. However as shown by \citet{vanEylen_eccentricity_small_planets_2018}, the eccentricity on multi-planet systems follow half-Gaussian distributions centred at 0 with $\sigma$=0.083. Hence, we fit for the semi-amplitude of the radial velocities allowing the eccentricities of the two outer planet (candidates) to follow a Truncated Normal Distributions centred at 0 with  $\sigma$=0.083 and a lower and upper limit at 0 and 1 respectively while the eccentricity of the inner planet remains fixed at 0 (as shown in \autoref{tab:fit_results_planet}). While we also fit for the systemic radial velocity and jitter as summarised in \autoref{tab:instr_fit}, we keep the period and mid-transit time for planet b and c fixed on the values we obtained from the photometric fit, while we use a uniform distribution for the period and mid-transit time of the outer planet candidate. This produces no significant eccentricities, but allows us to place upper limits (99.7\%) of 0.22 on eccentricity for planet c and the outer planet candidate. 
With this, we are able to significantly obtain masses of $4.4\substack{+1.6\\-1.5}$\,M$_\oplus$ for the innermost planet and Msin(i)=$26.8\substack{+7.0\\-7.5}$\,M$_\oplus$ for the outer planet candidate. For planet c, we are not able to recover a significant detection of the semi-amplitude and hence mass, however we are able to place an upper limit (99.7\%) on the mass of 14.04\,M$_\oplus$ and hence confirm the planetary nature of the signal. 


\begin{figure*}
    \centering
    \includegraphics[width=\linewidth]{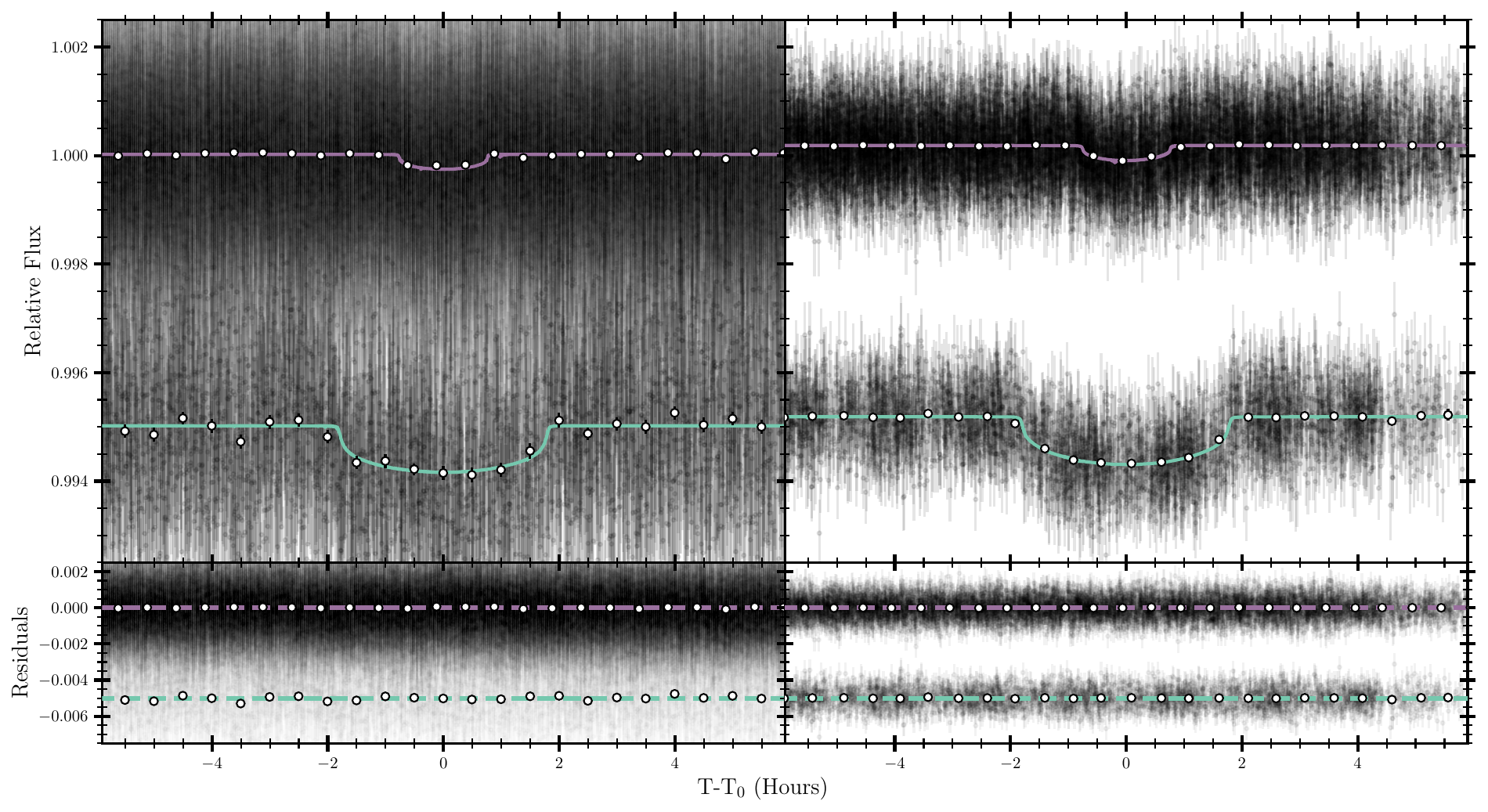}
    \caption{Phase folded best-fit transit models of TOI-4311\,b (purple) and TOI-4311\,c (green) with the underlying \textit{TESS} (left) and \textit{CHEOPS} (right) data in black in the top panel. The residuals are shown in the bottom. The data is binned into 30\,min showed by the white points.}
    \label{fig:transits}
\end{figure*}

\begin{figure*}
    \centering
    \includegraphics[width=\linewidth]{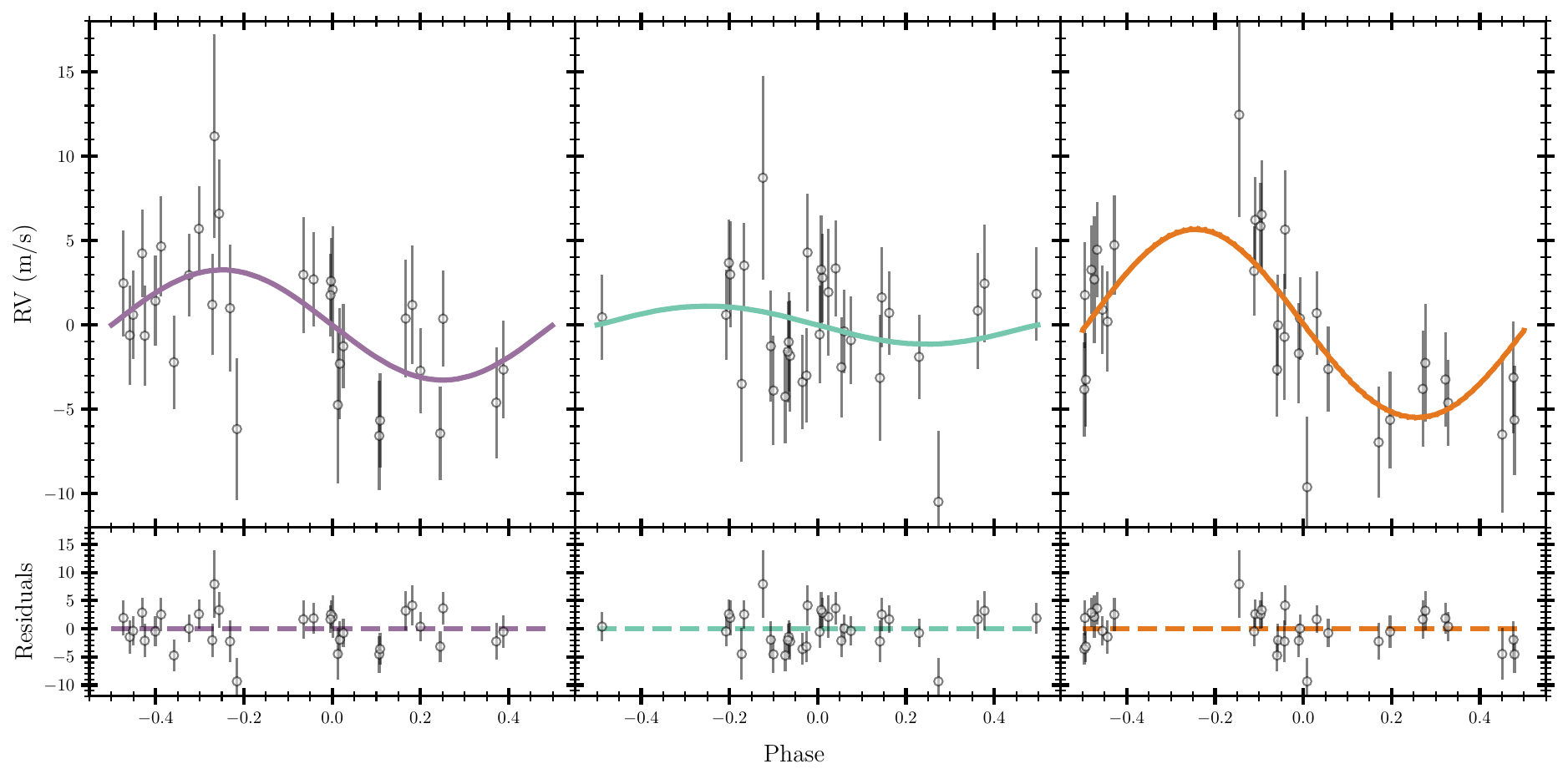}
    \caption{Phase-folded best fit RV curved of TOI-4311\,b (purple), TOI-4311\,c (green) and TOI-4311.03 (orange) with the underlying HARPS data in black in the top panel. The residuals are shown in the bottom. The time of inferior conjunction corresponds to phase=0.}
    \label{fig:rv_curves}
\end{figure*}

\begin{table*}
        \centering
        \caption{Fitted and Derived Planetary Parameters. Uniform distributions are noted by $\mathcal{U}$, normal distributions by $\mathcal{N}$, truncated normal distributions by $\mathcal{T}$ and fixed values are noted by $\mathcal{F}$ and the value. * denotes Msin(i).}
        \begin{tabular}{llrlrlr}
        \hline
        \hline
             &  \multicolumn{2}{c}{Planet b} &  \multicolumn{2}{c}{Planet c} & \multicolumn{2}{c}{Planet d}\\
             \hline
             Parameter  & prior & posterior & prior & posterior & prior & posterior \\
             \hline
             T$_0$ (BJD) & $\mathcal{N}(2458382.64,0.1)$ &  $2458382.6409\substack{+0.0019\\-0.0023}$ & $\mathcal{N}(2458394.74,0.1)$ &  $2458394.7242\substack{+0.0058\\-0.0057}$ & $\mathcal{U}(2460127,2460167)$ & $2460159.2\substack{+1.4\\-1.4}$ \\
             P (days) & $\mathcal{N}(0.99,0.1)$ & $0.9902738\substack{+0.0000029\\-0.0000030}$ & $\mathcal{N}(15.07,0.1)$ & $15.070425\substack{+0.000081\\-0.000075}$ & $\mathcal{U}(37,40)$ & $38.94\substack{+0.11\\-0.13}$\\
             R$_\text{P}$/R$_*$  & $\mathcal{U}(0,1)$ & $0.01495\substack{+0.00083\\-0.00087}$ & $\mathcal{U}(0,1)$ & $0.0268\substack{+0.0013\\-0.0012}$& - & - \\
             b  & $\mathcal{U}(0,1)$ & $0.37\substack{+0.13\\-0.15}$ & $\mathcal{U}(0,1)$ & $0.47\substack{+0.11\\-0.12}$ & - & - \\             
             K  (m/s) & $\mathcal{U}(0,100)$ & $3.3\substack{+1.1\\-1.0}$ & $\mathcal{U}(0,100)$ & $1.15\substack{+0.83\\-0.70}$ & $\mathcal{U}(0,100)$ & $5.71\substack{+0.72\\-0.59}$ \\
             ecc  & $\mathcal{F}(0)$ & - & $\mathcal{T}(0,0.083,0,1)$ & $<0.22$ & $\mathcal{T}(0,0.083,0,1)$ & $<0.22$ \\
             \hline
             \multicolumn{7}{c}{Derived Parameters}\\
             \hline
             a/R$_*$  & - & $4.629\substack{+0.096\\-0.099}$ & - & $28.43\substack{+0.59\\-0.61}$ & - & $53.5\substack{+1.1\\-1.2}$ \\
             $i$  (deg) & - & $85.4\substack{+1.8\\-1.6}$ & - & $89.05\substack{+0.24\\-0.21}$ & - & - \\
             R$_\text{P}$  (R$_\oplus$) & - & $1.376\substack{+0.077\\-0.080}$ & - & $2.47\substack{+0.12\\-0.11}$ & - & - \\
             a  (AU) & - & $0.01817\substack{+0.00036\\-0.00037}$ & - & $0.1116\substack{+0.0022\\-0.0023}$ & - & $0.2100\substack{+0.0042\\-0.0043}$ \\
             M$_\text{P}$  (M$_\oplus$) & - & $4.5\substack{+1.5\\-1.4}$ & - & $<13.58$ & - & $26.4\substack{+6.3\\-6.8}^*$\\
             $\rho_\text{P}$  (g/cm$^3$) & - & $9.3\substack{+3.8\\-3.0}$ & - & $<5.29$ & - & - \\
             T$_\text{eq}$  (K) & - & $1664\pm28$ & - & $671\pm11$ & - & $489.5\pm8.2$ \\
             S$_\text{P}$ (S$_\oplus$) & - & $1273\substack{+87\\-83}$ & - & $33.8\substack{+2.3\\-2.2}$ & - & $9.53\substack{+0.66\\-0.62}$\\
             \hline
             \hline
        \end{tabular}
        \label{tab:fit_results_planet}
\end{table*}

    

\subsection{Joint Fit}
Finally we run a joint fit of the photometric and RV data in \textsc{juliet} which we analysed individually before. We use the same priors as in the photometric analysis but additionally account for the third Keplerian signal, and hence fit for the semi-amplitudes of the three Keplerian signals, the jitter and systemic RV obtained by HARPS and follow the eccentricity prior distributions following \citet{vanEylen_eccentricity_small_planets_2018} as outlined in \autoref{sec:RV}. The priors and posteriors of this fit are shown in \autoref{tab:fit_results_planet}. This results in a precise radius and mass measurement for TOI-4311\,b of $1.376\substack{+0.077\\-0.080}$\,R$_\oplus$ and $4.5\substack{+1.5\\-1.4}$\,M$_\oplus$, a precise radius measurement of $2.47\substack{+0.12\\-0.11}$\,R$_\oplus$ and a mass upper limit of $<13.58$\,M$_\oplus$ for TOI-4311\,c and a minimum mass due to determining only Msin(i) from the RVs of $26.4\substack{+6.3\\-6.8}$\,M$_\oplus$ for the third planet candidate orbiting TOI-4311. This results in a precise density for the Ultra-Short Period planet, planet b, of $9.3\substack{+3.8\\-3.0}$\,g/cm$^3$. These small planets and candidates, that span the radius valley, range in equilibrium temperatures from 490\,K to 1664\,K. We show the phase-folded transits from the \textit{TESS} and \textit{CHEOPS} data in \autoref{fig:transits} and the resulting phase-folded RV curves in \autoref{fig:rv_curves}.

\section{Discussion}
\label{sec:discussion}

\subsection{A third planet in the system?}
\begin{figure}
    \centering
    \includegraphics[width=\linewidth]{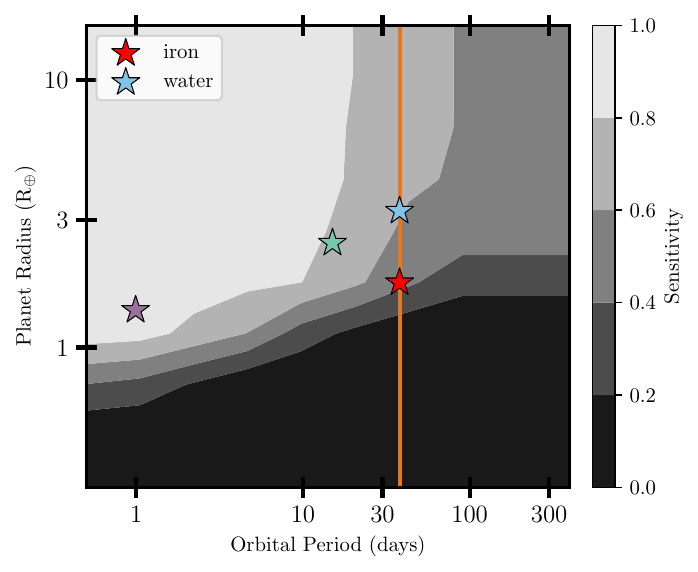}
    \caption{Sensitivity map of the available 4 \textit{TESS} sectors. White areas represent where planets should have been easily detected by \textit{TESS}, while the black areas show where \textit{TESS} is not able to find transiting planets with the available observations. The grey steps show the sensitivities in between including monotransits. TOI-4311\,b is shown by the purple star, TOI-4311\,c by the green star and the period of TOI-4311.03 is shown by the orange line since the radius is unknown. The computed radius following the relations by \citet{Zeng_Growth_Model_2019} for a composition of 100\% iron is shown in red and for 100\% H$_2$O-gaseous atmosphere at 500\,K in blue. }
    \label{fig:tess_sensitivity}
\end{figure}

\begin{figure*}
    \centering
    \includegraphics[width=\linewidth]{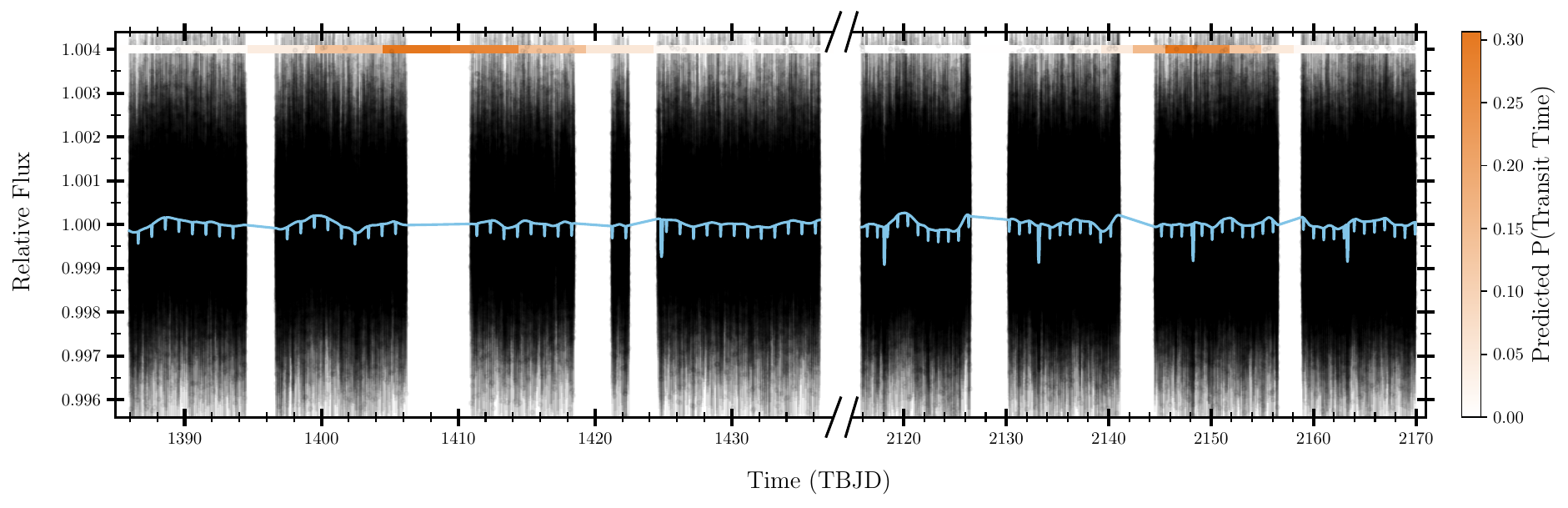}
    \caption{GP and transit model from the \textsc{juliet} fit in lightblue of the \textit{TESS} data in black. We show the posterior distribution and hence where the transit of the planet candidate is predicted to most likely fall by the intensity of the orange colour.}
    \label{fig:transit_planetd}
\end{figure*}

Since we identify an additional signal in the RVs that is well modelled by a Keplerian and is not indicated by stellar activity, it could arise from another planet in the system. In this scenario, we determine a minimum mass of $26.4\substack{+6.3\\-6.8}$\,M$_\oplus$ from the RVs. To confirm its planetary nature, we search for an additional signal of a transiting planet in the \textit{TESS} photometry. 
We perform a sensitivity analysis using the Transit Investigation and Recoverability Application \citep[TIaRA;][]{Rodel_TIaRA_2024,Eschen_PLATO_Sensitivities_2024}. We show the resulting sensitivity map in \autoref{fig:tess_sensitivity} including the detected planets by transit, TOI-4311\,b and TOI-4311\,c, as well as a vertical line for TOI-4311.03 since there is no measured radius. This analysis shows that TOI-4311\,b was easily found by \textit{TESS}, while TOI-4311\,c has a slightly lower chance of being found due to the limited baseline in time and its orbital period being longer than a \textit{TESS} orbit. As also found for TOI-4311\,c, this results in several transits falling into the gaps and only one transit being caught in the first two consecutive sectors. However, \textit{TESS} is still sensitive enough to find Sub-Neptunes at an orbital period of 15\,d orbiting TOI-4311.
For TOI-4311.03 we use the planet models of \citet{Zeng_Growth_Model_2019} and compute the predicted radii for the hypothetical cases that TOI-4311.03 was iron only (R=$1.749\substack{+0.078\\-0.101}$\,R$_\oplus$) or had a 100\% H$_2$O-gaseous atmosphere at 500\,K (R=$3.24\substack{+0.16\\-0.21}$\,R$_\oplus$). We highlight these two radii at the orbital period of planet d by the red and blue star in \autoref{fig:tess_sensitivity} respectively.
This places TOI-4311.03 in an area on the sensitivity map, where transiting planets can be detected but also missed depending on the transit times and coverage of these transits by \textit{TESS}. To check, whether transits of TOI-4311.03 could have been observed by \textit{TESS} we compute the transit times over the timespan of available \textit{TESS} data using the determined period and mid-transit time for this planet candidate obtained from our joint photometry RV analysis. We show this on top of the GP and transit model from \autoref{sec:photometry} in \autoref{fig:transit_planetd} to see if we can identify any of these potential transits. Since this signal was only detected by RVs, the uncertainties on period and mid-transit time are larger compared to planets that are detected by transit. Propagating the posteriors of the \textit{TESS} observations taken several years before the RVs results in an even wider distribution. This causes wide timespans of several days where the transit could have happened at the end of sector 3 or just at the beginning of sector 31. However, both of these timespans also cover observational gaps and hence we cannot rule out that this RV signal arises from an additional transiting planet. TOI-4311 will get observed though in \textit{TESS} sectors 97, 105, 106 and 108. As these sector times are known, we can compute the transit times for these sectors. Although we do not know the observational gaps in these sectors yet, these further observations will enable to rule out the transit or find it. However, even if there will not be a transit in the upcoming \textit{TESS} data, the signal can still be of planetary nature.

\autoref{fig:tess_sensitivity} also shows that planets below $\sim$1\,R$_\oplus$ at orbital periods between 1 and 15\,d could have been missed. Intriguingly, the two inner planets have a substantial gap in orbital periods and neither shows a high eccentricity. This could suggest that any planets in between TOI-4311\,b and c might be stable. However, we see no evidence of any additional transiting planets in the current data. The upcoming additional \textit{TESS} data could however reveal these and in general expand the discovery space to transiting planets of longer orbital period and smaller radii if these are present in the system. 

\begin{figure}
    \centering
    \includegraphics[width=0.99\linewidth]{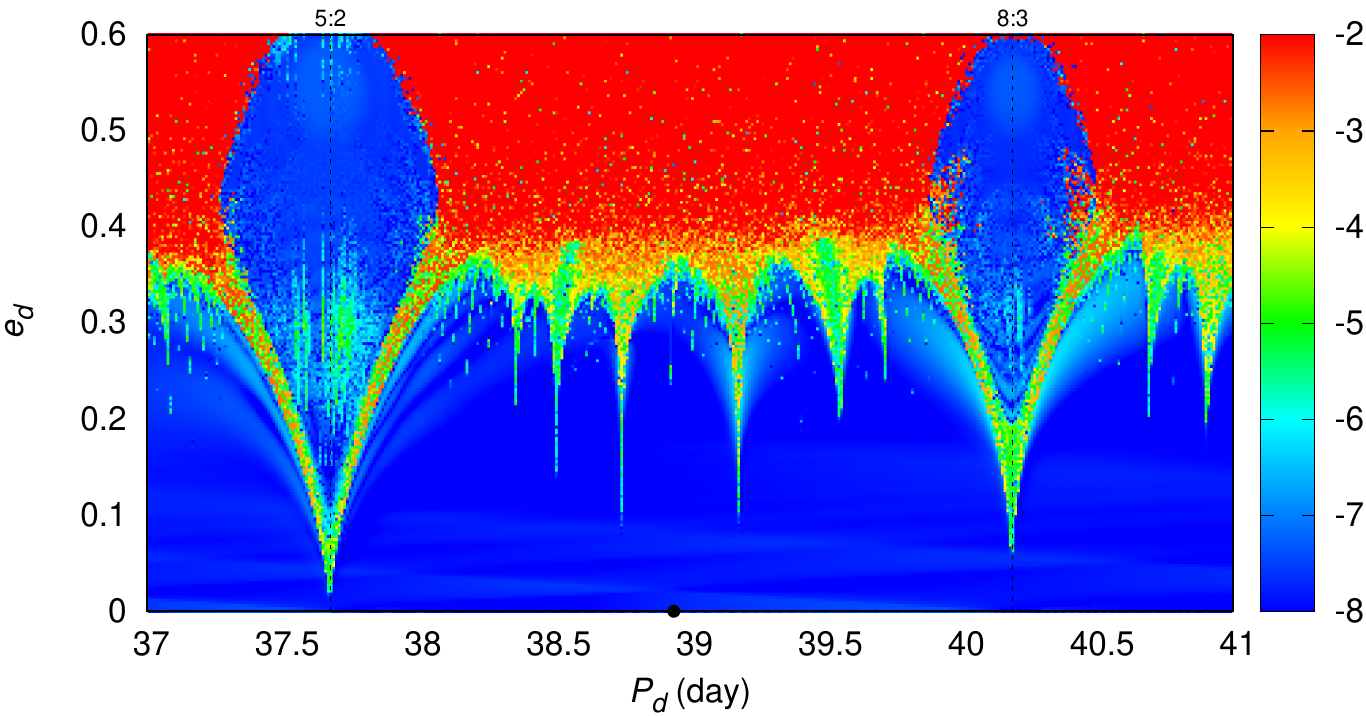} \\
    \includegraphics[width=1.00\linewidth]{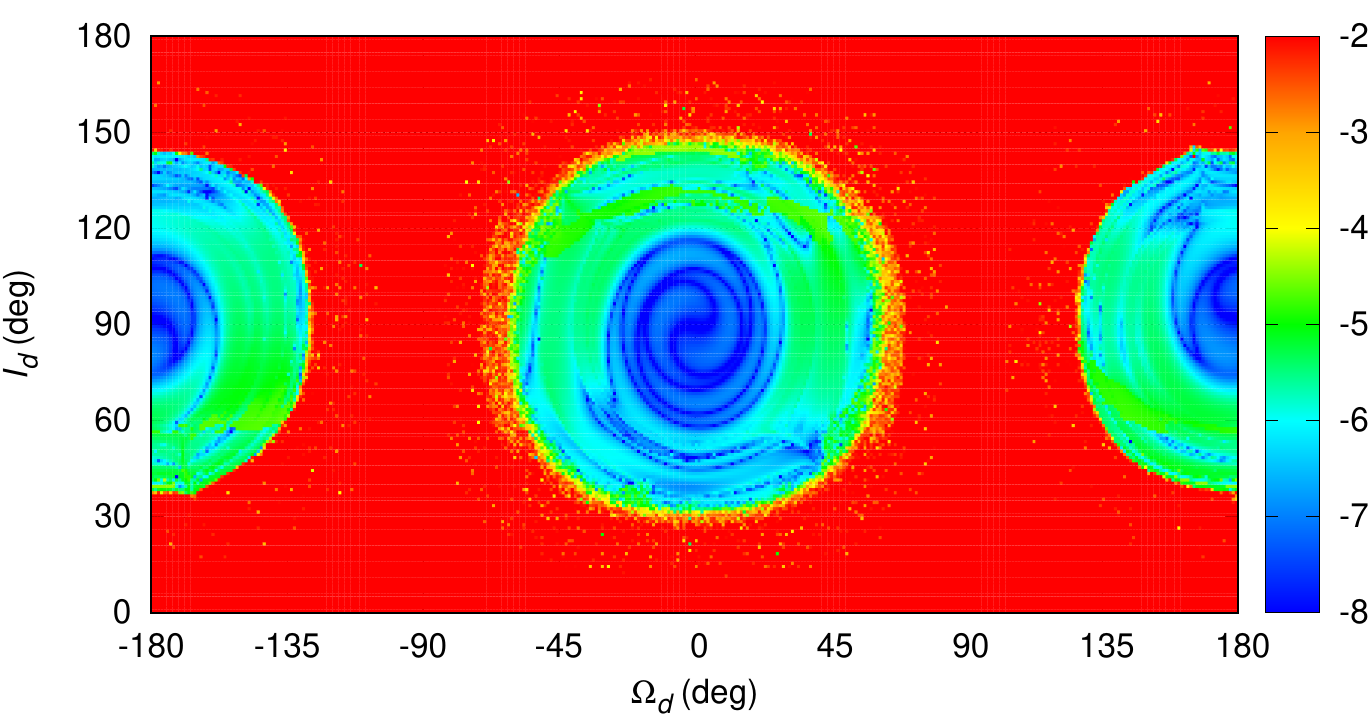}
    \caption{Stability analysis of the TOI-4311 planetary system focusing on planet\,d. We vary the orbital period $P_d$ and the eccentricity $e_d$ (\textit{top panel}) and the longitude of the node $\Omega_d$ and the inclination $I_d$ (\textit{bottom panel}), while keeping all the remaining orbital parameters fixed at their best-fit values (Tab.~\ref{tab:fit_results_planet}). Plots are colour-coded according to the stability indicator calculated from the frequency analysis of the mean longitude of TOI-4311.03. Red points correspond to highly unstable orbits, while blue points correspond to orbits that are likely to be stable on Gyr timescales. The vertical dashed line (\textit{top panel}) corresponds to the 5:2 and to the 8:3~MMR between planets c and d; the black dot at $P_d=38.9$\,d highlights the best fit solution from Tab.~\ref{tab:fit_results_planet}.}
    \label{fig:stabilityTOI43311d}
\end{figure}

Following \citet{correia2005,correia2010}, we also ran a stability analysis of the best-fit solution (Table~\ref{tab:fit_results_planet}). The system was integrated over $10^4$~yr for each initial condition and a stability indicator was calculated from the frequency analysis of the mean longitude \citep{laskar1990,laskar1993} of TOI-4311.03. We found that the three-planet system is stable, and lies in a region between the strong 5:2 and the 8:3 mean motion resonances between TOI-4311\,c and d (Fig.~\ref{fig:stabilityTOI43311d}, top panel). The system is also stable if we increase the eccentricity up to 0.3, which strengthens the possibility that the signal at 38.9~d is indeed a planet.
We also found that the inclination of TOI-4311.03, which is not constrained from RVs, must be within $30^\circ < I_d < 150^\circ$ (Fig.~\ref{fig:stabilityTOI43311d}, bottom panel). We thus conclude that $\sin I_d > 0.5$ and put an upper limit on planet-d true mass $M_d < 53\,M_\oplus$.

\subsection{Interior Structure Modelling}
\begin{figure}
    \centering
    \includegraphics[width=\linewidth]{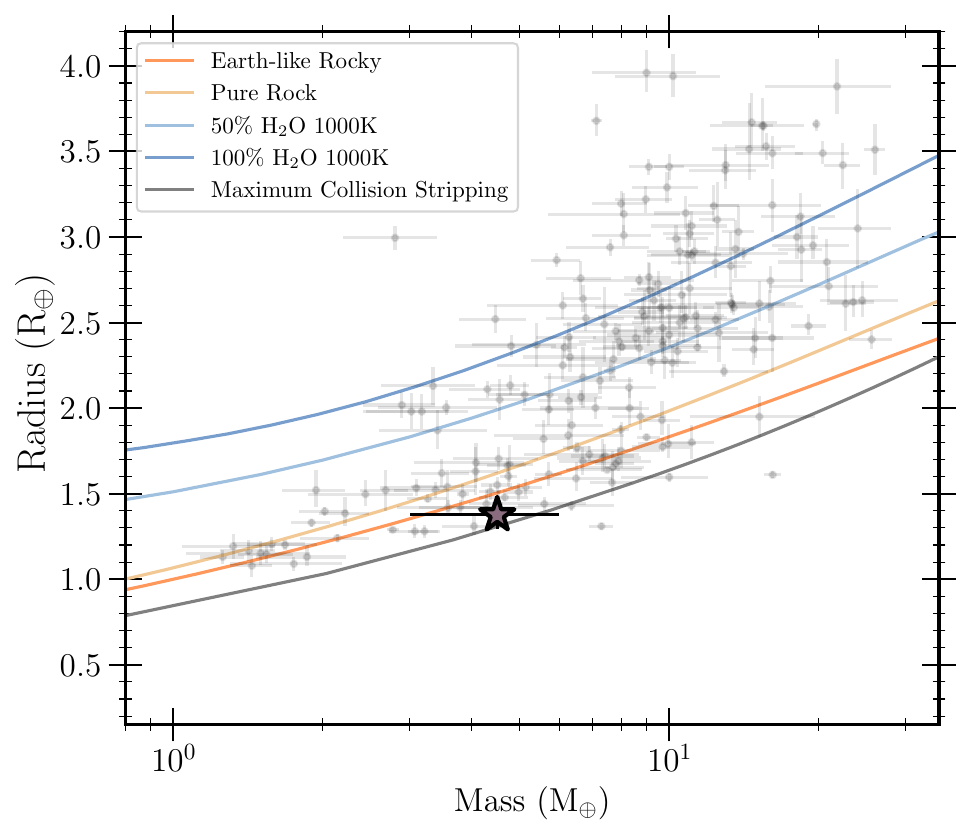}
    \caption{Comparison of the TOI-4311\,b to other well characterised planets below 4\,R$_\oplus$ on a mass-radius diagram. TOI-4311\,b is shown by the star in purple with the black border. We show compositional lines for Earth-like rocky, pure rock, 100\% water at 1000\,K and 50\% water at 1000\,K at the respected colour-coded lines following \citet{Zeng_Growth_Model_2019}. Additionally, we show the maximum collision stripping following \citet{Marcus_Giant_Impacts_Super_Earths_2010} by the gray line.}
    \label{fig:mass_radius}
\end{figure}

\begin{figure*}
    \centering
    \includegraphics[width=\linewidth]{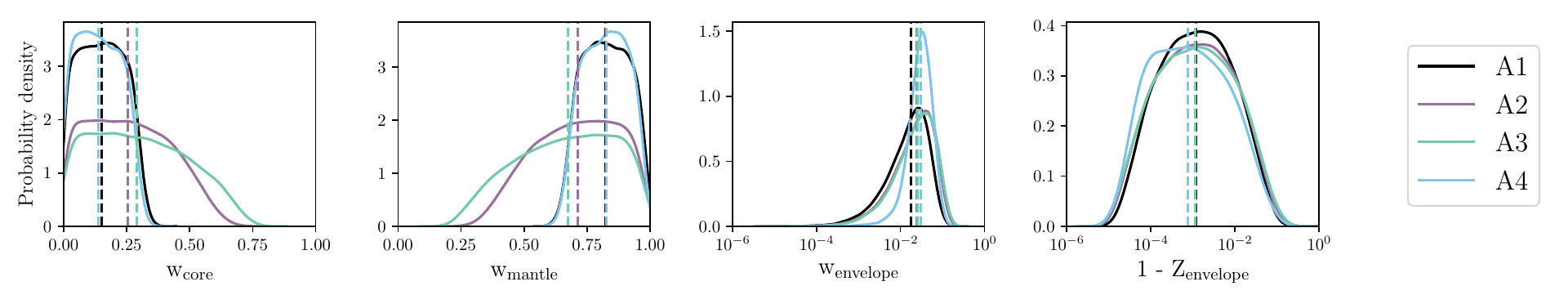}
    \caption{\textsc{planetic} posterior distributions for TOI-4311\,b. The different abundance priors are shown by the different colours. The median of each posterior is marked by the vertical dashed line.}
    \label{fig:plaNETic}
\end{figure*}

Using the \textit{TESS} and \textit{CHEOPS} photometry and HARPS spectra, we are able to determine a precise radius and mass for TOI-4311\,b. Combining the two, we compute a density of $9.3\substack{+3.8\\-3.0}$\,g/cm$^3$ for the planet. This in combination with the stellar parameters enables us to model its interior structure. We use \textsc{planetic} \citep{Egger_TOI469_plaNETic_2024}, a framework using Deep Neural Networks (DNNs) to determine the planets interior structure. Since \textsc{planetic}'s DNNs are trained for masses from 0.5-15\,M$_\oplus$, TOI-4311\,b is within its applicability. \textsc{planetic} models planets with up to three layers: a core, a mantle and a volatile layer mixed between water and hydrogen-helium. In addition to the planetary parameters \textsc{planetic} also accounts for stellar abundances as these are shown to be linked to the planetary composition/structure (see \autoref{sec:Compositional_link}). However, several scenarios for the planetary abundance priors are considered ranging from using the stellar abundances \citep{Thiabaud_elements_stars_planets_2015}, to iron-enriched priors \citet{Adibekyan_compositional_link_2021} or uniformly sampled priors with an upper limit of 75\% of Fe compared to the other two refractory elements. \citet{Eschen_TOI2345_2025} included an additional prior implementing \textsc{exoint} \citep{Wang_ExoInt_2019}, a tool that devolatises the stellar abundances into the theoretically predicted abundances for a rocky planet. We use all of these priors to model the interior structure of TOI-4311\,b. However, TOI-4311\,b has a high density compared to other planets on the mass-radius diagram shown in \autoref{fig:mass_radius}. Comparing it to the compositional lines from \citet{Marcus_Giant_Impacts_Super_Earths_2010} and \citet{Zeng_Growth_Model_2019}, accounting for TOI-4311\,b's uncertainties, it partially lies beyond the maximum collisional stripping line. Interior structures for such a dense planet are not covered within the \textsc{planetic} model. Hence we remove posterior samples of our interior structure modelling beyond this line. Since TOI-4311\,b is highly irradiated and likely does not contain a primordial atmosphere, we only use the water-rich prior as a purely H/He atmosphere is not physical for such a planet. 
We show our final results in \autoref{fig:plaNETic}. 
We find that TOI-4311\,b has a core mass fraction of $\sim$15\% for (devolatised) abundances. The $\sim$15\% core mass fraction implies that TOI-4311 b's high density can be achieved without an exceptionally iron-rich composition. Essentially, it's a mostly rocky (iron + silicate) planet with minimal gaseous envelope. The model suggests that high density does not automatically mean a Mercury-like iron enrichment, but could result from an Earth-like composition with no volatiles any more. Our \textsc{planetic} results also show that the atmospheric mass fraction of TOI-4311\,b is very low as expected for such a highly irradiated Super-Earth.

\subsection{TOI-4311\,b - a Super-Earth or Super-Mercury?}
\label{sec:Compositional_link}
Since stars and planets form from the same material \citep{Nielsen_Planet_Formation_2023}, their composition in refractory elements is expected to be similar, as found for the proto-sun and Earth \citep{Wang_Protosun_Abundances_2019}. This link has been suggested to exist for small exoplanets as well by \citet{Adibekyan_compositional_link_2021}, in which a sample of 22 planets that have precise mass, radius and abundances measurements is analysed. Comparing the planetary bulk density and stellar iron-to-silicate mass fraction, $f_{\text{iron}}^{\text{star}}$, which is a metric that represents the iron abundance of the host star, a compositional link was found. Their trend is also carried by the lack of small planets orbiting metal-poor stars, only including TOI-561\,b \citep[][]{Lacedelli_TOI561_2022} in their study. However, in recent years the sample of small planets around metal-poor stars has grown including TOI-1203\,b \citep{Gandolfi_TOI1203_2025} and K2-111\,b \citep[][]{Mortier_K2_111_2020}. Additionally, based on density \citet{Adibekyan_compositional_link_2021}, differentiate between Super-Earths and Super-Mercuries, suggesting the dearth between these two clusters arising from planet formation composition effects. 

\begin{figure}
    \centering
    \includegraphics[width=\linewidth]{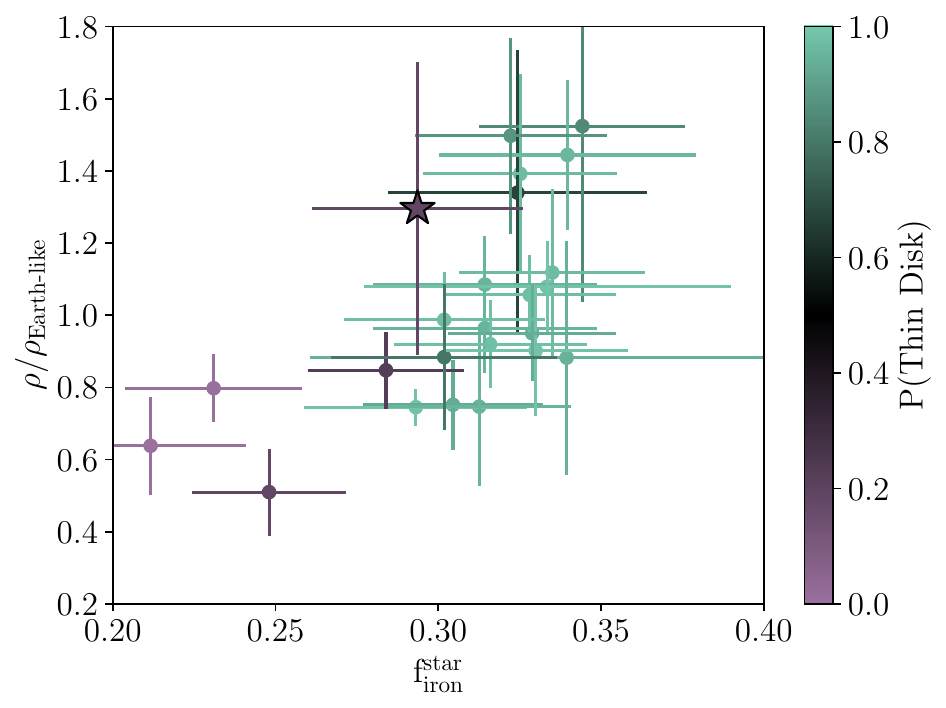}
    \caption{Stellar iron-to-silicate mass fraction against the planet's density normalised to an Earth-like composition for planets in \citet{Adibekyan_compositional_link_2021} and \citet{Mortier_K2_111_2020} and \citet{Gandolfi_TOI1203_2025}. The sample is colour-coded by the kinematic thin disk probability. TOI-4311\,b is highlighted by the star.}
    \label{fig:iron_fractions}
\end{figure}

Using the mass, radius and abundances reported in \citet{Adibekyan_compositional_link_2021} and the respective discovery papers, we compute the density of each planet, normalise it by the predicted density for an Earth-like planet for the given mass using the compositional lines computed by \citet{Zeng_Growth_Model_2019} and compute $f_{\text{iron}}^{\text{star}}$. Following these computations, we also compute the normalised density for TOI-4311\,b and $f_{\text{iron}}^{\text{star}}$ for TOI-4311 ($f_{\text{iron}}^{\text{star}}$=0.294+/-0.033). We place TOI-4311\,b on the diagram presented by \citet{Adibekyan_compositional_link_2021} in \autoref{fig:iron_fractions} and find that it lies in a unique region at a density between the Super-Mercuries and Super-Earths. Hence, TOI-4311\,b could bridge the gap between the two clusters, and test if the Super-Mercuries were sculpted via formation chemistry or impact-driven evolution \citep{Adibekyan_compositional_link_2021}. However, the uncertainties on the density are high and mainly arise from the high uncertainty on the mass. Hence more RVs are required to characterise TOI-4311\,b's mass and hence density more precisely which will allow to determine its position in \autoref{fig:iron_fractions} better. With this it would be clearer whether, TOI-4311\,b is a Super-Mercury at a very low $f_{\text{iron}}^{\text{star}}$ or the bridge between the Super-Earths and Super-Mercuries which could challenge the physical existence of these two populations.

Additionally, we compute the galactic kinematics and membership probabilities of each of these stars and colour-code the planets by their host star's kinematic thin disk probability. Targets with a low thin disk probability are shown in purple in \autoref{fig:iron_fractions}, while targets with a high thin disk probability are shown in green. This demonstrates a trend in the thin disk probability decreasing with $f_{\text{iron}}^{\text{star}}$ for the targets of the sample. This agrees with stars of the thin disk being chemically similar to the Sun ($f_{\text{iron}}^{\text{star}}$=0.332) \citep{Chen_PAST_I_2021}, while targets at a low $f_{\text{iron}}^{\text{star}}$ are mainly thick disk stars, which were formed in more metal-poor, $\alpha$-enhanced conditions \citep{Gondoin_chemical_abundances_history_milkyway_2024} resulting in a low value of $f_{\text{iron}}^{\text{star}}$. 

\subsection{TOI-4311's Galactic Membership}
While the formation of stars leaves imprints on their chemistry, the formation environment can also be reflected in its kinematics \citep{Quillen_GALAH_Hercules_2018}. As shown in \autoref{tab:stellarParam} and by $f_{\text{iron}}^{\text{star}}$ in \autoref{sec:Compositional_link}, TOI-4311 does not show any rare chemistry. 
Having computed the kinematic Galactic memberships in \autoref{sec:kinematic_analysis}, we find that TOI-4311 has a probability of 49\% of belonging to the Hercules stream, 33\% belonging to the thick disk and 17\% of belonging to the thin disk. Hence, TOI-4311 kinematically neither belongs clearly to the thin or thick disk of the Milky Way. However, its probability of 49\% of belonging to the Hercules stream is unique for a planet-hosting star. Plotting TOI-4311 in comparison to stars with clear membership probabilities ($>$50\%) (see \autoref{fig:toomre_divided}), highlights how TOI-4311 lies right at the boundary between the thin disk, thick disk and Hercules stream and could kinematically belong to either of these. 

Using the phase space density analysis, we also find inconclusive results.  TOI-4311 does have a very high phase space density when comparing it to stars with a similar velocity. However, this is likely attributed to its peculiar velocity being in plane. Hence TOI-4311 could be a star of the thin disk whose in-plane motion got excited. 

While the phase-space density analysis is mainly used to differentiate thin disk, thick disk and halo stars, the Hercules stream is not accounted for but could be an explanation of TOI-4311 kinematic behaviour. A variety of studies using different stellar samples have been analysing the origin and properties of this stream \citep{Bensby_Hercules_2007,Antoja_Hercules_2014,Ramay_Hercules_2016,Perez_Hercules_2017, Quillen_GALAH_Hercules_2018}. It is mainly found that stars of the Hercules stream range in their age and metallicity \citep{Bensby_Hercules_2007}. Hence, stars of the Hercules stream are identified through their kinematics, mainly their low value of the galactic velocity U. The Hercules stream is thought to have a dynamical origin and could be caused by perturbations from the Galactic bar. Studies using e.g. the GALAH survey \citep{Buder_GALAH_DR3_Overview_2021} have found that the Hercules stream can be identified more clearly for metal-rich stars \citep{Quillen_GALAH_Hercules_2018}. Their limit to identifying the Hercules stream based on metallicity is at [Fe/H]=-0.1. Due to these metallicity cuts, \citet{Ramay_Hercules_2016} and \citet{Quillen_GALAH_Hercules_2018} find that the Hercules stream could be dominated by stars with thin disc chemistry.
TOI-4311 with a metallicity of [Fe/H]=-0.07$\pm$0.04 lies within this range and hence lies within these chemical boundaries as well. However, the 49\% kinematic probability and the chemistry of TOI-4311 make it challenging to fully claim its membership to either galactic population. For this, a more detailed chemical characterisation of TOI-4311 is required as well as further studies of the Hercules stream to understand its stars and origins fully. 
We hence identify TOI-4311 to not clearly kinematically belong to any population. While its kinematics hint slightly towards the Hercules stream, its chemistry shows that it could belong to the thin disk or Hercules stream.

\section{Conclusion}

We report the discovery and characterisation of the TOI-4311 system using \textit{TESS}, \textit{CHEOPS}, and HARPS data. TOI-4311 is transited by an Ultra-Short Period Super-Earth at 0.99\,d and a Sub-Neptune at $\sim$15\,d. We find an additional planet candidate at $\sim$38\,d in the RVs that cannot be attributed to stellar activity. Investigating the \textit{TESS} data, we do not find any transits of this planet candidate. We estimate transit depths for iron- and water-rich bodies of that mass and show that \textit{TESS} would be sensitive to detect a planet of this size if it was transiting. However, our analysis finds that the transit could have been missed due to gaps. More photometric data are needed to reveal this. 

The precise radius and mass characterisation of TOI-4311\,b allows us to compare it to other well-characterised small planets around compositionally-diverse stars. We find that this planet lies just at the boundary between Super-Earths and Super-Mercuries which could challenge compositional formation theories. A better mass characterisation would allow us to constrain the planet's density more clearly and test these models.

Finally, we analysed the kinematics of TOI-4311 which places it at the border of the Galactic Hercules stream and therefore it is kinematically an interesting system due to a different formation environment.

In summary, TOI-4311 is a rare example of a (likely) thick-disk or Hercules-stream star hosting multiple small planets. Its ultra-short-period planet has an unusually high density but not an iron-enriched composition, while the systems galactic properties tie into broader trends of planet occurrence in different stellar populations. This system provides a bridge between chemical/dynamical Galactic history and exoplanetary science.

\section*{Acknowledgements}
CHEOPS is an ESA mission in partnership with Switzerland with important contributions to the payload and the ground segment from Austria, Belgium, France, Germany, Hungary, Italy, Portugal, Spain, Sweden, and the United Kingdom. The CHEOPS Consortium would like to gratefully acknowledge the support received by all the agencies, offices, universities, and industries involved. Their flexibility and willingness to explore new approaches were essential to the success of this mission. CHEOPS data analysed in this article will be made available in the CHEOPS mission archive (\url{https://cheops.unige.ch/archive_browser/}). 
This paper made use of data collected by the TESS mission and are publicly available from the Mikulski Archive for Space Telescopes (MAST) operated by the Space Telescope Science Institute (STScI). Funding for the TESS mission is provided by NASA’s Science Mission Directorate. We acknowledge the use of public TESS data from pipelines at the TESS Science Office and at the TESS Science Processing Operations Center. Resources supporting this work were provided by the NASA High-End Computing (HEC) Program through the NASA Advanced Supercomputing (NAS) Division at Ames Research Center for the production of the SPOC data products.
YNEE acknowledges support from a Science and Technology Facilities Council (STFC) studentship, grant number ST/Y509693/1. 
TWi acknowledges support from the UKSA and the University of Warwick. 
ACC acknowledges support from STFC consolidated grant number ST/V000861/1, and UKSA grant number ST/X002217/1. 
AJM gratefully acknowledges support from the Swedish National Space Agency (grant 2023-00146).
JAE acknowledges support through the European Space Agency (ESA) Research Fellowship programme in Space Science. 
DG gratefully acknowledges financial support from the CRT foundation under Grant No. 2018.2323 ‘Gaseous or rocky? Unveiling the nature of small worlds’. 
This work was also supported by FCT - Fundação para a Ciência e a Tecnologia through national funds by grants reference UID/04434/2025. 
O.D.S.D. is supported in the form of work contract (DL 57/2016/CP1364/CT0004) funded by national funds through FCT. 
S.G.S. acknowledge support from FCT through FCT contract nr. CEECIND/00826/2018 and POPH/FSE (EC). 
The Portuguese team thanks the Portuguese Space Agency for the provision of financial support in the framework of the PRODEX Programme of the European Space Agency (ESA) under contract number 4000142255. 
ACo, ADe, BEd, KGa, and JKo acknowledge their role as ESA-appointed CHEOPS Science Team Members. 
ABr was supported by the SNSA. 
BAk and MLe acknowledge support of the Swiss National Science Foundation under grant number PCEFP2\_194576. 
YAl acknowledges support from the Swiss National Science Foundation (SNSF) under grant 200020\_192038. 
RAl, DBa, EPa, IRi, and EVi acknowledge financial support from the Agencia Estatal de Investigación of the Ministerio de Ciencia e Innovación MCIN/AEI/10.13039/501100011033 and the ERDF “A way of making Europe” through projects PID2021-125627OB-C31, PID2021-125627OB-C32, PID2021-127289NB-I00, PID2023-150468NB-I00 and PID2023-149439NB-C41. 
SCCB acknowledges the support from Fundação para a Ciência e Tecnologia (FCT) in the form of work contract through the Scientific Employment Incentive program with reference 2023.06687.CEECIND and DOI 10.54499/2023.06687.CEECIND/CP2839/CT0002. 
LBo, VNa, IPa, GPi, RRa, GSc, and TZi acknowledge support from CHEOPS ASI-INAF agreement n. 2019-29-HH.0. 
CBr and ASi acknowledge support from the Swiss Space Office through the ESA PRODEX program. 
This project was supported by the CNES. 
ADe acknowledges financial support from the Swiss National Science Foundation (SNSF) for project 200021\_200726. 
B.-O. D. acknowledges support from the Swiss State Secretariat for Education, Research and Innovation (SERI) under contract number MB22.00046. 
This project has received funding from the Swiss National Science Foundation for project 200021\_200726. It has also been carried out within the framework of the National Centre of Competence in Research PlanetS supported by the Swiss National Science Foundation under grant 51NF40\_205606. The authors acknowledge the financial support of the SNSF. 
MF and CMP gratefully acknowledge the support of the Swedish National Space Agency (DNR 65/19, 174/18). 
M.G. is F.R.S.-FNRS Research Director. 
MNG is the ESA CHEOPS Project Scientist and Mission Representative. BMM is the ESA CHEOPS Project Scientist. KGI was the ESA CHEOPS Project Scientist until the end of December 2022 and Mission Representative until the end of January 2023. All of them are/were responsible for the Guest Observers (GO) Programme. None of them relay/relayed proprietary information between the GO and Guaranteed Time Observation (GTO) Programmes, nor do/did they decide on the definition and target selection of the GTO Programme. 
CHe acknowledges financial support from the Österreichische Akademie der Wissenschaften. 
CHe acknowledges the European Union H2020-MSCA-ITN-2019 under Grant Agreement no. 860470 (CHAMELEON). 
Calculations were performed using supercomputer resources provided by the Vienna Scientific Cluster (VSC). 
K.W.F.L. was supported by Deutsche Forschungsgemeinschaft grants RA714/14-1 within the DFG Schwerpunkt SPP 1992, Exploring the Diversity of Extrasolar Planets. 
This work was granted access to the HPC resources of MesoPSL financed by the Region Ile de France and the project Equip@Meso (reference ANR-10-EQPX-29-01) of the programme Investissements d'Avenir supervised by the Agence Nationale pour la Recherche. 
PM acknowledges support from STFC research grant number ST/R000638/1. 
This work was also partially supported by a grant from the Simons Foundation (PI Queloz, grant number 327127). 
NCSa acknowledges funding by the European Union (ERC, FIERCE, 101052347). Views and opinions expressed are however those of the author(s) only and do not necessarily reflect those of the European Union or the European Research Council. Neither the European Union nor the granting authority can be held responsible for them. 
This work was funded by the European Union (ERC, FIERCE, 101052347). Views and opinions expressed are however those of the author(s) only and do not necessarily reflect those of the European Union or the European Research Council. Neither the European Union nor the granting authority can be held responsible for them. This work was also supported by FCT - Fundação para a Ciência e a Tecnologia through national funds by these grants: UIDB/04434/2020 DOI: 10.54499/UIDB/04434/2020, UIDP/04434/2020 DOI: 10.54499/UIDP/04434/2020, PTDC/FIS-AST/4862/2020, UID/04434/2025. 
Gy.M.Sz. thanks the SNN-147362 and the ADVANCED-153410 of the National Research, Development and Innovation Office (NKFIH, Hungary), and the ESA PRODEX Experiment Agreements No. 4000137122 No. 4000149203.
V.V.G. is an F.R.S-FNRS Research Associate. 
JV acknowledges support from the Swiss National Science Foundation (SNSF) under grant PZ00P2\_208945. 
NAW acknowledges UKSA grant ST/R004838/1.

\section*{Data Availability}
The data underlying this article will be made available in the \textit{CHEOPS} mission archive \href{https://cheops.unige.ch/archive_browser/}{https://cheops.unige.ch/archive$\_$browser/}. This paper includes data collected by the \textit{TESS} mission, which is publicly available from the Mikulski Archive for Space Telescopes (MAST) at the Space Telescope Science Institute (STScI) \href{https://mast.stsci.edu}{https://mast.stsci.edu}.

\bibliographystyle{mnras}
\bibliography{example}

\vspace{1cm}
\noindent 
$^{1}$Department of Physics, University of Warwick, Gibbet Hill Road, Coventry CV4 7AL, United Kingdom \\
$^{2}$Centre for Exoplanet Science, SUPA School of Physics and Astronomy, University of St Andrews, North Haugh, St Andrews KY16 9SS, UK \\
$^{3}$Lund Observatory, Dept. of Astronomy and Theoretical Physics, Lund University, Box 43, 22100 Lund, Sweden \\
$^{4}$European Space Agency (ESA), European Space Research and Technology Centre (ESTEC), Keplerlaan 1, 2201 AZ Noordwijk, The Netherlands \\
$^{5}$Leiden Observatory, University of Leiden, Einsteinweg 55, 2333 CA Leiden, The Netherlands \\
$^{6}$Space sciences, Technologies and Astrophysics Research (STAR) Institute, Université de Liège, Allée du 6 Août 19C, 4000 Liège, Belgium \\
$^{7}$Dipartimento di Fisica, Università degli Studi di Torino, via Pietro Giuria 1, I-10125, Torino, Italy \\
$^{8}$Institute of Space Research, German Aerospace Center (DLR), Rutherfordstrasse 2, 12489 Berlin, Germany \\
$^{9}$Instituto de Astrofisica e Ciencias do Espaco, Universidade do Porto, CAUP, Rua das Estrelas, 4150-762 Porto, Portugal \\
$^{10}$Departamento de Fisica e Astronomia, Faculdade de Ciencias, Universidade do Porto, Rua do Campo Alegre, 4169-007 Porto, Portugal \\
$^{11}$Space Research Institute, Austrian Academy of Sciences, Schmiedlstrasse 6, A-8042 Graz, Austria \\
$^{12}$CFisUC, Departamento de Física, Universidade de Coimbra, 3004-516 Coimbra, Portugal \\
$^{13}$Instituto de Astrofísica de Canarias, Vía Láctea s/n, 38200 La Laguna, Tenerife, Spain \\
$^{14}$Departamento de Astrofísica, Universidad de La Laguna, Astrofísico Francisco Sanchez s/n, 38206 La Laguna, Tenerife, Spain \\
$^{15}$Department of Astronomy, Stockholm University, AlbaNova University Center, 10691 Stockholm, Sweden \\
$^{16}$Observatoire astronomique de l'Université de Genève, Chemin Pegasi 51, 1290 Versoix, Switzerland \\
$^{17}$Center for Space and Habitability, University of Bern, Gesellschaftsstrasse 6, 3012 Bern, Switzerland \\
$^{18}$Space Research and Planetary Sciences, Physics Institute, University of Bern, Gesellschaftsstrasse 6, 3012 Bern, Switzerland \\
$^{19}$Admatis, 5. Kandó Kálmán Street, 3534 Miskolc, Hungary \\
$^{20}$Depto. de Astrofísica, Centro de Astrobiología (CSIC-INTA), ESAC campus, 28692 Villanueva de la Cañada (Madrid), Spain \\
$^{21}$INAF, Osservatorio Astronomico di Padova, Vicolo dell'Osservatorio 5, 35122 Padova, Italy \\
$^{22}$INAF, Osservatorio Astrofisico di Torino, Via Osservatorio, 20, I-10025 Pino Torinese To, Italy \\
$^{23}$Centre for Mathematical Sciences, Lund University, Box 118, 221 00 Lund, Sweden \\
$^{24}$Aix Marseille Univ, CNRS, CNES, LAM, 38 rue Frédéric Joliot-Curie, 13388 Marseille, France \\
$^{25}$ARTORG Center for Biomedical Engineering Research, University of Bern, Bern, Switzerland \\
$^{26}$ELTE Gothard Astrophysical Observatory, 9700 Szombathely, Szent Imre h. u. 112, Hungary \\
$^{27}$INAF, Istituto di Astrofisica e Planetologia Spaziali, via del Fosso del Cavaliere 100, 00133 Roma, Italy \\
$^{28}$SRON Netherlands Institute for Space Research, Niels Bohrweg 4, 2333 CA Leiden, Netherlands \\
$^{29}$Centre Vie dans l’Univers, Faculté des sciences, Université de Genève, Quai Ernest-Ansermet 30, 1211 Genève 4, Switzerland \\
$^{30}$Leiden Observatory, University of Leiden, PO Box 9513, 2300 RA Leiden, The Netherlands \\
$^{31}$Department of Space, Earth and Environment, Chalmers University of Technology, Onsala Space Observatory, 439 92 Onsala, Sweden \\
$^{32}$National and Kapodistrian University of Athens, Department of Physics, University Campus, Zografos GR-157 84, Athens, Greece \\
$^{33}$Astrobiology Research Unit, Université de Liège, Allée du 6 Août 19C, B-4000 Liège, Belgium \\
$^{34}$Department of Astrophysics, University of Vienna, Türkenschanzstrasse 17, 1180 Vienna, Austria \\
$^{35}$Institute for Theoretical Physics and Computational Physics, Graz University of Technology, Petersgasse 16, 8010 Graz, Austria \\
$^{36}$NASA Ames Research Center, Moﬀett Field, CA 94035, USA \\
$^{37}$Konkoly Observatory, Research Centre for Astronomy and Earth Sciences, 1121 Budapest, Konkoly Thege Miklós út 15-17, Hungary \\
$^{38}$ELTE E\"otv\"os Lor\'and University, Institute of Physics, P\'azm\'any P\'eter s\'et\'any 1/A, 1117 Budapest, Hungary \\
$^{39}$IMCCE, UMR8028 CNRS, Observatoire de Paris, PSL Univ., Sorbonne Univ., 77 av. Denfert-Rochereau, 75014 Paris, France \\
$^{40}$Institut d'astrophysique de Paris, UMR7095 CNRS, Université Pierre \& Marie Curie, 98bis blvd. Arago, 75014 Paris, France \\
$^{41}$Astrophysics Group, Lennard Jones Building, Keele University, Staffordshire, ST5 5BG, United Kingdom \\
$^{42}$European Space Agency, ESA - European Space Astronomy Centre, Camino Bajo del Castillo s/n, 28692 Villanueva de la Cañada, Madrid, Spain \\
$^{43}$ETH Zurich, Department of Physics, Wolfgang-Pauli-Strasse 2, CH-8093 Zurich, Switzerland \\
$^{44}$INAF, Osservatorio Astrofisico di Catania, Via S. Sofia 78, 95123 Catania, Italy \\
$^{45}$Kavli Institute for Astrophysics and Space Research, Massachusetts Institute of Technology, Cambridge, MA, USA \\
$^{46}$Weltraumforschung und Planetologie, Physikalisches Institut, University of Bern, Gesellschaftsstrasse 6, 3012 Bern, Switzerland \\
$^{47}$Dipartimento di Fisica e Astronomia "Galileo Galilei", Università degli Studi di Padova, Vicolo dell'Osservatorio 3, 35122 Padova, Italy \\
$^{48}$Cavendish Laboratory, JJ Thomson Avenue, Cambridge CB3 0HE, UK \\
$^{49}$German Aerospace Center (DLR), Markgrafenstrasse 37, 10117 Berlin, Germany \\
$^{50}$Institut fuer Geologische Wissenschaften, Freie Universitaet Berlin, Malteserstrasse 74-100,12249 Berlin, Germany \\
$^{51}$Institut de Ciencies de l'Espai (ICE, CSIC), Campus UAB, Can Magrans s/n, 08193 Bellaterra, Spain \\
$^{52}$Institut d'Estudis Espacials de Catalunya (IEEC), 08860 Castelldefels (Barcelona), Spain \\
$^{53}$European Space Agency (ESA), European Space Operations Centre (ESOC), Robert-Bosch-Str. 5, D-64293 Darmstadt, Germany \\
$^{54}$Institute of Astronomy, University of Cambridge, Madingley Road, Cambridge, CB3 0HA, United Kingdom \\
$^{55}$Department of Physics, Engineering and Astronomy, Stephen F. Austin State University, 1936 North St, Nacogdoches, TX 75962, USA

\appendix

\section{\textit{TESS} Photometry}
\begin{figure*}
    \centering
    \includegraphics[width=\linewidth]{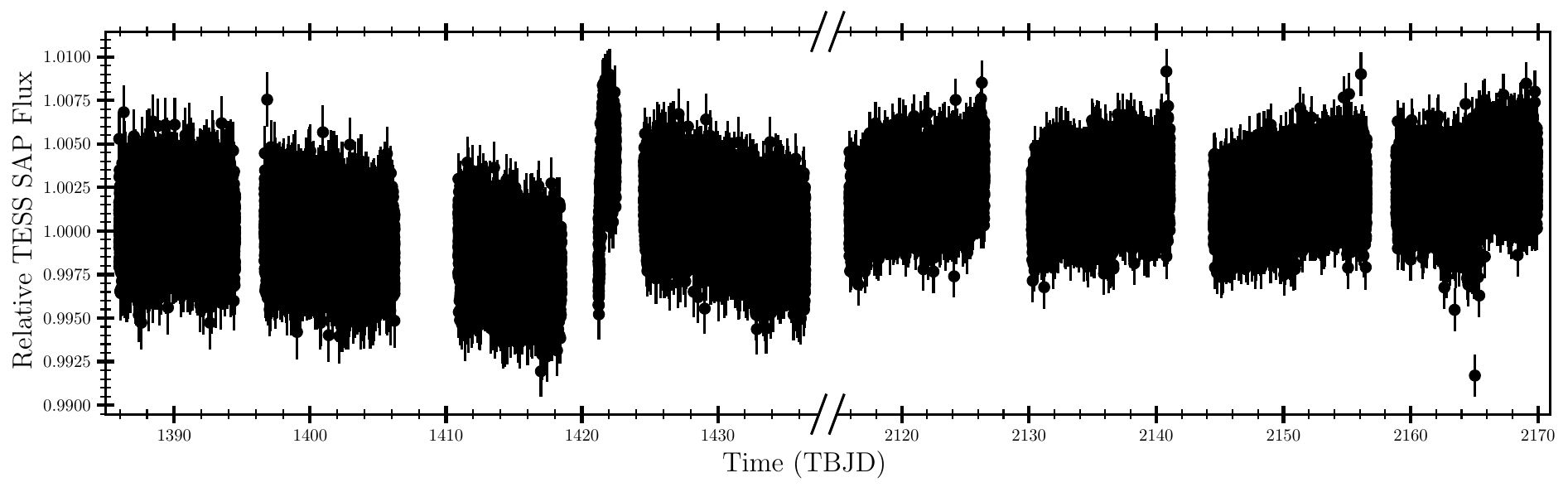}
    \caption{\textit{TESS} SAP flux from sectors 3,4,30 and 31.}
    \label{fig:tess_sap_lightcurve}
\end{figure*}

\begin{figure*}
    \centering
    \includegraphics[width=\linewidth]{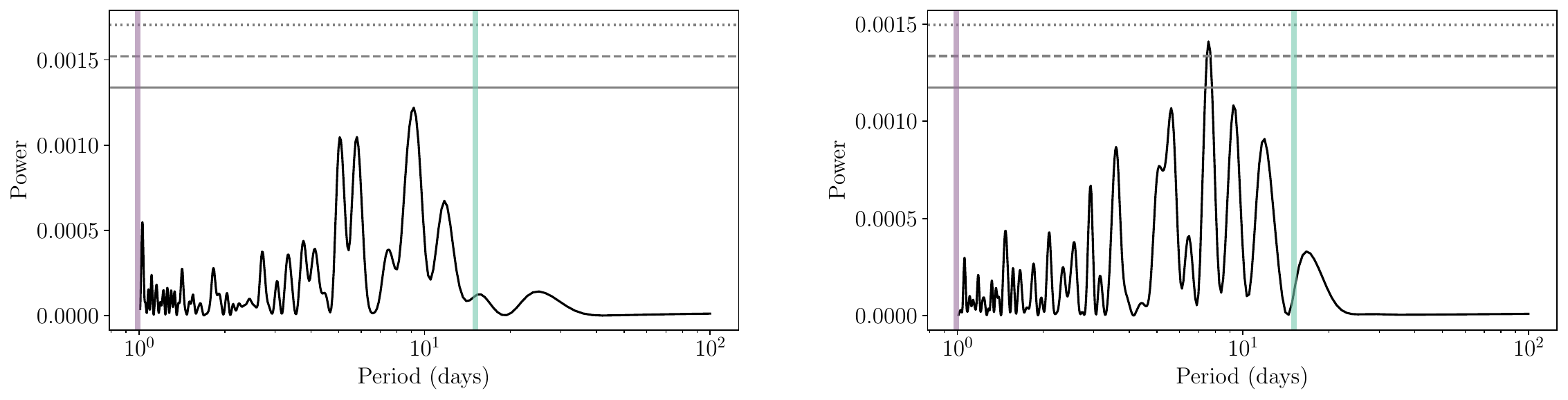}
    \caption{Lomb-Scargle periodograms of \textit{TESS} sectors 3 and 4 (left) and 30 and 31 (right).}
    \label{fig:tess_periodogram}
\end{figure*}

\section{\textit{CHEOPS} Photometry}
\begin{figure*}
    \centering
    \includegraphics[width=0.3\linewidth]{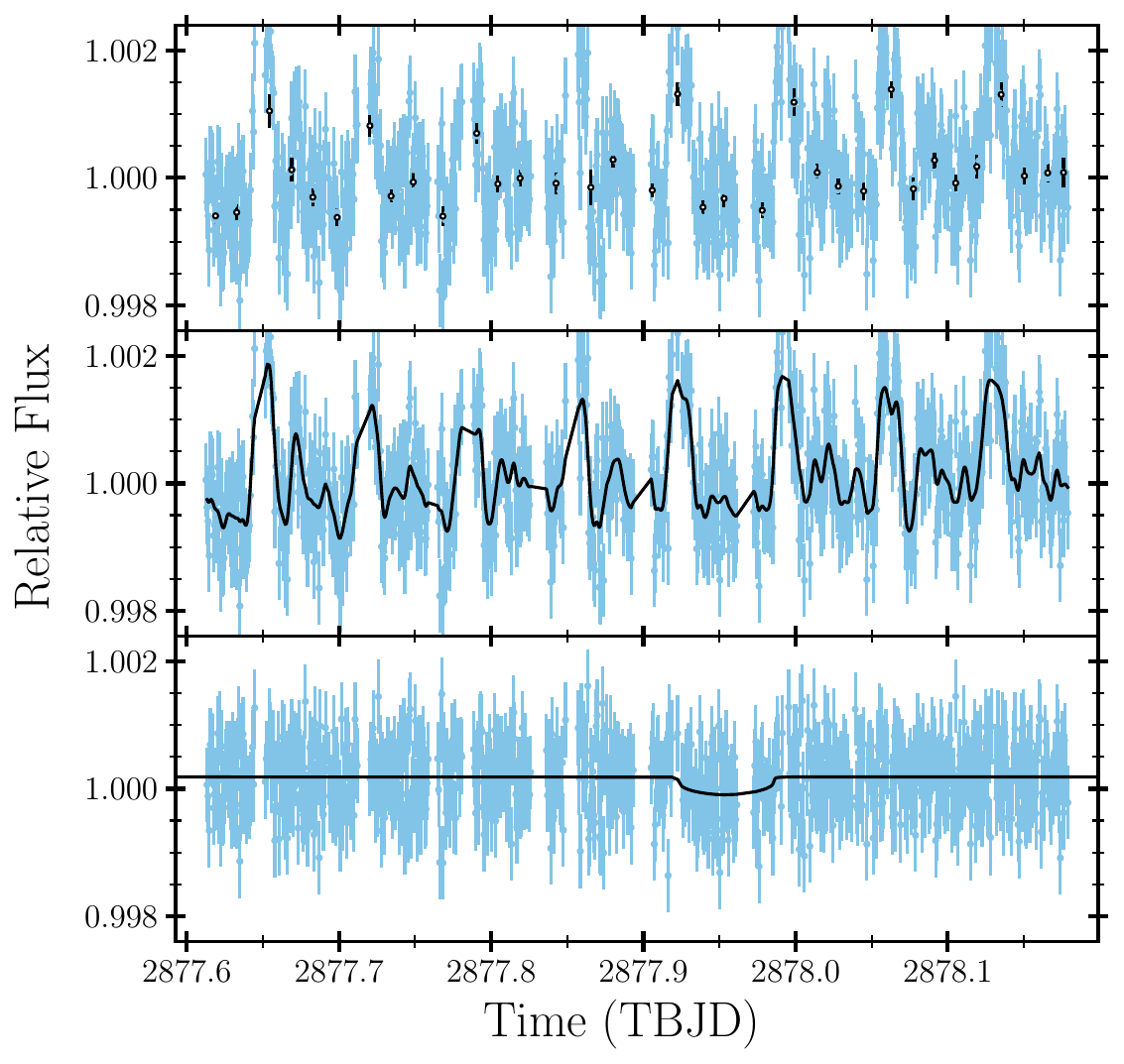}
    \quad
    \includegraphics[width=0.3\linewidth]{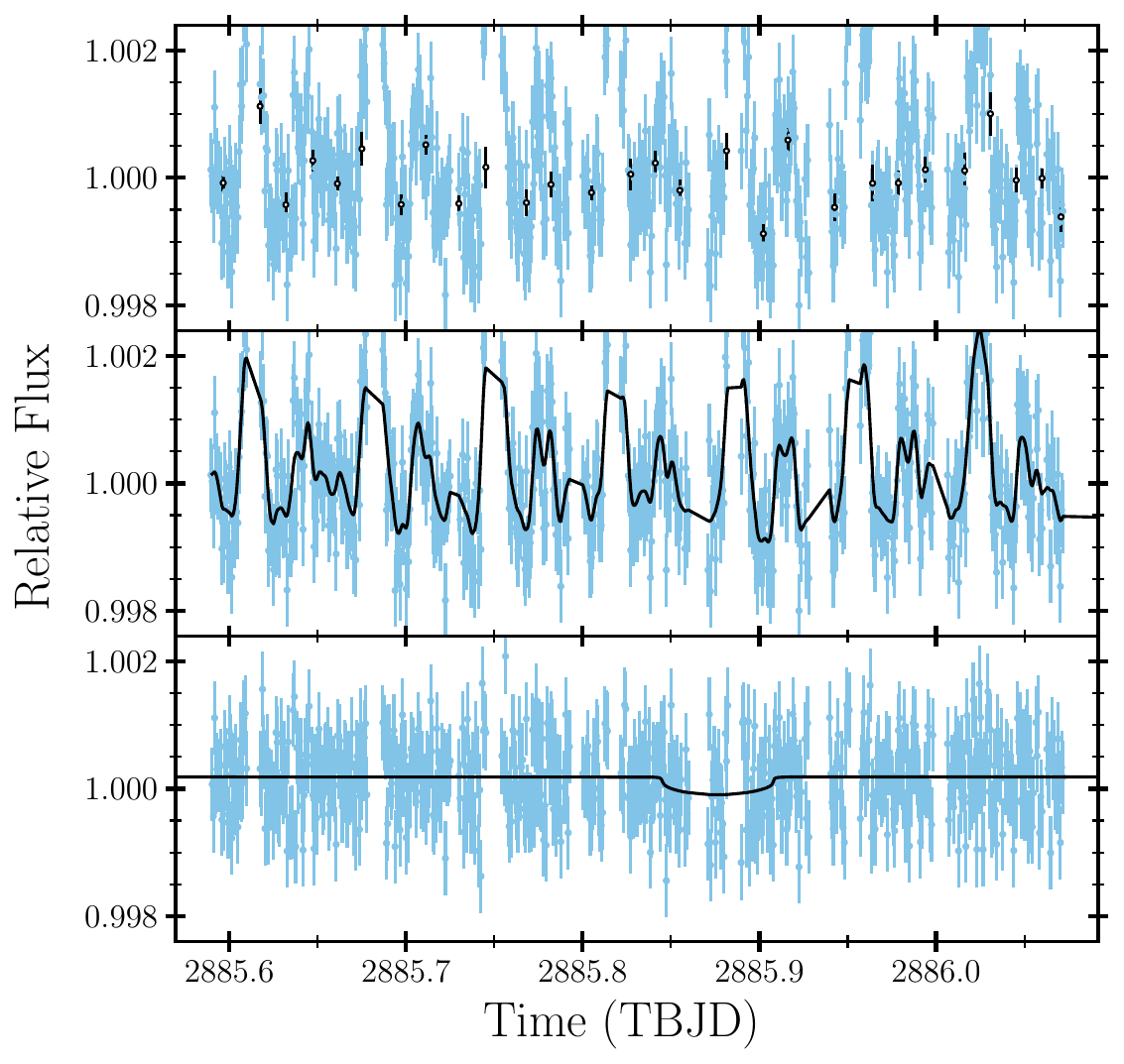}
    \quad
    \includegraphics[width=0.3\linewidth]{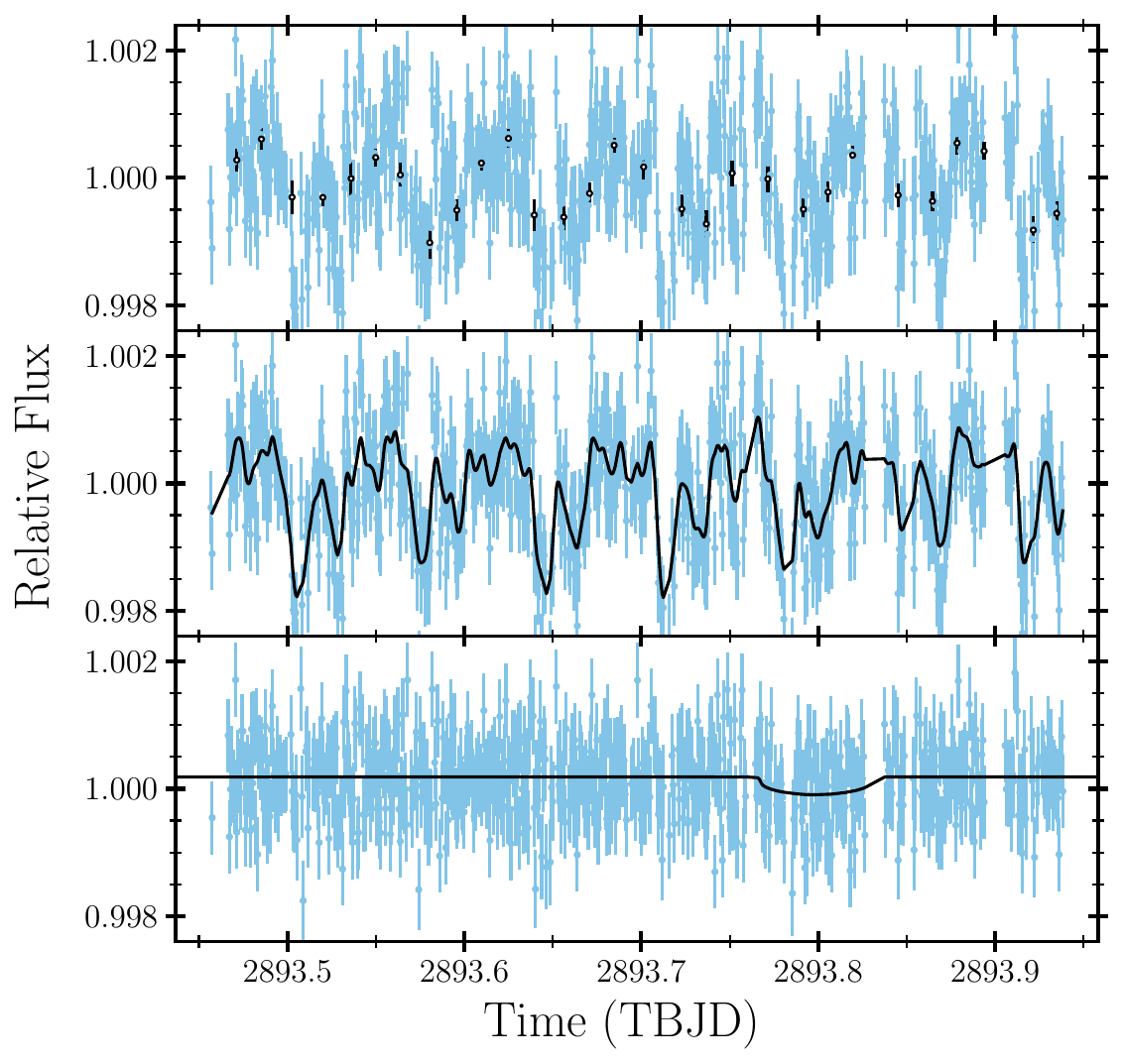}
    \quad
    \includegraphics[width=0.3\linewidth]{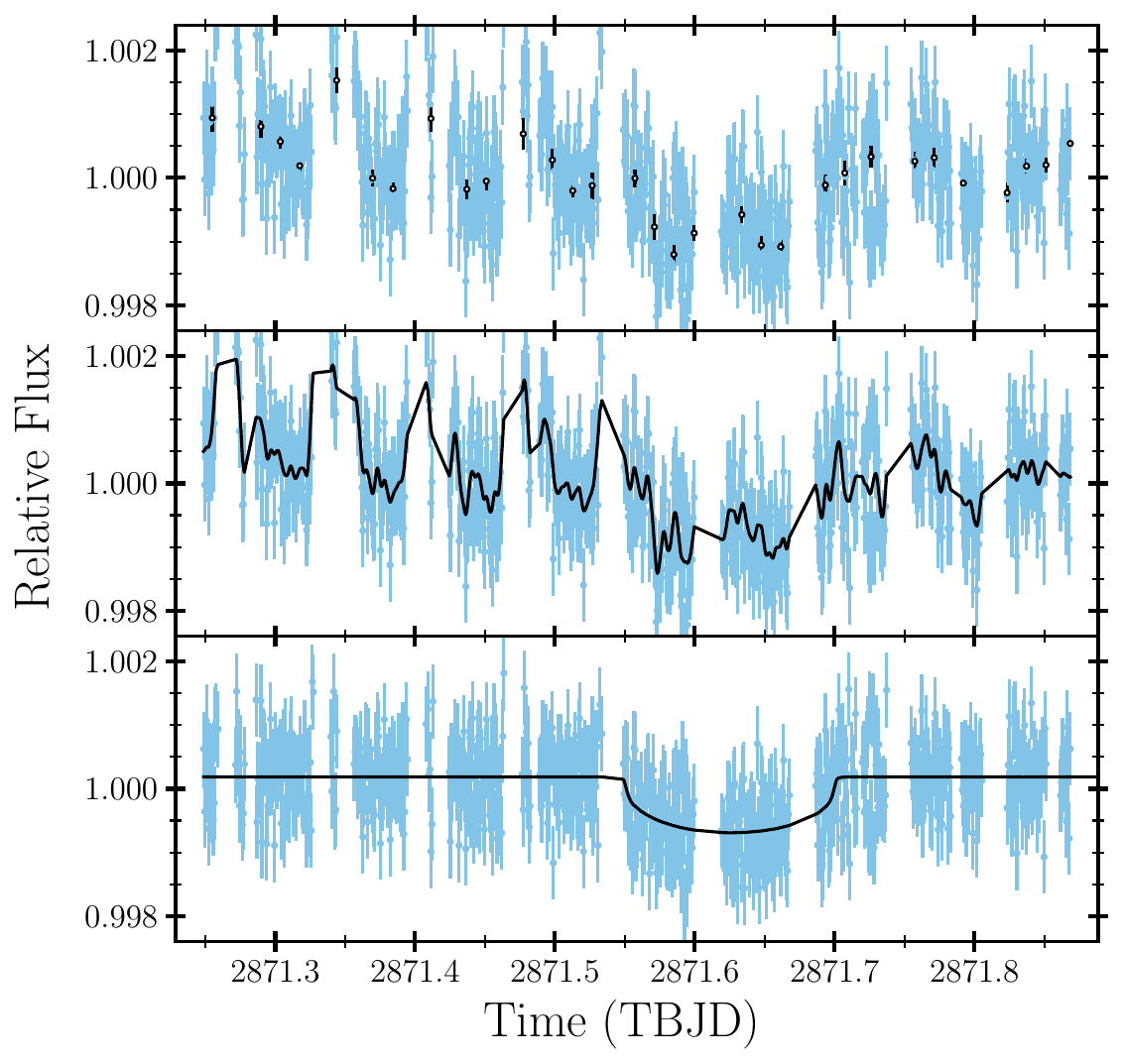}
    \quad
    \includegraphics[width=0.3\linewidth]{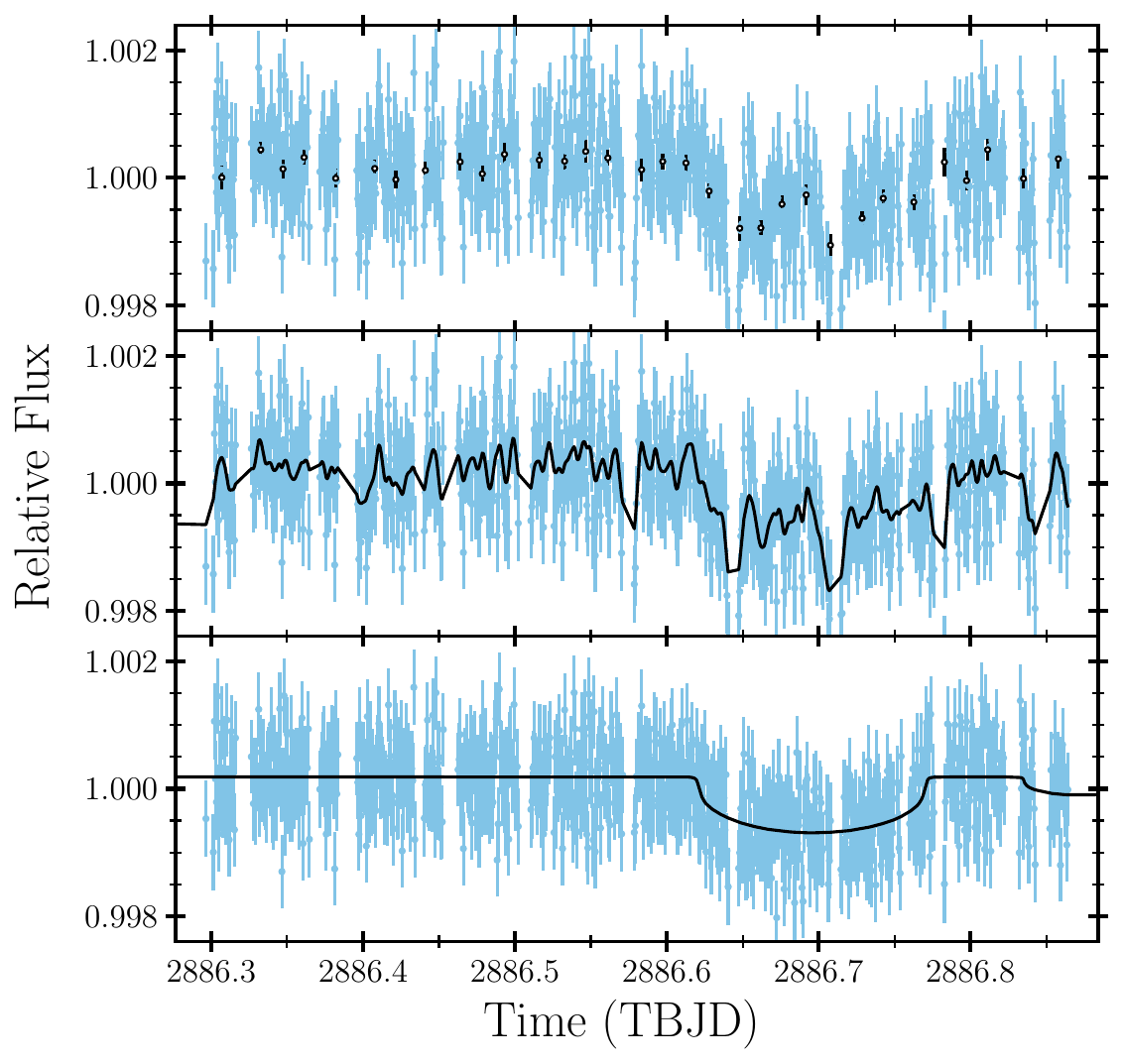}
    \quad
    \includegraphics[width=0.3\linewidth]{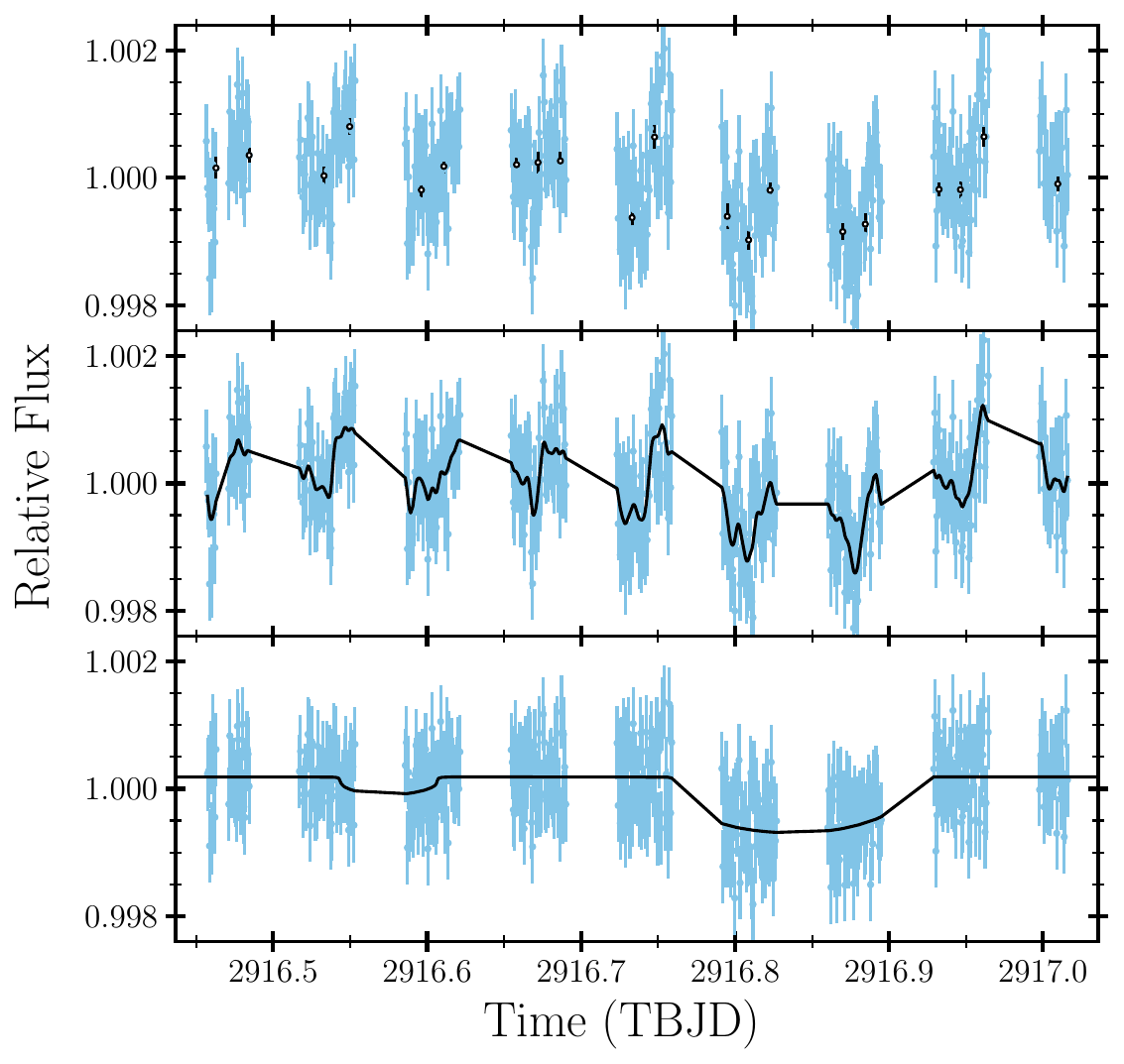}
    \quad
    \includegraphics[width=0.3\linewidth]{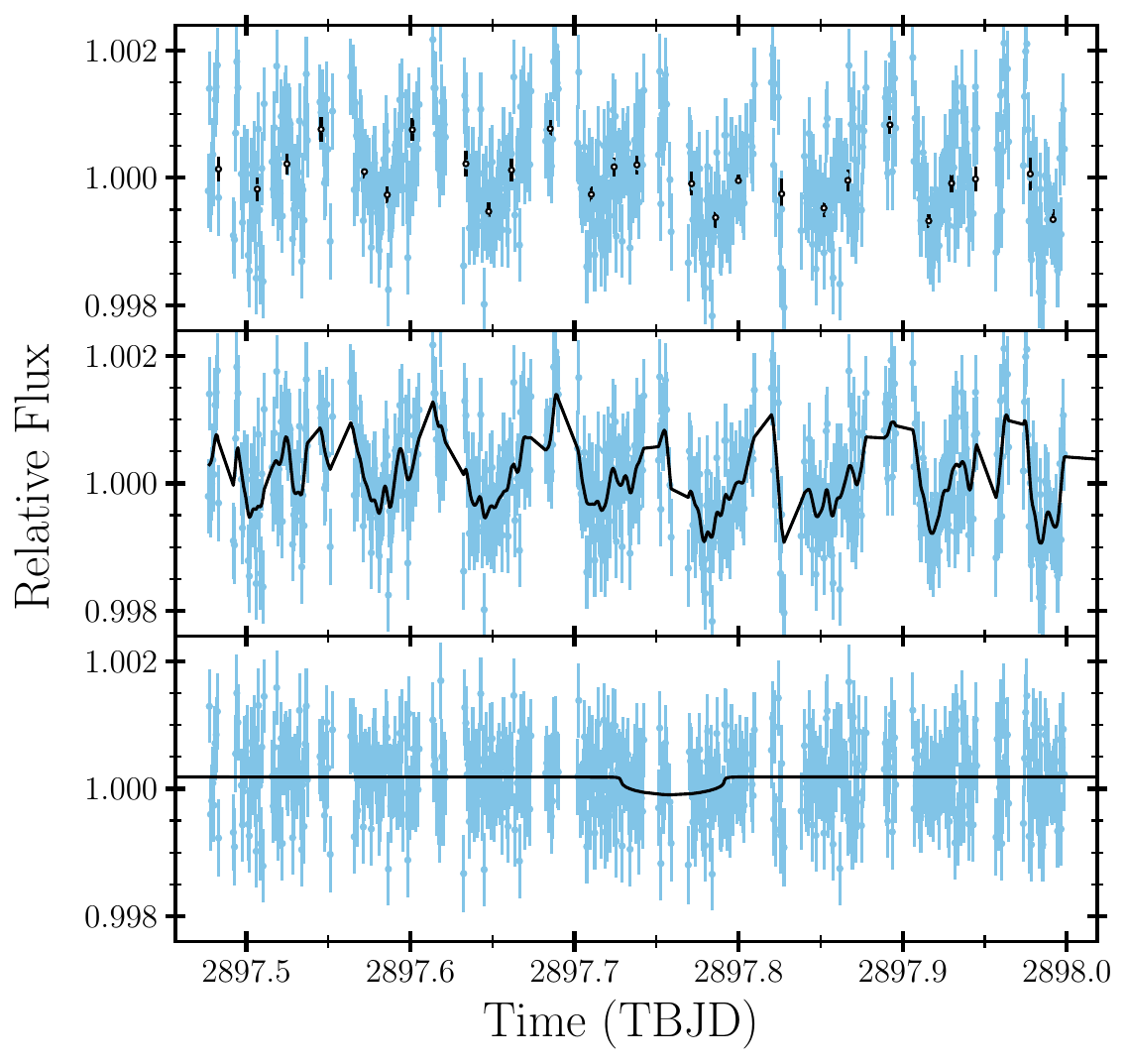}
    \quad
    \includegraphics[width=0.3\linewidth]{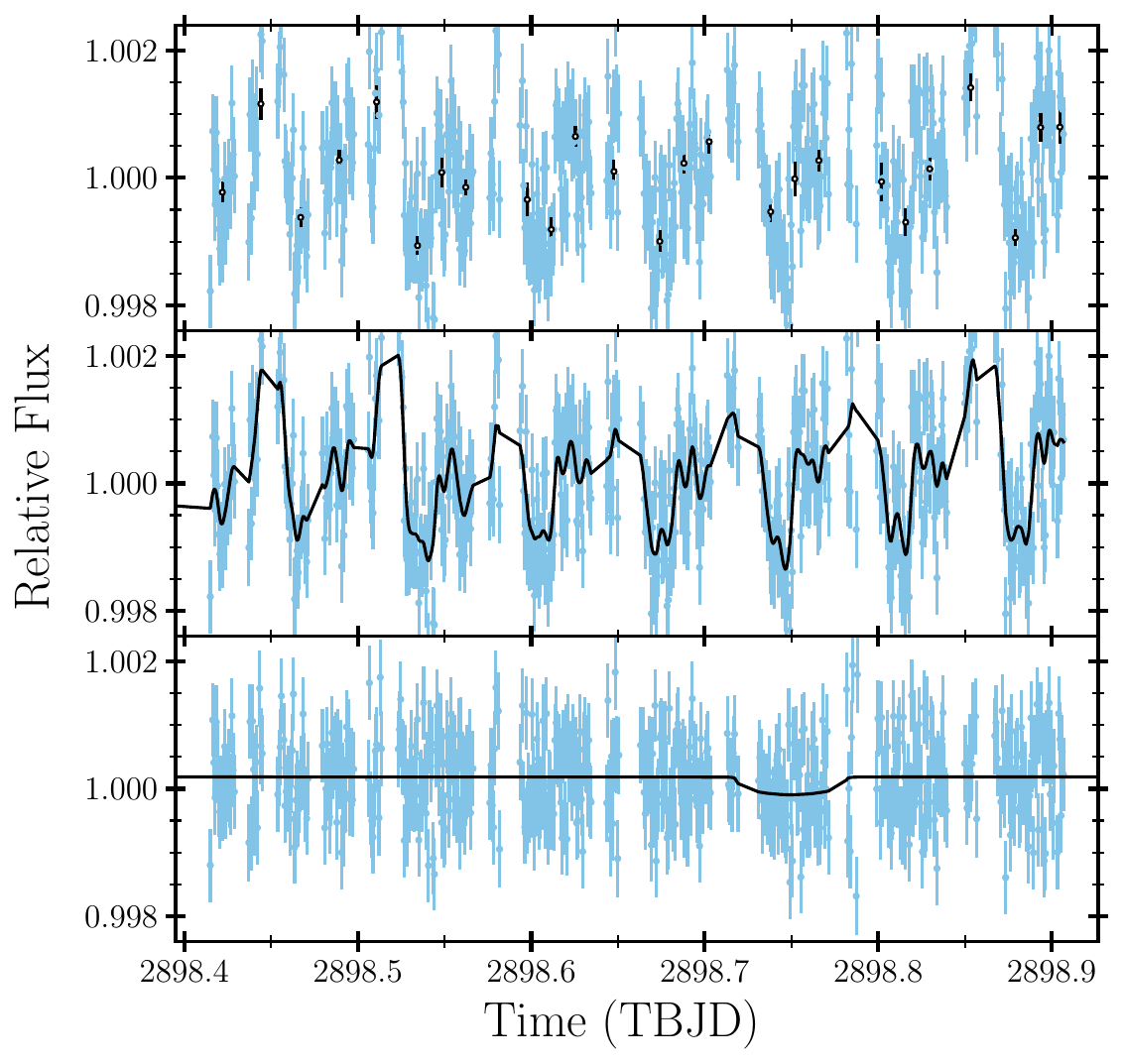}
    \quad
    \includegraphics[width=0.3\linewidth]{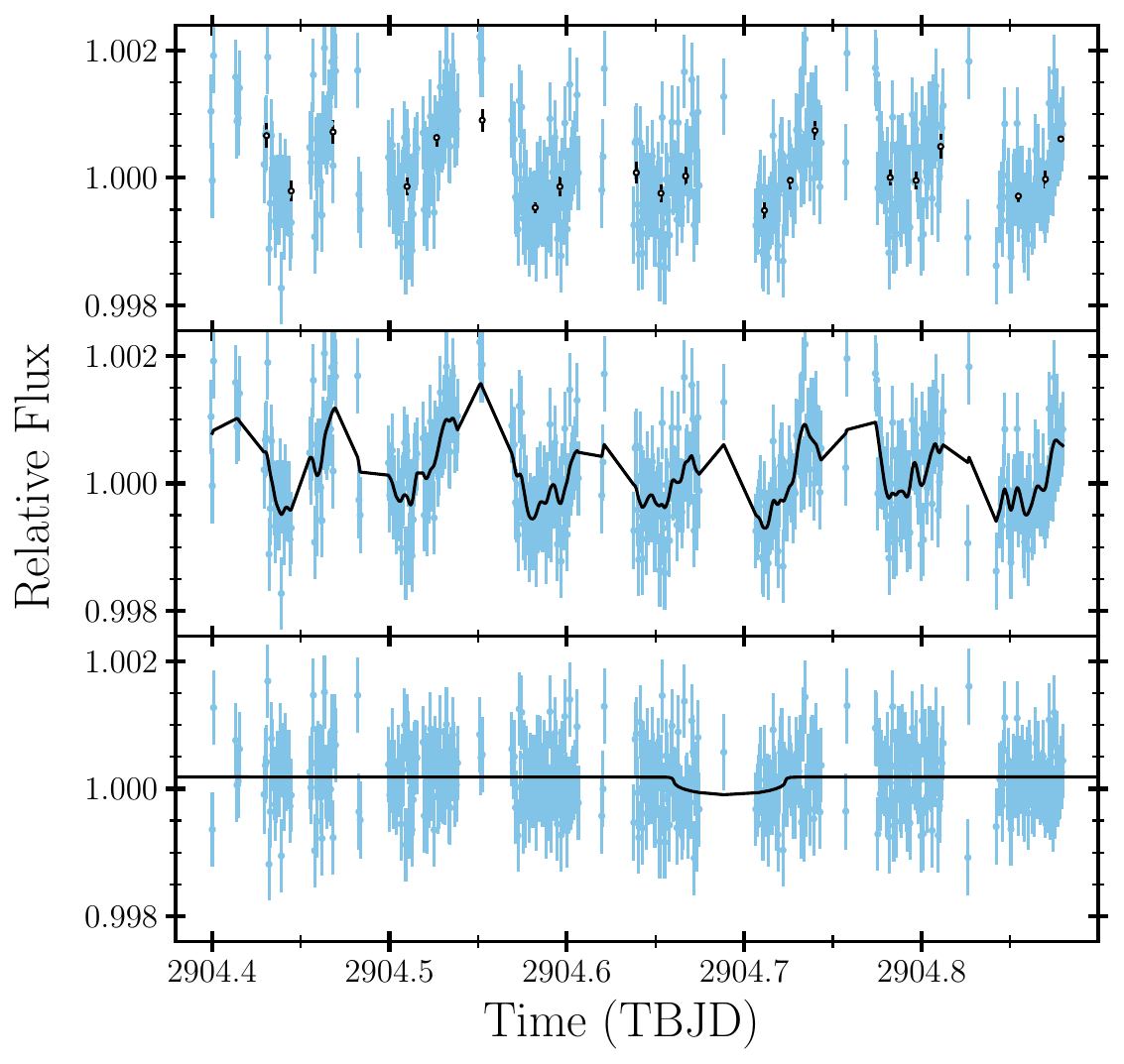}
    \caption{\textit{CHEOPS} visits within XGal. The top row shows the data, the second row the GP and transit model and the third row shows the transit model with the GP model removed from the data. }
    \label{fig:cheops120054}
\end{figure*}

\begin{figure*}
    \centering
    \includegraphics[width=0.3\linewidth]{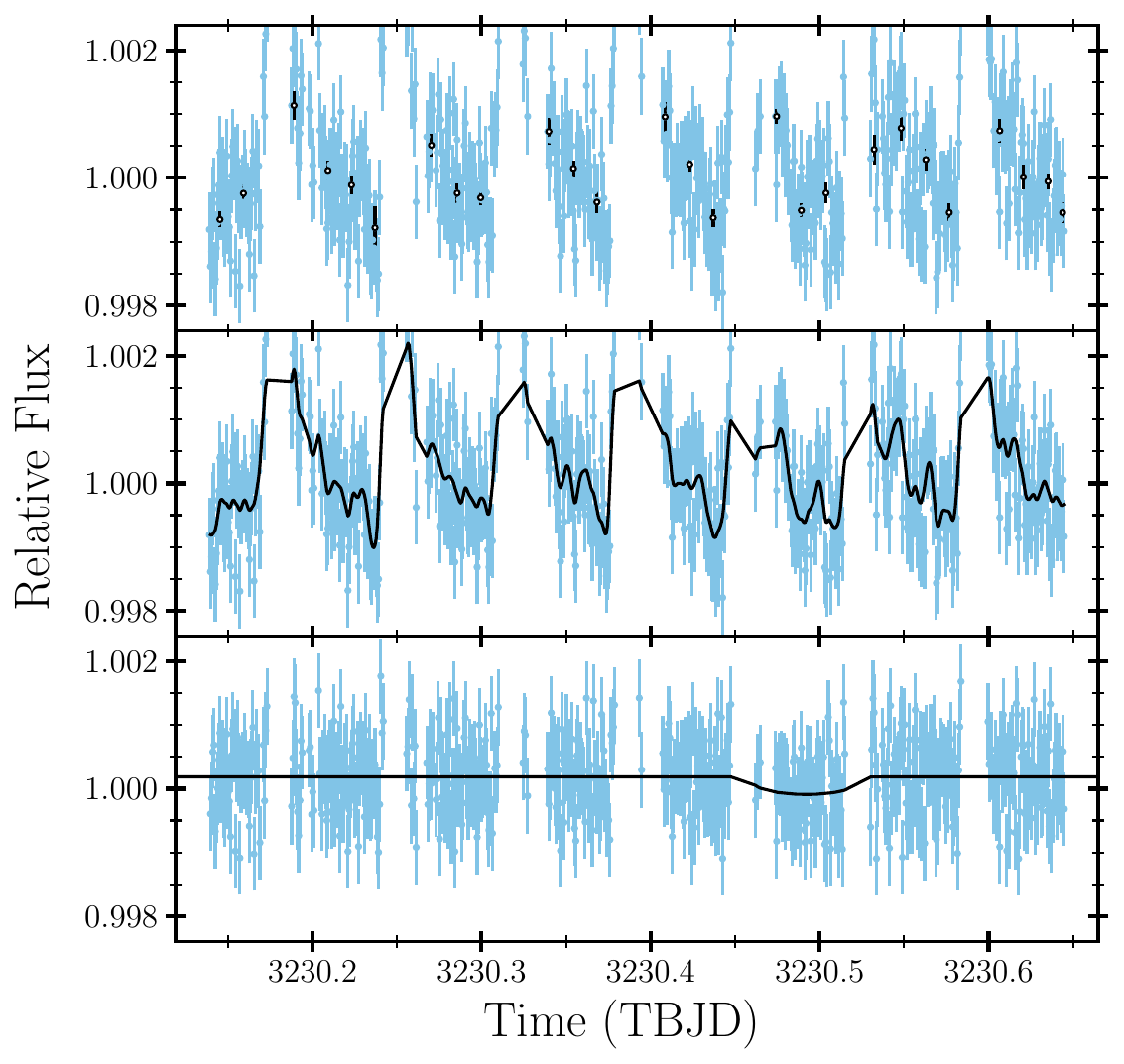}
    \quad
    \includegraphics[width=0.3\linewidth]{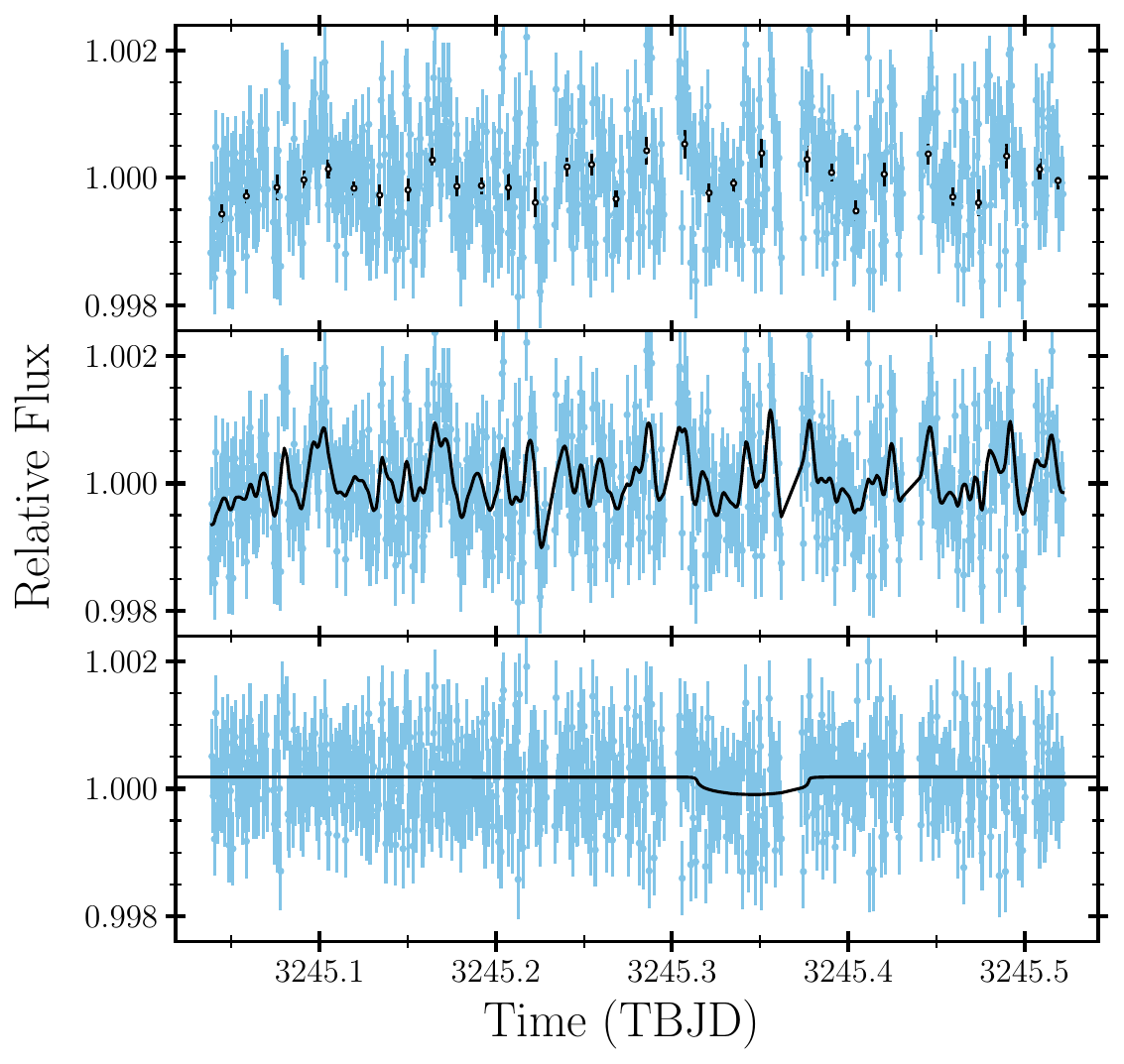}
    \quad
    \includegraphics[width=0.3\linewidth]{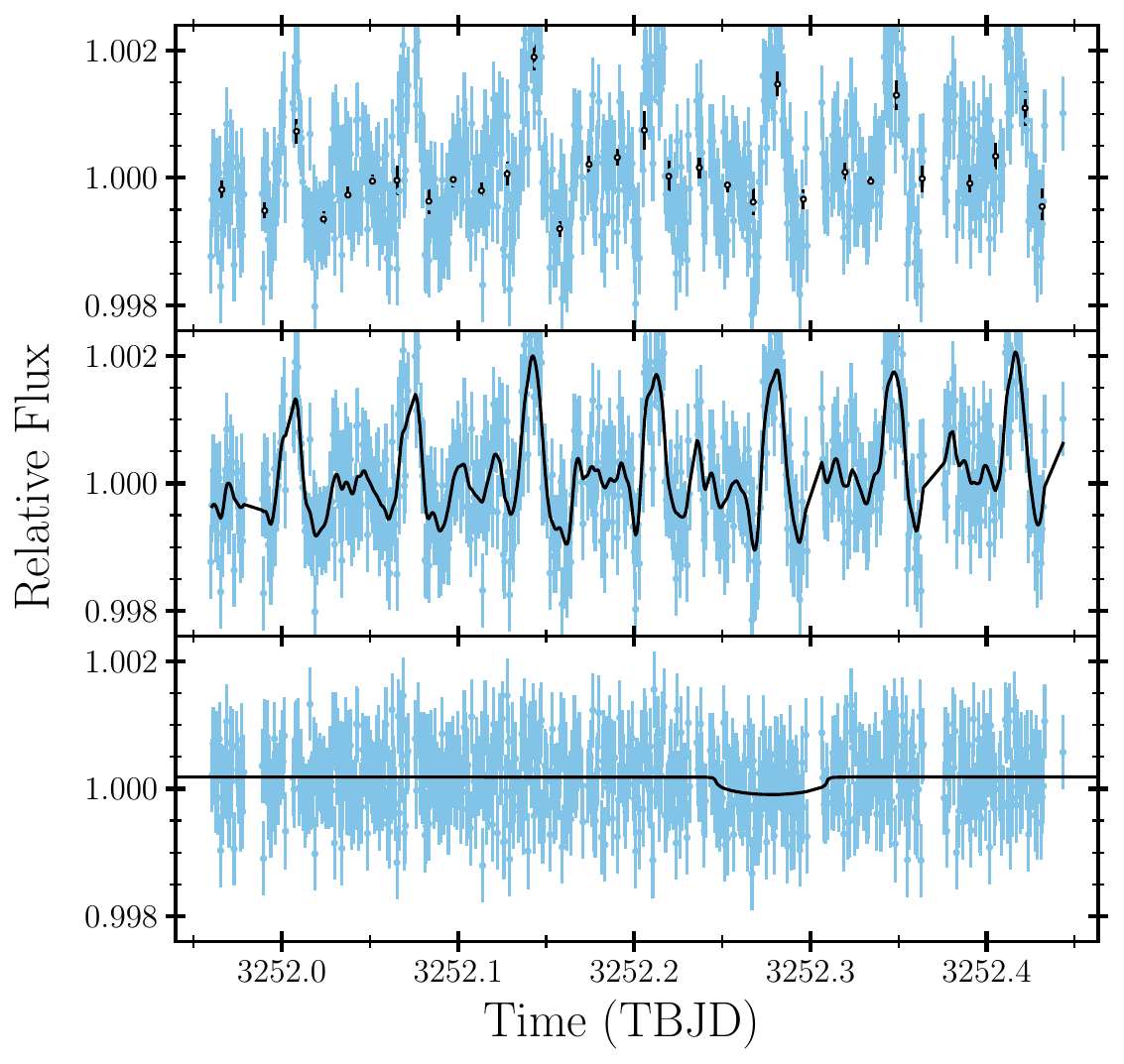}
    \quad
    \includegraphics[width=0.3\linewidth]{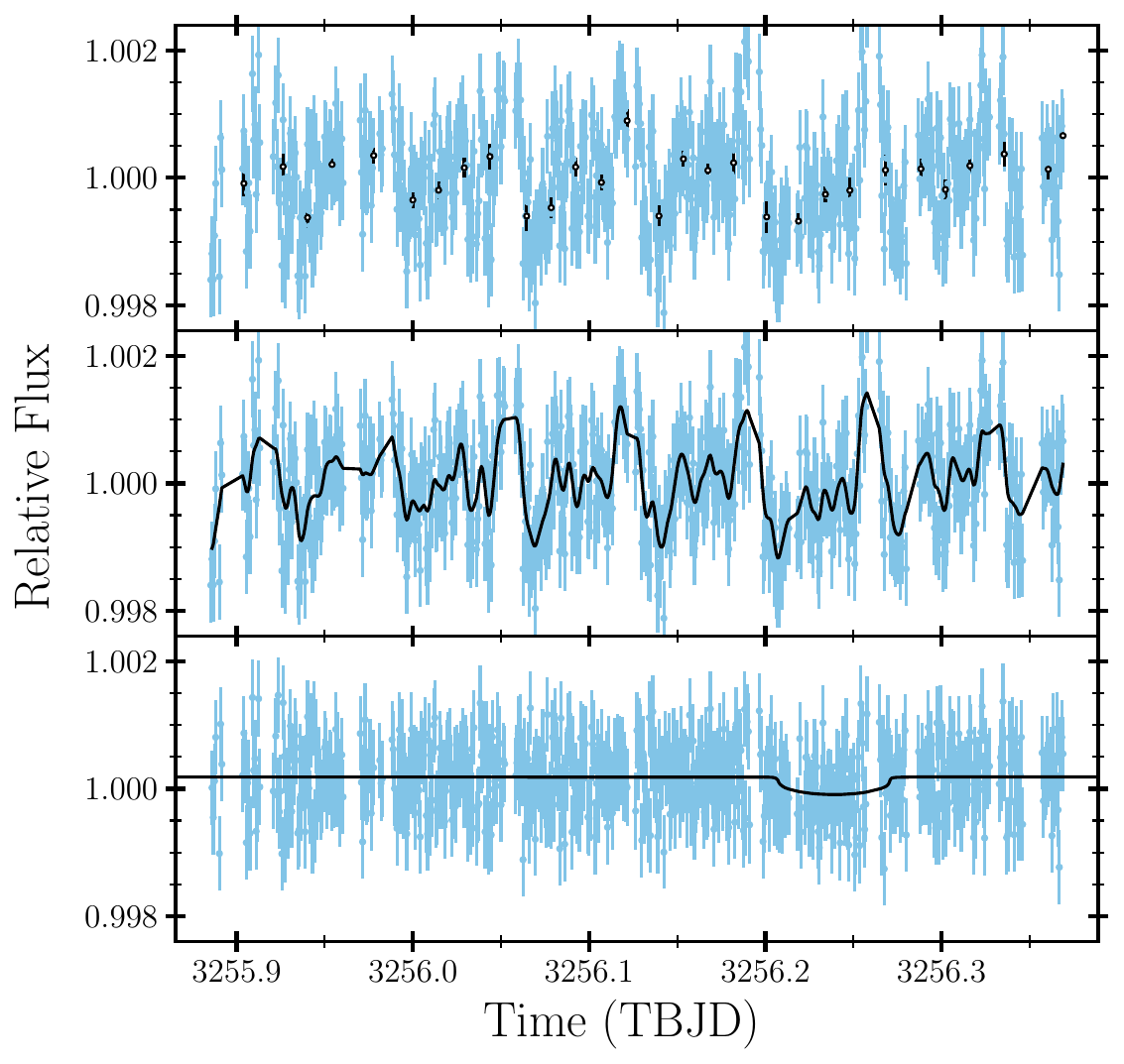}
    \quad
    \includegraphics[width=0.3\linewidth]{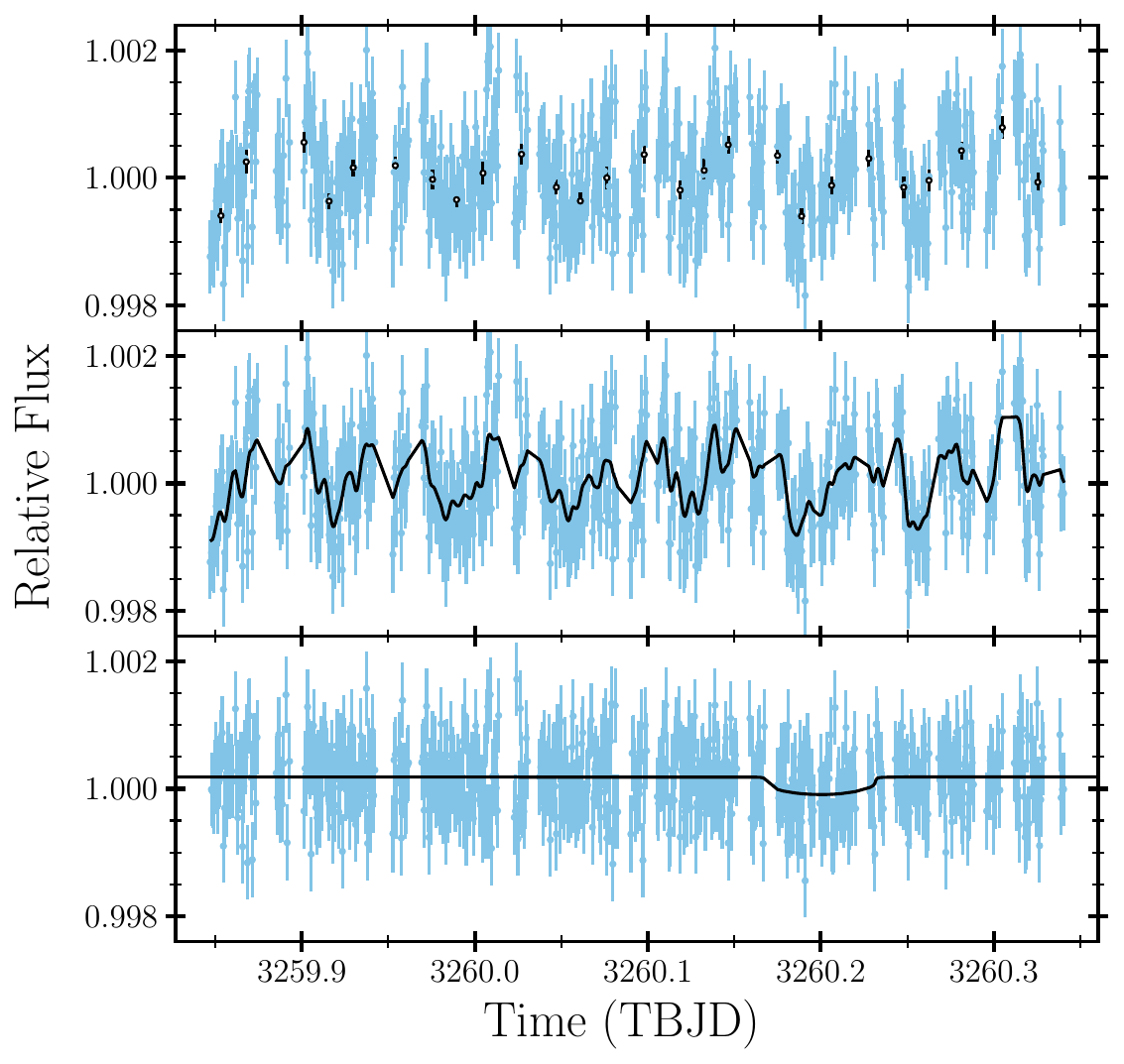}
    \quad
    \includegraphics[width=0.3\linewidth]{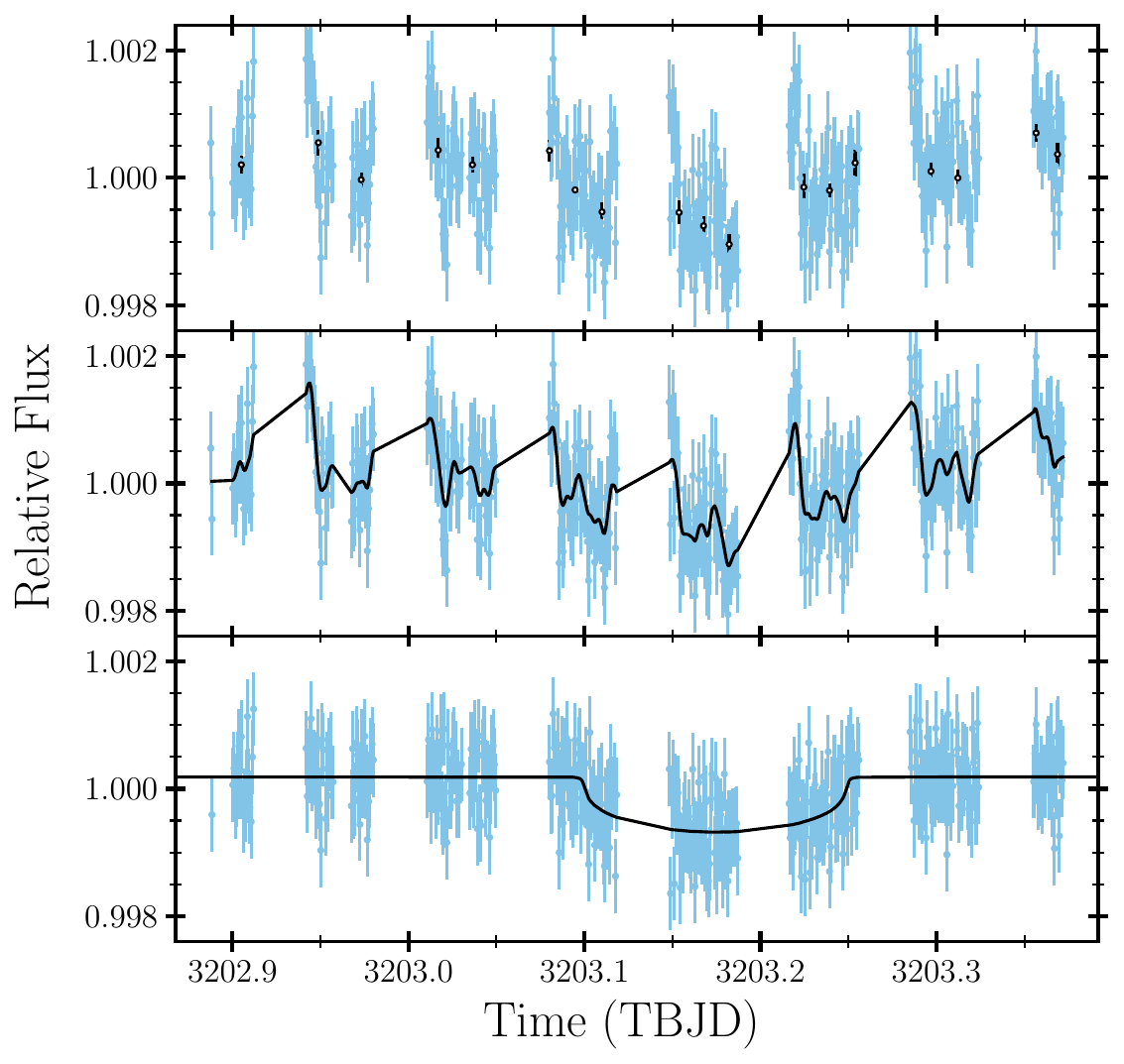}
    \quad
    \includegraphics[width=0.3\linewidth]{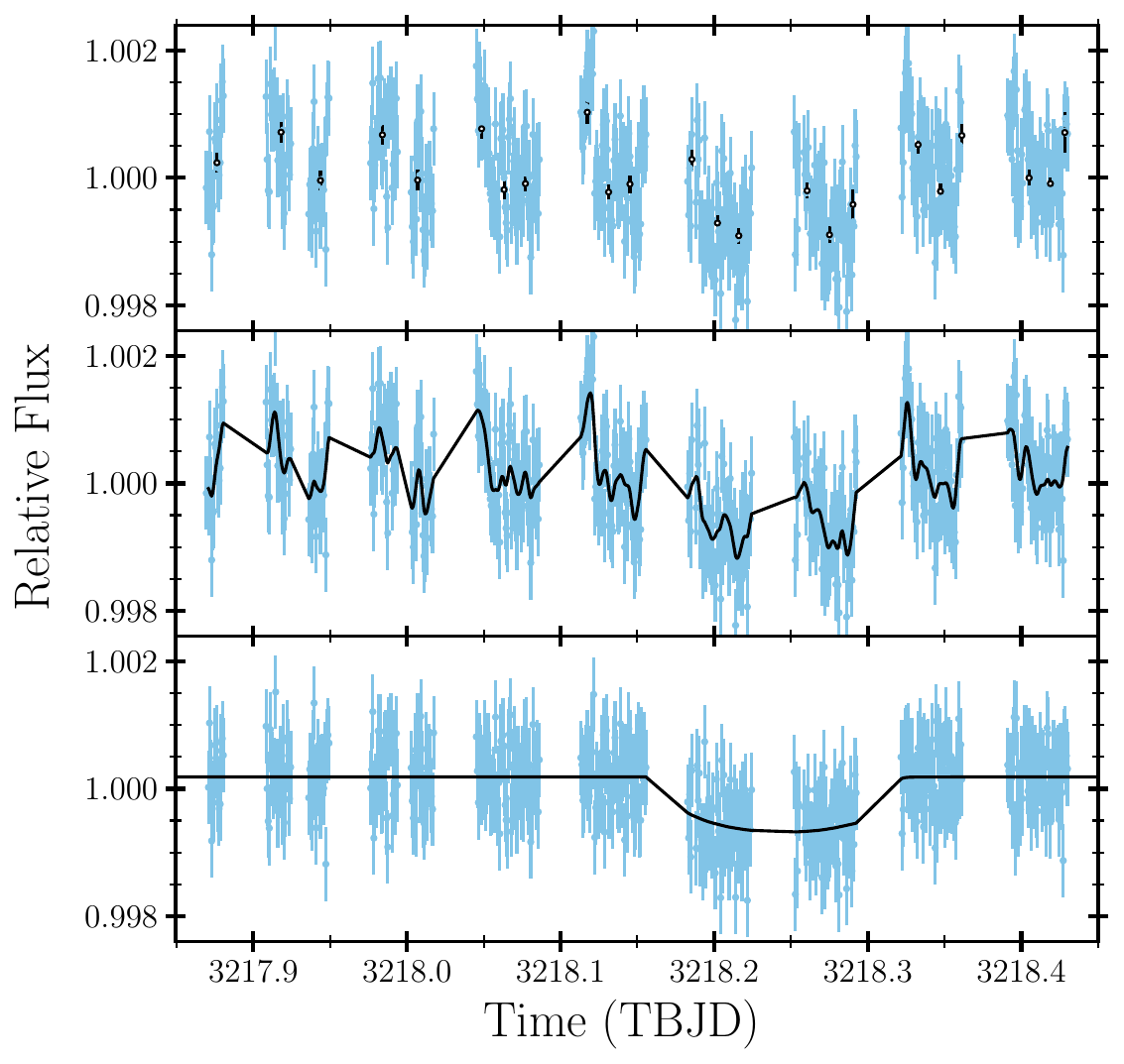}
    \quad
    \includegraphics[width=0.3\linewidth]{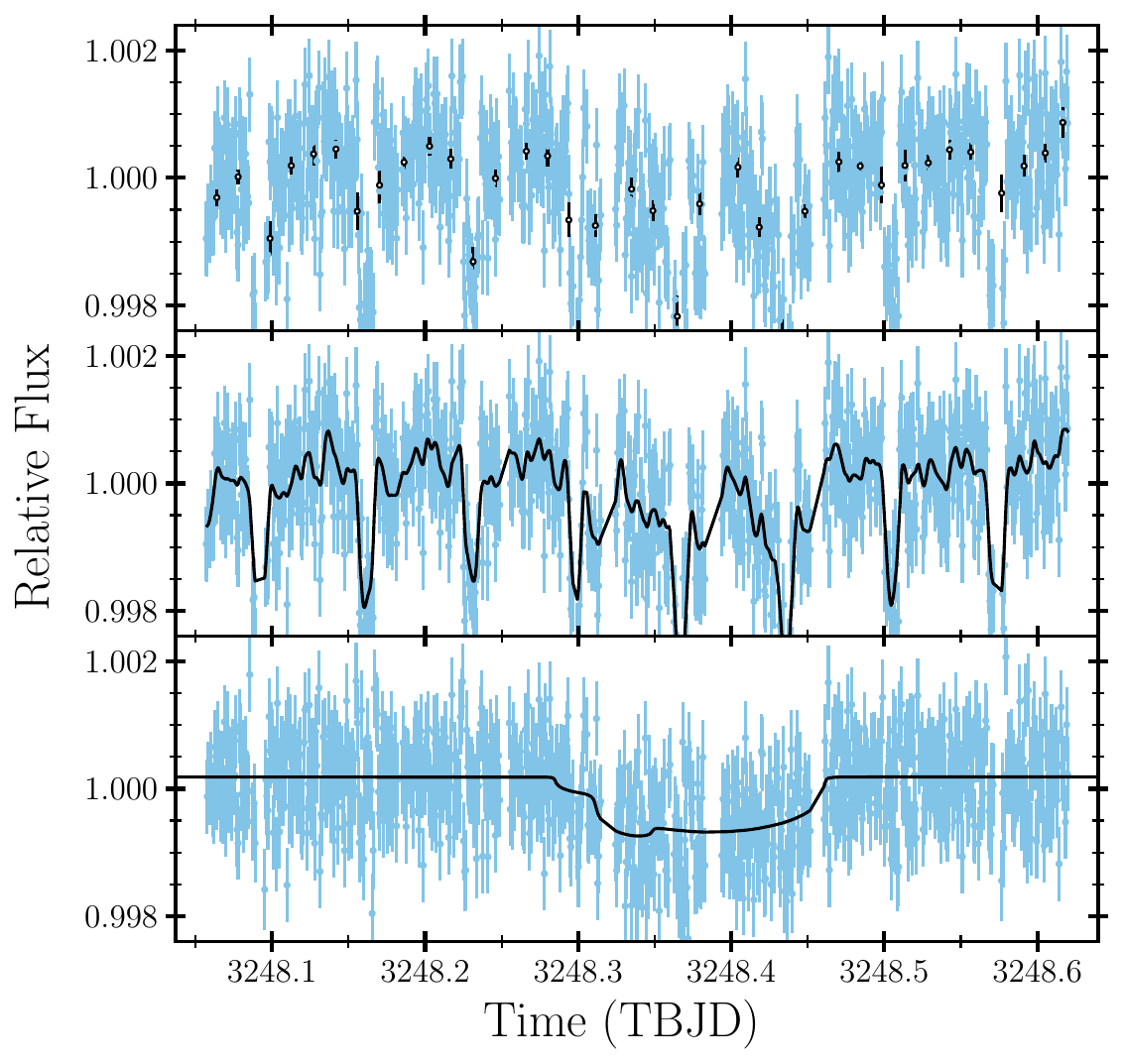}
    \quad
    \includegraphics[width=0.3\linewidth]{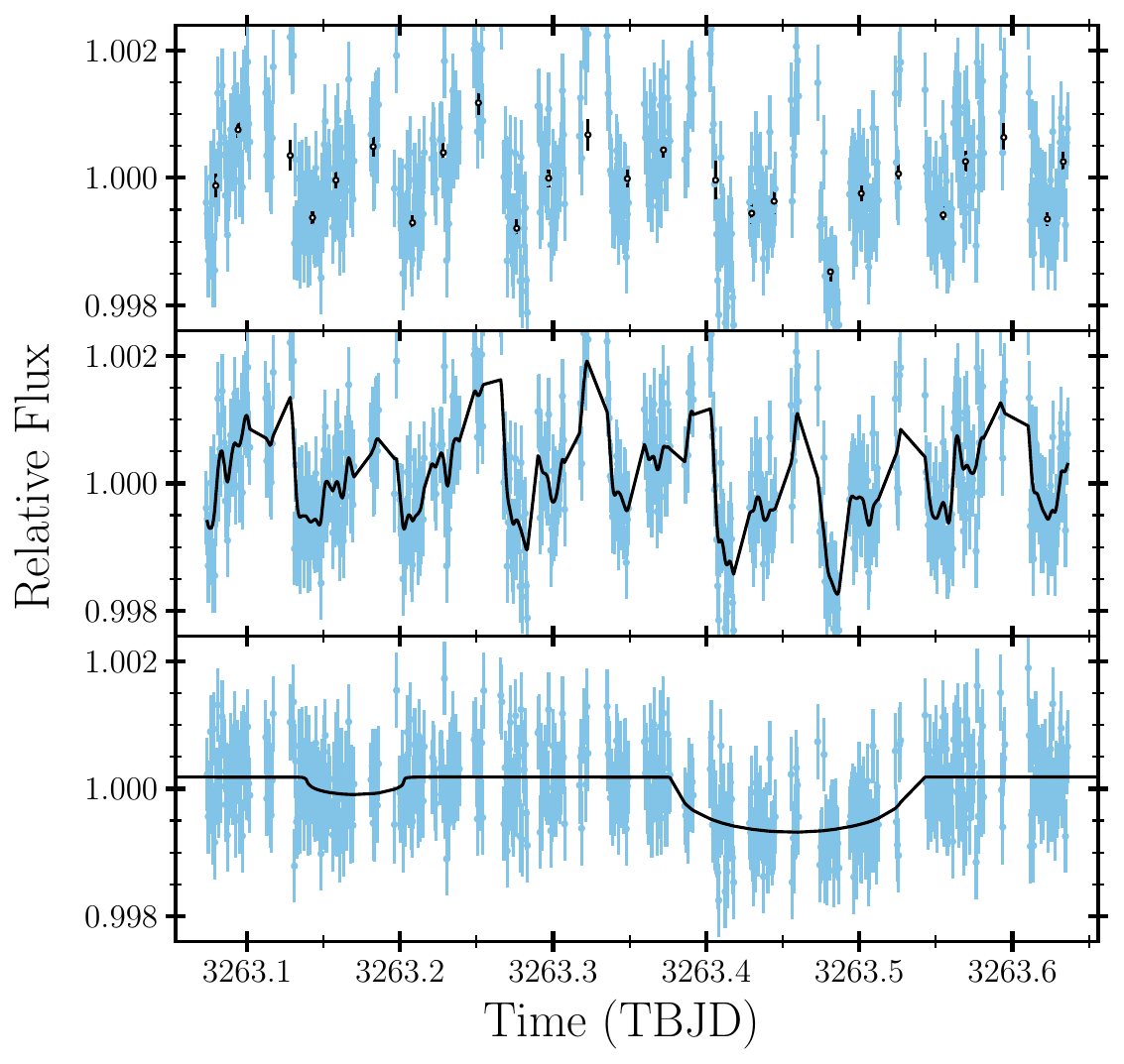}
    \quad
    \includegraphics[width=0.3\linewidth]{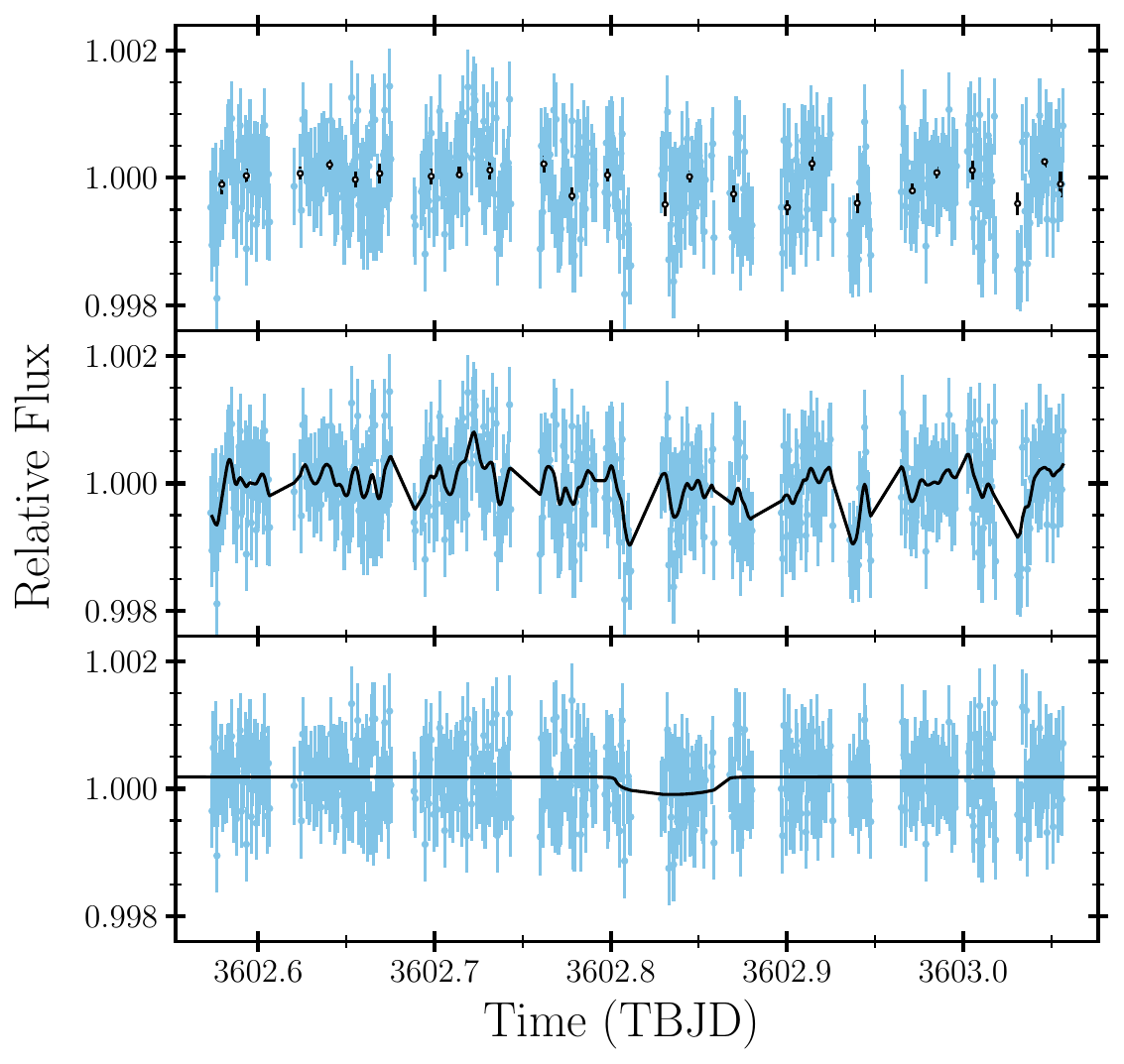}
    \quad
    \includegraphics[width=0.3\linewidth]{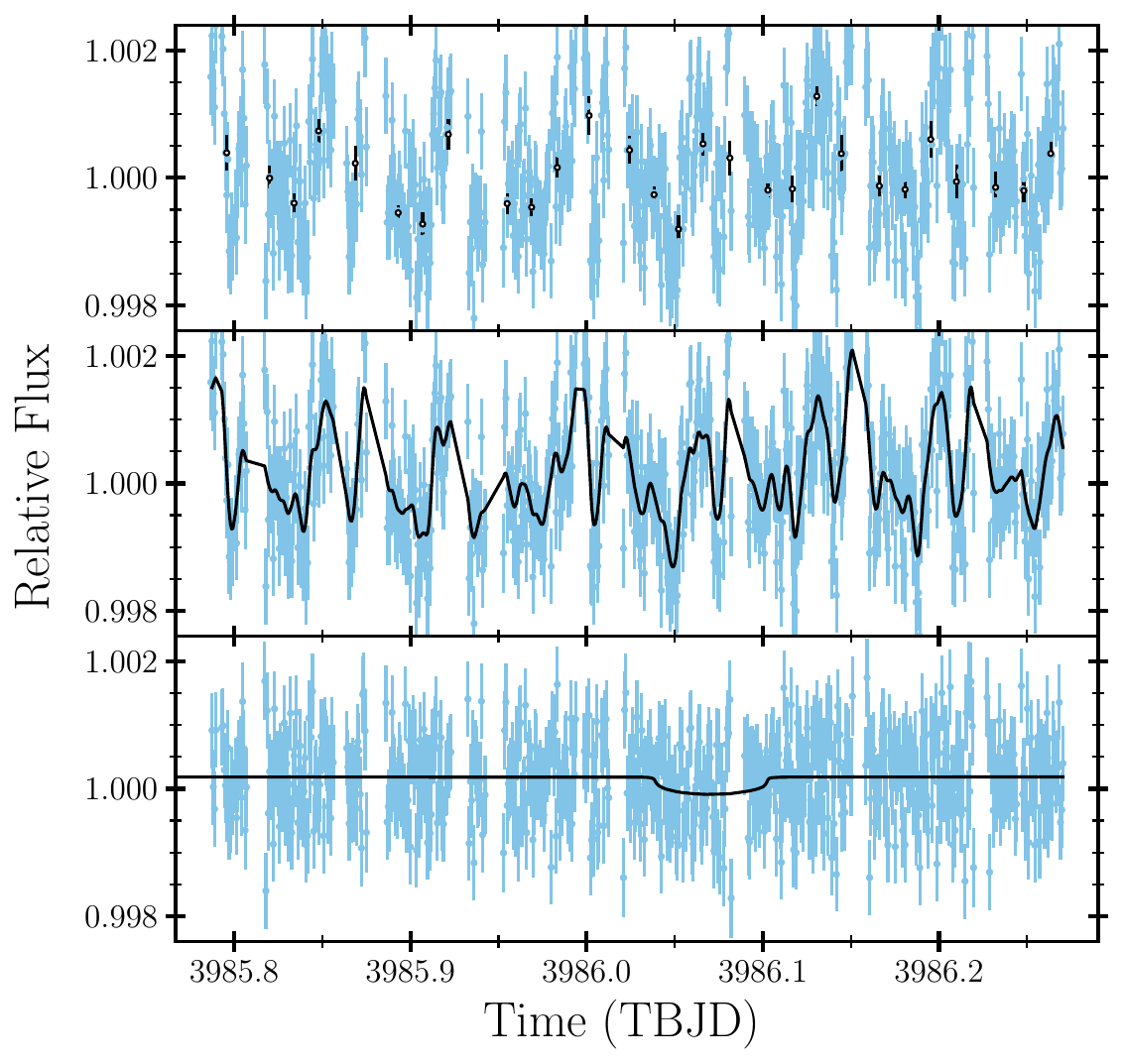}
    \caption{\textit{CHEOPS} visits within YGal. The top row shows the data, the second row the GP and transit model and the third row shows the transit model with the GP model removed from the data. }
    \label{fig:cheops140065}
\end{figure*}

\section{Long-Term Photometry}
\begin{figure*}
    \centering
    \includegraphics[width=\linewidth]{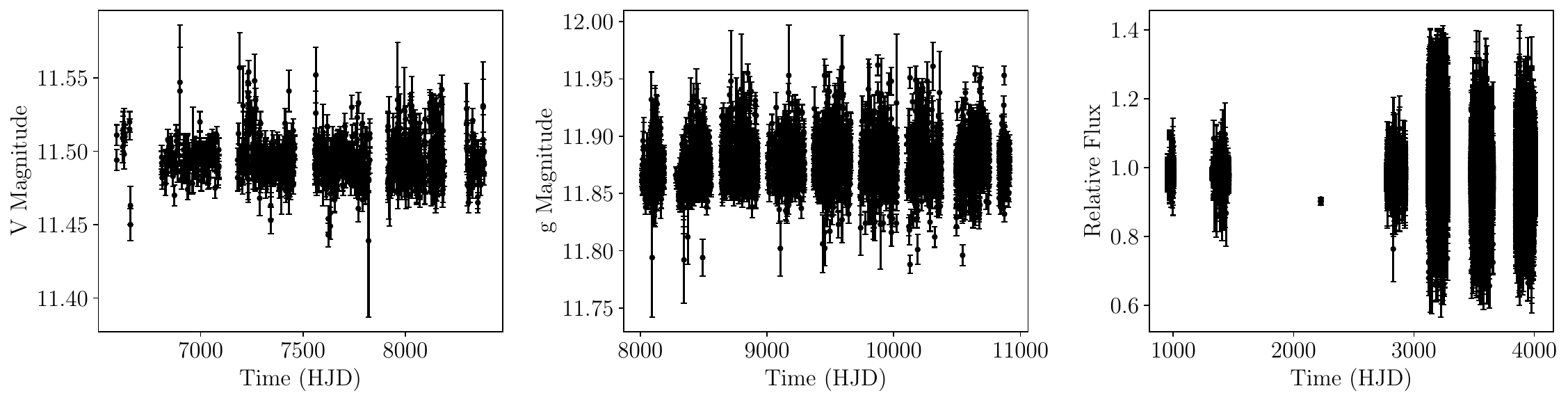}
    \caption{ASAS-SN and WASP photometry. Left: ASAS-SN V-band. Middle: ASAS-SN g-band. Right: WASP.}
    \label{fig:asas_sn_wasp_lightcurves}
\end{figure*}

\section{Priors and Fitted Instrumental and Stellar Parameters}
\begin{table*}
    \caption{Stellar and Instrumental Parameters of the joint fit in \textsc{juliet}. Uniform distributions are noted by $\mathcal{U}$, normal distributions by $\mathcal{N}$ and Log-uniform distributions by $\mathcal{L}$.}
    \centering
    \begin{tabular}{lccc}
    \hline \hline
        Parameter & Unit & Prior & Posterior  \\
        \hline
        $\rho_*$ & (kg/m$^3$) & $\mathcal{N}(1913,122)$ & $1920\pm110$ \\
        GP$_\rho$ \textit{TESS} & (d) & $\mathcal{L}(0.001,1000)$ & $0.70\substack{+0.18\\-0.15}$ \\
        GP$_\sigma$ \textit{TESS} & (ppm) & $\mathcal{L}(0.000001,1000000)$ & $0.0001052\substack{+0.0000102\\-0.0000095}$\\
        GP$_\rho$ \textit{CHEOPS} & (d) & $\mathcal{L}(0.001,1000)$ & $0.00440\substack{+0.00022\\-0.00019}$ \\
        GP$_\sigma$ \textit{CHEOPS} & (ppm) & $\mathcal{L}(0.000001,1000000)$ & $0.000648\substack{+0.000014\\-0.000015}$\\
        q$_1$ \textit{TESS} & - & $\mathcal{U}(0.0,1.0)$ & $0.51\substack{+0.29\\-0.26}$\\
        q$_2$ \textit{TESS} & - & $\mathcal{U}(0.0,1.0)$ & $0.51\substack{+0.30\\-0.31}$\\
        q$_1$ \textit{CHEOPS} & - & $\mathcal{U}(0.0,1.0)$ & $0.62\substack{+0.25\\-0.29}$\\
        q$_2$ \textit{CHEOPS} & - & $\mathcal{U}(0.0,1.0)$ & $0.52\substack{+0.29\\-0.31}$\\
        jitter \textit{TESS} & (ppm) & $\mathcal{L}(0.1,1000)$ & $1.6\substack{+7.9\\-1.4}$\\
        jitter \textit{CHEOPS} & (ppm) & $\mathcal{L}(0.1,100000)$ & $208\pm15$\\
        offset \textit{TESS} & - & $\mathcal{N}(0,0.1)$ & $-0.000022\substack{+0.000015\\-0.000014}$\\
        offset \textit{CHEOPS} & - & $\mathcal{N}(0,0.1)$ & $-0.000186\substack{+0.000022\\-0.000021}$\\
        jitter HARPS & (m/s) & $\mathcal{L}(0.001,100)$ & $2.04\substack{+0.72\\-0.59}$\\
        offset HARPS & (m/s) & $\mathcal{U}(52000,52500)$ & $52240.62\pm0.52$\\
        \hline \hline
    \end{tabular}
    \label{tab:instr_fit}
\end{table*}

\bsp	
\label{lastpage}
\end{document}